\documentclass[final,twocolumn,twoside,journal]{IEEEtran}

\usepackage{cite,acronym}
\usepackage{amssymb,psfrag,graphicx,subfigure,textcomp,bm, mathrsfs}
\usepackage{dsfont,stfloats, url}
\usepackage[dvips]{color}
\usepackage{psfrag}
\usepackage[cmex10]{amsmath}

\usepackage{multirow}
\usepackage[table]{xcolor}

\IEEEoverridecommandlockouts

\newcommand\V[1]  { \mathbf{#1} }
\newcommand\B[1]  { \boldsymbol{#1} }
\newcommand\tr {\text{tr}}
\newcommand\T  { ^\text{T} }

\newcommand\NA  { { \mathcal{N}_\text{a} } }
\newcommand\Na  { { N_\text{a} }}
\newcommand\NB { { \mathcal{N}_\text{b} } }
\newcommand\Nb { { {N}_\text{b} } }

\newcommand\NT  { { \mathcal{N}_\text{t} } }
\newcommand\Nt  { { N_\text{t} }}
\newcommand\NR { { \mathcal{N}_\text{r} } }
\newcommand\Nr { { {N}_\text{r} } }
\newcommand\PW { \V{x} }

\newcommand\Pact { \mathscr{P}_{\hspace{-0.3mm}\text{A}}}
\newcommand\Ppas { \mathscr{P}_{\text{P}}}

\newcommand\PactR { \mathscr{P}_{\hspace{-0.3mm}\text{A-R}}}
\newcommand\PpasR { \mathscr{P}_{\text{P-R}}}

\newcommand\PactRu { \overline{\mathscr{P}}^M_{\hspace{-0.4mm}\text{A-R}}}
\newcommand\PactRl { \underline{\mathscr{P}}^M_{\hspace{0mm}\text{A-R}}}

\newcommand\wPact { \widetilde{\mathscr{P}}_{\hspace{-0.3mm}\text{A}} }
\newcommand\wPactR { \widetilde{\mathscr{P}}_{\hspace{-0.3mm}\text{A-R}} }

\newcommand\Ang { \mathcal{B}}
\newcommand\Chn { \mathcal{C}}

\newcommand\matfigscale {1}

\newtheorem{lemma}{Lemma}
\newtheorem{proposition}{Proposition}
\newtheorem{remark}{Remark}

\acrodef{SOCP}{second-order cone program}
\acrodef{SDP}{semidefinite program}
\acrodef{RV}{random variable}
\acrodef{FIM}{Fisher information matrix}
\acrodef{EFIM}{equivalent Fisher information matrix}
\acrodef{SPEB}{squared position error bound}
\acrodef{WNL}{wireless network localization}
\acrodef{RNL}{radar network localization}
\acrodef{TOA}{time-of-arrival}
\acrodef{RSS}{received signal strength}
\acrodef{RF}{radio frequency}
\acrodef{PAV}{power allocation vector}

\acrodef{RCS}{radar cross section}
\acrodef{RC}{ranging coefficient}
\acrodef{ERC}{equivalent ranging coefficient}
\acrodef{LAR}{localization accuracy requirement}
\acrodef{SOC}{second-order cone}

\acrodef{LHS}{left-hand side}
\acrodef{RHS}{right-hand side}

\acrodef{MSE}{mean squared error}
\acrodef{mDPEB}{maximum direction position error bound}

\usepackage[level=0]{wgroup_message}

\newcommand{\paperTitle}{Power Optimization for Network Localization}

\begin{document}

% input catechism here
% { \onecolumn
% \input{Catechism.tex}
% }
% \twocolumn

% \title{Efficient and Robust Power Allocation for\\Wireless Network Localization}
% \title{Localization Power Optimization:\\A Unifying Framework}
% \title{A Unifying Framework for Localization Power Optimization}
% \title{A Unifying Power Optimization Framework for Active and Passive Localization}

% \title{A Unifying Framework for Optimal\\Active and Passive Localization}
% \title{Optimal Active and Passive Localization}
% \title{Robust Active and Passive Localization}
% \title{A Unifying Framework for Optimal Active and Passive Localization}

\title{\paperTitle}

\author{Yuan Shen,~\IEEEmembership{Student~Member,~IEEE,} Wenhan Dai,~\IEEEmembership{Student~Member,~IEEE,} and Moe Z.~Win,~\IEEEmembership{Fellow,~IEEE}
\thanks{Manuscript received November 23, 2012; revised June 14, 2013; accepted July 17, 2013; approved by {\scshape IEEE/ACM Transactions on Networking} Editor Y.~Liu. Date of publication Month Date, 2013; date of current version Month Date, 2013. This research was supported, in part, by 
the Air Force Office of Scientific Research under Grant FA9550-12-0287,
			the Office of Naval Research under Grant N00014-11-1-0397, 
			and
			the MIT Institute for Soldier Nanotechnologies.} 
\thanks{The authors are with the Laboratory for Information and Decision Systems (LIDS), Massachusetts Institute of Technology, 77 Massachusetts Avenue, Cambridge, MA 02139 USA (e-mail: \{shenyuan, whdai, moewin\}@mit.edu).}
\thanks{Color versions of one or more of the figures in this paper are available online at http://ieeexplore.ieee.org.}
\thanks{Digital Object Identifier 10.1109/TNET.20xx.xxxxxxx}
}

\maketitle 
%\IEEEpeerreviewmaketitle

\setcounter{page}{1}

% \markboth{PLEASE DO NOT DISTRIBUTE WITHOUT THE WRITTEN CONSENT OF THE AUTHORS}{Shen \emph{\MakeLowercase{et al.}}: Network Power Allocation for Active and Passive Localization}

\markboth{IEEE/ACM Transactions on Networking, Vol.~X, No.~Y, Month~2013}{Shen \emph{\MakeLowercase{et al.}}: \paperTitle}

\acresetall

% range-based localization 

\begin{abstract}
% Wireless network localization will enable numerous location-based applications in various sectors.
Reliable and accurate localization of {mobile} objects is essential for many applications in wireless networks. {In range-based localization, the position of the object can be inferred using the distance measurements from wireless signals exchanged with active objects or reflected by passive ones.} 
% with respect to reference nodes.
% The most common approach for network localization is to determine the object positions using distance measurements with respect to reference nodes, 
% where the measurements can be obtained by signals either exchanged with active objects or reflected by passive objects. 
{Power allocation for ranging signals} is important since it affects not only network lifetime and throughput but also localization accuracy. In this paper, we establish a unifying optimization framework for power allocation in both active and passive localization networks.
% through the examples of wireless network localization (WNL) and multi-radar target localization (MTL).
%Our framework unifies the power optimization for active and passive localization. 
% conventionally recognized
In particular, we first determine the functional properties of the localization accuracy metric, which enable us to transform the power allocation problems into \acp{SOCP}. We then propose the robust counterparts of the problems in the presence of parameter uncertainty and develop asymptotically optimal and efficient near-optimal \ac{SOCP}-based algorithms.
% develop a sequence of SOCPs that yield asymptotically optimal solutions. Moreover, we develop efficient algorithms for the robust case based on relaxation methods. 
Our simulation results validate the efficiency and robustness of the proposed algorithms.
% in terms of the topology matrix expressed in a low dimensional subspace. 
% present an optimization formulation to minimize the total transmit power while achieving given localization accuracy requirement. 

% Based on this property, we prove that the formulation is a second-order conic program (SOCP). 

% Moreover, we develop a robust scheme to tackle the uncertainties in network parameters. 
% power allocation schemes to improve localization accuracy
% investigate the power allocation problem for network localization to improve localization efficiency. We first present an optimization formulation that minimizes the transmit power subject to given accuracy requirement, and prove that the formulation is a second-order conic program (SOCP). We then propose a robust counterpart to tackle the uncertainties in network parameters. Our simulation results validate the efficiency and robustness of the proposed method.
\end{abstract}

\acresetall


\acresetall

\section{Introduction}\label{sec:introduction}

\mynote{Location-awareness is important}

% PahLiMak:02,

\begin{figure}[t]
	\vspace{2mm}
	\hspace*{3mm}
	
	% \psfrag{WNL}[c][][0.7]{(a) Wireless Network Localization}
	% 	\psfrag{MRL}[c][][0.7]{(b) Radar Network Localization}
	% 	\psfrag{Anchor}[c][][0.6]{\hspace{-1mm} Anchor}
	% 	\psfrag{Agent}[c][][0.6]{\hspace{-4mm} Agent}
	% 	
	% 	\psfrag{Target}[c][][0.6]{\hspace{-1mm} Target}
	% 	\psfrag{Tx}[c][][0.6]{\hspace{0mm} Tx}
	% 	\psfrag{Rx}[c][][0.6]{\hspace{2mm} Rx}
	
	\includegraphics[width=1\columnwidth,draft=false]{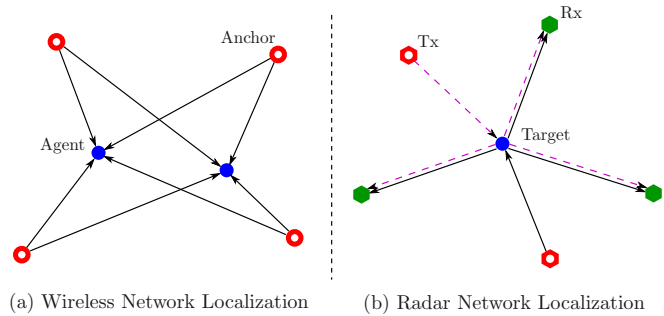}
	\vspace*{-6mm}
	\caption{Network deployments for \ac{WNL} and \ac{RNL}: (a) wireless network with four anchors and two agents; (b) radar network with two transmit and three receive antennas. %Both 1-agent and 2-agent networks are considered.
	}\label{fig:wnl_mtl_fig}
\end{figure}

\IEEEPARstart{N}{etwork localization} of active and passive objects is essential for many location-based applications in commercial, military, and social sectors \cite{WinConMazSheGifDarChi:J11, GezTiaGiaKobMolPooSah:05, SayTarKha:05, CafStu:98, GodHaiBlu:10, GodPetPoo:11, DarDErRobSibWin:J10 , PaoGioChiMinMon:08}. Contemporary localization techniques can be classified into two main categories, i.e., range-based and range-free techniques. The former locate the object using distance/angle measurements \cite{WinConMazSheGifDarChi:J11, GezTiaGiaKobMolPooSah:05, SayTarKha:05, CafStu:98}, and the latter using connectivity or fingerprint information \cite{BahPad:00, ShaRumZhaFro:04, HeHuaBluStaAbd:03, LiLiu:10}. {Compared to range-free ones}, range-based techniques are more suited and hence widely employed for high-accuracy localization despite the hardware complexity.
% Range-based techniques determine the positions of the objects through ranging with respect to reference nodes that have known positions. 
Active or passive localization refers to the rang-based techniques that utilize distance/angle measurements from wireless signals exchanged with active objects or reflected by passive ones, respectively. Two corresponding examples are \ac{WNL} \cite{WinConMazSheGifDarChi:J11, GezTiaGiaKobMolPooSah:05, SayTarKha:05, CafStu:98} and \ac{RNL} \cite{DarDErRobSibWin:J10, GodHaiBlu:10, GodPetPoo:11, PaoGioChiMinMon:08} (see Fig.~\ref{fig:wnl_mtl_fig}).

% to add afterwards
% HaiBluCim:08, 

\mynote{wireless network localization}

% KhaKarMou:09, YuMonRabCheOpp:06, ChiConVer:01, FonNicVan:10

Wireless networks have been employed for active localization since they are capable of providing accurate position information in GPS-challenged environments \cite{GezTiaGiaKobMolPooSah:05, WinConMazSheGifDarChi:J11, CafStu:98, SayTarKha:05,  KhaKarMou:10, VerDarMazCon:B08, ConDarZua:08, SheMazWin:J12, DenPieAbo:06, RabOppDen:06}. Locating a mobile node (agent) in such networks can be accomplished by using the range measurements between the agent and nodes at known positions (anchors). The ranges can be estimated from the \ac{TOA} or \ac{RSS} of the signals transmitted from the anchors to the agent \cite{SheWin:J10a, SheWymWin:J10, MazLorBah:J10, JouDarWin:J08, QiKobSud:06,  DarConFerGioWin:J09}. Localization accuracy in wireless networks is determined by the network topology and the accuracy of the range measurements, where the latter depends on the signal  bandwidth, channel condition, and transmit power \cite{SheWin:J10a}.
%\footnote{More precisely, the accuracy of the inter-node range measurements depends on , , and multipath propagation condition \cite{SheWin:J10a}.} 
Hence, power allocation in \ac{WNL} is important not only for the conventionally recognized lifetime and throughput \cite{MesPooSch:07} but also for agent localization accuracy.

% , GorKinKymRubWanZus:10

\mynote{radar network localization}
% BliFor:10, , CasNanDar:11, BarCon:11, BliForDavFawRabHorKra:09, BarGueCon:11

% HaiBluCim:08,

Radar networks have been studied for passive localization since they can enhance target detection and localization capability by exploiting the spatial diversity of target's \ac{RCS} \cite{Che:98, GodPetPoo:11, FisHaiBluCimChiVal:06, DarDErRobSibWin:J10, BliFor:03,  ForBli:B08, LiSto:07,  GodHaiBlu:10,  BarConGio:10}.\footnote{The positions of antennas in radar networks can vary from collocated to widely separated.} The target can be located using the \ac{TOA} range measurements from the transmit to receive antennas via the reflection of the target. Localization accuracy in radar networks depends on the network topology, signal bandwidth, target \ac{RCS}, and transmit power \cite{GodPetPoo:11}. Hence, power allocation and management in \ac{RNL} is crucial not only for low-probability-of-intercept (LPI) capability \cite{Sko:B00} but also for target localization accuracy.%, especially in the case of mobile platforms operating in long duration missions. 

% Moreover, transmit power management is crucial for many military operations in hostile environments due to strict low-probability-of-intercept (LPI) requirement \cite{Sko:B00}.

% which is subject to limited power resource.
% This implies the importance of power management since it affects the target localization accuracy and efficiency. 

% Wireless network localization is a key enabler for reliable and high-accuracy location-based applications and services in commercial, military, and social sectors \cite{PahLiMak:02, GezTiaGiaKobMolPooSah:05, SheWin:J10a}. 

% especially where mobile nodes are subject to limited power resources,

{The main task of power allocation for network localization is to achieve the optimal tradeoffs between localization accuracy and energy consumption.} Such a task is commonly accomplished using optimization methods, which have played a {significant} role in maximizing communication and networking performance under limited resources \cite{AhuMagOrl:93, KelMauTan:98, BoyVan:B04, LuoYu:06, Fos:01, She:01, BerBroCar:11}. One can formulate the power allocation problem for network localization by {constraining} either the localization error or total transmit power and minimizing the other.
% Such a task can be formulated as optimization problems, where one can fix either the localization error or total transmit power to minimize the other.
%; for instance, one may minimize the total transmit power for a given localization error, or minimize the localization error for a given total transmit power. These optimization problems inevitably 
Solving these problems requires the knowledge of network parameters, which in practice are subject to uncertainty. Ignoring such uncertainty will lead to sub-optimal or even infeasible solutions \cite{BerBroCar:11, GhaOusLeb:98, BenNem:98, QueWinChi:J10}. Hence, two fundamental questions related to power allocation in network localization are:
% Such a task can be formulated as optimization problems involving network parameters, which in practice are subject to uncertainty.
\begin{enumerate}
	\item how to minimize the total transmit power while satisfying the localization requirement;
	% minimize the transmit power by efficient allocation to achieve a given accuracy requirement?
	\item how to guarantee the localization requirement in the presence of parameter uncertainty.
\end{enumerate}
Answers to these questions {will provide insights into the essence of network localization and} enable the design of robust network algorithms.

% lead to practical power allocation schemes for localization networks as well as new insights for the design of robust network algorithms.

% insights into the robust design .
% for network localization, as well as provide insights into robust 
% facilitate the design of power allocation for efficient network localization. 

Several formulations have been proposed {for power allocation in different localization scenarios using the performance metrics based on the information inequality \cite{SheWin:C08a, LiSheZhaWin:J13, GodPetPoo:11, DaiSheWin:C12}.} For WNL, the optimal power allocation solution was determined in closed forms for specific network topologies \cite{SheWin:C08a} and was obtained by an \ac{SDP} for general network topologies \cite{LiSheZhaWin:J13}. For \ac{RNL}, a suboptimal power allocation algorithm was developed via a relaxation technique \cite{GodPetPoo:11}. Most studies assume perfect knowledge of the network parameters with the exception of \cite{LiSheZhaWin:J13}, in which a robust formulation was proposed to cope with small parameter uncertainty and a suboptimal solution was obtained through relaxation. However, the performance loss from the relaxation was not quantified since the optimal solution of the robust formulation remains unknown.

In this paper, we investigate the optimal power allocation problem for reliable and accurate network localization, aiming to minimize the total transmit power for a given localization requirement. Our work also encompasses robust counterparts to cope with the parameter uncertainty. 
% in the presence of parameter uncertainty
% develop a unifying optimization framework for power allocation in active and passive localization, . 
% transform the power allocation problem in the form of efficient second-order cone programs (SOCPs). 
%Our framework also encompasses robust optimization counterparts for the scenarios with parameter uncertainties. 
% In this paper, we first determine the property of the SPEB in terms of the topology matrix expressed in a low dimensional subspace. Based on the property, we prove that the power optimization problem can be equivalently transformed into a second-order conic program (SOCP), which is a subclass of SDP and has more efficient solvers than SDP \cite{BoyVan:04}. Moreover, we develop a robust scheme that retains the SOCP structure to tackle the network parameter uncertainties, where the robust formulation naturally reduces to the non-robust one when the uncertainties vanish.
The main contributions of this paper are as follows.
\begin{itemize}
	\item We establish a unifying {optimization} framework for power allocation in both active and passive localization networks %to optimize the power efficiency and localization accuracy, 
% which unifies for active and passive localization 
through \ac{WNL} and \ac{RNL} examples.
	\item We determine the functional properties of the localization accuracy metric {and transform} the power allocation problems into \acp{SOCP}.
	% \item We propose robust power optimization formulations also in the form of conic programs to tackle parameter uncertainties, which naturally reduce to their non-robust counterparts when the uncertainties vanish.
	\item We propose a robust power allocation formulation that guarantees the localization requirement in the presence of parameter uncertainty {over large ranges}. %and develop asymptotically optimal \ac{SOCP}-based algorithms. %, which converge sublinearly with the number of constraints. 
	% derive sequential bounds for the worst-case SPEB, 
	% and prove its convergence to the optimal solution.
	\item We develop {asymptotically optimal and efficient near-optimal \ac{SOCP}-based algorithms for the robust formulation, and characterize the convergence rate of the asymptotic algorithms to the optimal solution.} %and characterize their performance losses to the optimal solution.
	% The efficient algorithms retain the form of SOCPs and naturally reduce to their non-robust counterparts when the uncertainty vanishes. 
\end{itemize}

The rest of the paper is organized as follows. In Section \ref{sec:system_model}, we introduce the system models and formulate the power allocation problems. In Section \ref{sec:perfect_case}, we present the properties of the localization accuracy metric and show that the power allocation problems can be transformed into \acp{SOCP}. In Section \ref{sec:robust_case}, we present robust formulations for the case with parameter uncertainty and develop asymptotically optimal and efficient near-optimal algorithms. 
%In Section \ref{sec:efficient_algorithm}, we propose relaxation methods to develop efficient algorithms. 
In Section \ref{sec:discussion}, we give some comments and discussions on the results. Finally, the performance of the proposed algorithms is evaluated by simulations in Section \ref{sec:simulation}, and conclusions are drawn in the last section.

\emph{Notation}: %$\mathbb{E}_{\V{x}}\{\cdot\}$ denotes the expectation operator with respect to the random vector $\V{x}$; 
%$[\,\cdot\,]\T$, $[\,\cdot\,]^\dagger$, and $\tr\{\cdot\}$ denote the transpose, Hermitian transpose, and trace operation, respectively;
$\mathbb{S}_{+}^n$ denotes the set of $n\times n$ positive-semidefinite matrices; matrices $\V{A}\succeq\V{B}$ denotes that $\V{A}-\V{B}$ is positive semidefinite; vectors $\V{x}\succeq\V{y}$ denotes that all elements of $\V{x}-\V{y}$ are nonnegative; 
% $\|\V{x}\|$ denotes the Euclidean norm of the vector $\V{x}$; 
$\V{1}_n\in\mathbb{R}^n$ denotes a column vector with all 1's, $\V{0}_n\in\mathbb{R}^n$ denotes a column vector with all 0's, and $\V{I}_n \in\mathbb{R}^{n\times n}$ denotes an identity matrix, where the subscript $n$ will be omitted if clear in the context; vector $\V{u}(\phi):= [\,\cos\phi~\sin\phi\,]\T$; matrix $\V{J}_\text{r}(\phi) := \V{u}(\phi) \,\V{u}(\phi)\T$; and we define the functions 
\begin{align*}
	\V{c}(\B\phi) & := [\, \cos\phi_{1}\;\;\cos\phi_{2}\;\;\cdots\;\;\cos\phi_{n}\,]\T \\
	\V{s}(\B\phi) & := [\, \sin\phi_{1}\;\;\sin\phi_{2}\;\;\cdots\;\;\sin\phi_{n}\,]\T
\end{align*}
where $\B\phi = [\,\phi_1~~\phi_2~~\cdots~~\phi_n\,]\T$.

% \begin{figure*}
% 	[!b] \vspace*{4pt} 
% 	\hrulefill
% 	\normalsize%
% 	\setcounter{MYtempeqncnt}{\value{equation}} 
% 	\setcounter{equation}{5} 
% 	\begin{align}\label{eq:lambda_matrix}
%  		\B\Lambda_k = \Matrix{ccccc}{\sin^2(\phi_{k1}-\phi_{k1}) & \sin^2(\phi_{k1}-\phi_{k2}) & \cdots & \sin^2(\phi_{k1}-\phi_{k\Nb}) \\ 
% 		\sin^2(\phi_{k2}-\phi_{k1}) &  \sin^2(\phi_{k2}-\phi_{k2}) & \cdots & \sin^2(\phi_{k2}-\phi_{k\Nb}) \\ 
% 		\vdots  & &\ddots \\
% 		\sin^2(\phi_{k\Nb}-\phi_{k1}) &  \sin^2(\phi_{k\Nb}-\phi_{k2}) & \cdots & \sin^2(\phi_{k\Nb}-\phi_{k\Nb})}
%  	\end{align}
% 	\setcounter{equation}{\value{MYtempeqncnt}} \vspace*{-6pt} 
% \end{figure*}

\section{Problem Formulation}\label{sec:system_model}

In this section, we introduce the system models, present the performance metric, and formulate the power allocation problems for \ac{WNL} and \ac{RNL}.

\subsection{System Models}\label{sec:sys_wnl}

{We first introduce the system models for \ac{WNL} and \ac{RNL}.}

\subsubsection*{Wireless Network Localization} Consider 2-D wireless localization using a location-aware network with $\Nb$ anchors and $\Na$ agents [see Fig.~\ref{fig:wnl_mtl_fig}(a)]. The sets of anchors and agents are denoted by $\NB$ and $\NA$, respectively. The position of node $k$ is denoted by $\V{p}_k \in \mathbb{R}^2$ for $k\in\NB\cup\NA$, and the angle and distance from nodes $k$ to $j$ are denoted by $\phi_{kj}$ and $d_{kj}$, respectively. The anchors are the nodes with known positions, whereas the agents are mobile nodes aiming to infer their positions based on the \ac{TOA} range measurements from the anchors \cite{SheWin:J10a}.\footnote{Note that one-way \ac{TOA}-based ranging requires network synchronization, but round-trip \ac{TOA}-based ranging and \ac{RSS}-based ranging can circumvent the synchronization requirement. We consider synchronous networks and the broadcast mode for anchor transmission in this paper, and the results can be extended to asynchronous networks.
% Our framework is also applicable to networks employing the request-respond mode, which is suitable for mobile ad hoc networks without synchronization
} 

% The range measurement between agent $k$ and anchor $j$ is denoted by $\V{z}_{kj}$, following the probability distribution $f(\V{z}_{kj}|\|\V{p}_k-\V{p}_j\|,\B\kappa_{kj})$, where $\B\kappa_{kj}$ is the nuisance measurement parameters, such as wireless channel parameters.
% \footnote{For simplicity, we consider that the ranging and communication are interference-free through medium access control.}

The equivalent lowpass waveform received at agent $k$ from anchor $j$ is modeled as \cite{SheWin:J10a}
\begin{align}\label{eq:Model_Multipath}
    r_{k}(t) = \sum_{j\in\NB} \frac{\sqrt{x_{j}}}{d_{kj}^{\,\beta}}\, % \sum_{l=1}^{L_{kj}} 
\alpha_{kj} \, s_j(t - \tau_{kj} ) + z_{kj}(t)
\end{align}
where $x_{j}$ is the transmit power of anchor $j$ measured at 1\thinspace m away from the transmitter, $\beta$ is the amplitude loss exponent, $\{s_j(t)\}_{j\in\NB}$ is a set of the orthonormal transmit waveforms,\footnote{
%The transmit waveforms are orthonormal in the sense that $\int s_i(t) s_j^\dagger(t) dt = 1$ for $i=j$, and $\int s_i(t-\tau_{ki}^{(l)}) s_j^\dagger(t-\tau_{kj}^{(l)}) dt = 0$ for $i\neq j$. 
The orthogonality can be obtained through medium access control and/or waveform design. When only approximate orthogonality is obtained in practice, the methods developed in the paper can still serve as a general design principle and yield near-optimal power allocation solution.} 
%$L_{kj}$ is the number of multipath components of the channel from anchor $j$ to agent $k$, 
$\alpha_{kj}$ and $\tau_{kj}$ are the amplitude gain and propagation delay, respectively, and $z_{kj}(t)$ represents the observation noise, modeled as additive white complex Gaussian processes. % with two-sided power spectral density $N_0/2$. %, and $[ \, 0, T_{\text{ob}})$ is the observation interval.
The relationship between the delay and agent's position is
\begin{align*}
	\tau_{kj} = \frac{1}{c} \, \|\V{p}_k-\V{p}_j\| 
\end{align*}
where $c$ is the propagation speed of the signal.%, and $b_{kj}^{(l)} \geq 0$ is a range bias induced by non-line-of-sight propagation ($b_{kj}^{(1)}=0$ if the first path is line-of-sight).

% The task of WNL is to estimate 
The agents' positions $\{\V{p}_k\}_{k\in\NA}$ are inferred using the measurements $\{r_{k}(t)\}_{k\in\NA}$. Since the channel parameters are also unknown, the complete set of unknown deterministic parameters is given by $\B\theta = \{\V{p}_k,\,\alpha_{kj}\}_{k\in\NA,\, j\in\NB}$.%, where $\B\kappa_k = \big\{\{\alpha_{kj},\,b_{kj}\}: j\in\NB,\, l=1,2,\ldots,L_{kj}\big\}$ includes the nuisance multipath parameters associated with the channels to agent $k$.

% the unknown deterministic parameters are given in $\B\theta = \big\{\{\V{p}_k,\,\B\kappa_k\}:k\in\NA\big\}$, where $\B\kappa_k = \big\{\{\alpha_{kj}^{(l)},\,b_{kj}^{(l)}\}: j\in\NB,\, l=1,2,\ldots,L_{kj}\big\}$ includes all the nuisance multipath parameters, based on the measurement set $\big\{r_{k}(t):k\in\NA \big\}$.

% \subsection{Radar Network Localization}\label{sec:sys_mtl}

\subsubsection*{Radar Network Localization} Consider 2-D target localization using a radar network with $\Nt$ transmit and $\Nr$ receive antennas [see Fig.~\ref{fig:wnl_mtl_fig}(b)]. The sets of transmit and receiver antennas are denoted by $\NT$ and $\NR$, respectively. The position of antenna $k$ is known and denoted by $\V{p}_k \in \mathbb{R}^2$ for $k\in\NT\cup\NR$, and the position of the target is denoted by $\V{p}_0 \in \mathbb{R}^2$.\footnote{{We consider single-target localization for notational brevity, and the proposed methods are applicable to multi-target cases. Note that one needs to deal with the target association problem in multi-target cases.}} The angle from antenna $k$ to the target is given by $\psi_k$ for $k\in\NR$ or $\varphi_k$ for $k\in\NT$, and the corresponding distance is given by $d_{k}$. The radar network aims to locate the target based on the \ac{TOA} range measurements from the transmit to receive antennas via the reflection of the target \cite{GodHaiBlu:10}.

The equivalent lowpass waveform received at antenna $k$ from the transmit antennas is modeled as%\footnote{The received waveform model assumes that the background clutters are removed for simplicity. However, the structure of (\ref{eq:EFIM_radar}) remains the same for models involving clutters.} 
\cite{GodHaiBlu:10}
\begin{align}\label{eq:Model_MIMO}
    r_{k}(t) = \sum_{j\in\NT} \frac{\sqrt{x_{j}}}{d_{k}^{\,\beta} d_{j}^{\,\beta}}\, \alpha_{kj} \, s_j\left(t - \tau_{kj} \right) + z_{kj}(t)
\end{align}
where $x_j$ is the transmit power of antenna $j$, 
%$\beta$ is the amplitude loss exponent, 
$\{s_j(t)\}_{j\in\NT}$ is a set of the orthonormal transmit waveforms, and $\alpha_{kj}$ and $\tau_{kj}$ are the amplitude gain and propagation delay.\footnote{The amplitude gain integrates the effect of the phase offsets between the transmit and receive antennas as well as that of the point scatters of the extended target \cite{GodHaiBlu:10}.} 
% characterizes the effect of target \ac{RCS} on the phase and amplitude of the received signal, 
% $z_{kj}(t)$ represents the observation noise, modeled as additive white complex Gaussian processes. 
Then the relationship between the delay and target's position is
\begin{align*}
	\tau_{kj} = \frac{1}{c} ( \|\V{p}_k-\V{p}_0\|+\|\V{p}_j-\V{p}_0\| )\,.
	%\frac{\|\V{p}_k-\V{p}_0\|+\|\V{p}_j-\V{p}_0\|}{c} \,.
\end{align*}

The target's position $\V{p}_0$ is estimated using the measurements $\{r_{k}(t)\}_{k\in\NR}$ by noncoherent processing. Since channel parameters are also unknown, the complete set of unknown deterministic parameters is given by  $\B\theta = \{\V{p}_0,\,\alpha_{kj}\}_{k\in\NR,\,j\in\NT}$.

% We next formulate the power optimization problem for network localization.

\subsection{Performance Metric}\label{sec:sys_performance_metric}

We now present the performance metric for localization accuracy as a function of the \ac{PAV} denoted by
\begin{align*}
	\V{x} = [\,x_{1}~~x_{2}~~\cdots~~x_n\,]\T
\end{align*}
where $n=\Nb$ for \ac{WNL}
% \footnote{We here consider the broadcast mode for anchor transmission in \ac{WNL}. Our framework is also applicable to networks employing the request-respond mode, which is suitable for mobile ad hoc networks without synchronization.} 
and $n=\Nt$ for \ac{RNL}. For conciseness, we only present the notions for WNL, as they are applicable to \ac{RNL} analogously. 

% and $\V{x} = [\,x_{1}~~x_{2}~~\cdots~~x_n\,]\T$ denotes the \ac{PAV}, in which $x_j$ is the power allocated to anchor $j$ and $n=\Nb$ for WNL.
% \footnote{For \ac{RNL}, $x_j$ is the power allocated to transmit antenna $j$ and $n=\Nt$.}

% We denote the power allocation vector (PAV) as
% \begin{align}\label{eq:pav_def}
% 	\V{x} = [\,x_{1}~~x_{2}~~\cdots~~x_n\,]\T
% \end{align}
% where $x_j$ is the power allocated to anchor $j$ and $n=\Nb$ for WNL, or the power allocated to transmit antenna $j$ and $n=\Nt$ for RNL.

Localization accuracy can be quantified by the \ac{MSE} of the position estimator. Let $\hat{\V{p}}_k$ be an unbiased position estimator for agent $k$ in WNL, then the \ac{MSE} matrix of $\hat{\V{p}}_k$ satisfies 
\begin{align*}%\label{eq:def_SPEB2}
    \mathbb{E}_{\V{r}} \big\{ ( \hat{\V{p}}_k - \V{p}_k )
    (\hat{\V{p}}_k - \V{p}_k)\T \big\}
    \succeq  \V{J}_\text{e}^{-1}(\V{p}_k;\V{x}) % \left[ \V{J}_{\B\theta}^{-1} \right]_{\V{p}_k} 
\end{align*}
where $\V{J}_\text{e}(\mathbf{p}_k;\V{x})$ is the \ac{EFIM}\footnote{The EFIM for a subset of parameters reduces the dimension of the original \ac{FIM}, while retaining all the necessary information to derive the information inequality for these parameters \cite{SheWin:J10a}.} for $\V{p}_k$ \cite{SheWin:J10a}.
%$\V{r}$ is the vector consisting of the Karhunen-Lo\`{e}ve (KL) expansion coefficients of the measurements $\{r_{k}(t):k\in\NA \}$, 
% $\V{J}_{\B\theta}$ is the Fisher information matrix (FIM) for $\B\theta$, and $[ \V{J}_{\B\theta}^{-1} ]_{\V{p}_k}$ denotes the square submatrix on the diagonal of $\V{J}_{\B\theta}^{-1}$ corresponding to $\V{p}_k$. To characterize $[ \V{J}_{\B\theta}^{-1} ]_{\V{p}_k}$, we can derive the equivalent FIM (EFIM) for $\V{p}_k$, i.e., $\V{J}_\text{e}(\mathbf{p}_k) = \big([ \V{J}_{\B\theta}^{-1} ]_{\V{p}_k}\big)^{-1}$, directly from $\V{J}_{\B\theta}$ \cite{SheWin:J10a}.\footnote{The EFIM for a subset of parameters reduces the dimension of the original FIM, while retaining all the necessary information to derive the information inequality for these parameters \cite{SheWin:J10a}.} 
Consequently, the MSE of the position estimate $\mathbb{E}_{\V{r}} \big\{ \|\hat{\V{p}}_k - \V{p}_k\|^2 \big\}$ is bounded below by the \ac{SPEB}, defined as \cite{SheWin:J10a}
\begin{equation}\label{eq:SPEB}
	\mathcal{P}(\mathbf{p}_k;\V{x}) := \tr \big\{\V{J}_\text{e}^{-1}(\V{p}_k;\V{x})\big\}
\end{equation}
and hence we adopt the \ac{SPEB} as the performance metric for WNL. More discussion on the \ac{SPEB} is given in Section \ref{sec:dis_speb}. We next present the \acp{EFIM} for \ac{WNL} and \ac{RNL}.

% Since power allocation is the focus of the paper, 

% Then we express the SPEB and EFIM explicitly as functions of the PAV by $\mathcal{P}(\mathbf{p}_k;\V{x})$ and $\V{J}_\text{e}(\V{p}_k;\V{x})$. 

%Since the SPEB can be characterized by the EFIM, 

% More discussion about the SPEB is given in Section \ref{sec:discussion}.

% In WNL, the mean squared error (MSE) matrix of an unbiased position estimator $\hat{\V{p}}_k$ for agent $k$ 

% Since only the position $\V{p}_k$ is the parameter of interest in $\B\theta$, we can derive the equivalent Fisher information matrix (EFIM) $\V{J}_\text{e}(\mathbf{p}_k)$ for $\V{p}_k$ based on $\V{J}_{\B\theta}$ \cite{SheWin:J10a}, in the sense that $\V{J}_\text{e}(\mathbf{p}_k) = \big([ \V{J}_{\B\theta}^{-1} ]_{\V{p}_k}\big)^{-1}$.

% Then, we adopt the SPEB as a performance metric, given by
% \begin{equation}\label{eq:SPEB}
% 	\mathcal{P}(\mathbf{p}_k) := \tr \left\{\V{J}_\text{e}^{-1}(\V{p}_k)\right\}
% \end{equation}
% which is a lower bound on the MSE of the position $\mathbb{E}_{\V{r}} \left\{ \|\hat{\V{p}}_k - \V{p}_k\|^2 \right\}$.

\begin{proposition}\label{pro:efim_wnl}
The \ac{EFIM} for the position of agent $k$ in \ac{WNL} based on (\ref{eq:Model_Multipath}) is given by\footnote{{Although the derivation in \cite{SheWin:J10a} is based on the received wideband waveforms, the structure of the \ac{EFIM} is observed for general range-based localization systems \cite{JouDarWin:J08, MazLorBah:J10,QiKobSud:06}.}}
\begin{equation}\label{eq:EFIM}
	\V{J}_\text{e}(\mathbf{p}_k;\V{x}) = \!\sum_{j\in\NB} x_{j} \, \xi_{kj} \cdot  \V{J}_\text{r}(\phi_{kj})
\end{equation}
where the \ac{ERC} $\xi_{kj} = {\zeta_{kj}}/{d_{kj}^{\,2\beta}}$, in which the \ac{RC} $\zeta_{kj}\ge 0$ is determined by the channel parameters, signal bandwidth, and noise power.
\end{proposition}

\begin{IEEEproof}
Refer to \cite{SheWin:J10a} for the detailed derivation.
\end{IEEEproof}

\begin{proposition}\label{pro:efim_rnl}
The \ac{EFIM} for the position of the target in \ac{RNL} based on (\ref{eq:Model_MIMO}) is given by
\begin{align}\label{eq:EFIM_radar}
	\V{J}_\text{e}(\mathbf{p}_0;\V{x}) = \!\sum_{j\in\NT} \sum_{k\in\NR} x_{j}\, \xi_{kj} \cdot \V{J}_\text{r}(\phi_{kj})
\end{align}
where $\phi_{kj}={(\psi_k+\varphi_j)}/{2}$ and the ERC
\begin{align}\label{eq:coeff_xi}
	\xi_{kj} = \frac{4\,\zeta_{kj}}{d_{k}^{\,2\beta} d_{j}^{\,2\beta}} \cos^2\Big(\frac{\psi_k-\varphi_j}{2}\Big)
\end{align}
in which the \ac{RC} $\zeta_{kj}\ge 0$ is determined by the channel parameters, signal bandwidth, and noise power.
\end{proposition}

\begin{IEEEproof}
Similar to the derivation in \cite{GodPetPoo:11}, the \ac{EFIM} based on (\ref{eq:Model_MIMO}) can be derived as 
\begin{align}\label{eq:EFIM_radar_1}
	\V{J}_\text{e}(\mathbf{p}_0;\V{x}) & = \sum_{j\in\NT}  \frac{x_{j}}{d_{j}^{\,2\beta}} \, \sum_{k\in\NR} \frac{\zeta_{kj}}{d_{k}^{\,2\beta}} \,  \cdot \V{u}_{kj} \,\V{u}_{kj}\T
\end{align}
where $\V{u}_{kj} = \V{u}(\psi_k)+\V{u}(\varphi_j)$. It can be shown using the property of the trigonometric functions that
\begin{align}\label{eq:EFIM_radar_2}
	\V{u}_{kj} = 2\cos\Big(\frac{\psi_k-\varphi_j}{2}\Big) \cdot \V{u}\Big(\frac{\psi_k+\varphi_j}{2}\Big)
\end{align}
and by substituting (\ref{eq:EFIM_radar_2}) into (\ref{eq:EFIM_radar_1}), we obtain (\ref{eq:EFIM_radar}).
\end{IEEEproof}

\begin{remark}
The propositions show that the \acp{EFIM} for \ac{WNL} and \ac{RNL} have a canonical form as a weighted sum of rank-one matrices $\V{J}_\text{r}(\phi_{kj})$. These matrices in (\ref{eq:EFIM}) and (\ref{eq:EFIM_radar}) respectively characterize the network topology of the anchors and agent for WNL and that of the transmit/receive antennas and target for \ac{RNL}. 
% Such common forms of the \acp{EFIM} permits unifying development of power allocation problems for \ac{WNL} and \ac{RNL}.
% the theoretical analysis and algorithm design 
% The similarity of the EFIMs for WNL and RNL permits unifying 
\end{remark}

\begin{remark}
% Note that narrowband transmission considered in Section \ref{sec:sys_wnl} serves as an example, and the proposed methods are applicable to network localization using any transmission technology. In general, the transmission technology only affects the \acp{RC} but not the structure of the \acp{EFIM}. Moreover, the waveform model for \ac{RNL} in (\ref{eq:Model_MIMO}) assumes that the background clutters are removed and the Doppler shifts are corrected for simplicity. Such a simplification will not affect the structure of the \ac{EFIM}, and only the \acp{RC} need to be adjusted if more comprehensive waveform models are used.
Note that specific transmission technology and waveform model are considered in Section II-A to derive the \acp{EFIM}. However, the analytical methods and algorithms developed in this paper are applicable to network localization using general transmission technologies and waveform models, which only affect the \acp{RC} but not the structure of the \acp{EFIM}. For instance, the waveform model (\ref{eq:Model_MIMO}) for \ac{RNL}  assumes that the background clutters are removed and the Doppler shifts are corrected for simplicity; nevertheless, such a simplification does not affect the structure of the \ac{EFIM}.
\end{remark}

% serve as an example of typical localization systems.

% For notational convenience, we define the equivalent RC (ERC) as 
% \begin{align}\label{eq:coeff_xi}
% 	\xi_{kj} := \begin{cases}
% 	{\zeta_{kj}}/{d_{kj}^{\,2\beta}}\,, & \qquad k\in\NA,\, j\in\NB\;\; (\text{WNL}) \\
% 	{4\,\cos^2\big(\frac{\psi_k-\varphi_j}{2}\big) \,\zeta_{kj}}/{d_{k}^{\,2\beta} d_{j}^{\,2\beta}} \,,  & \qquad k\in\NR,\,j\in\NT\;\;(\text{MTL}) 
% 	\end{cases}
% \end{align}
% and then write the EFIMs for WNL and MTL respectively as
% \begin{align}\label{eq:EFIM_common}
% 	\V{J}_\text{e}(\mathbf{p}_k;\V{x}) & = \sum_{j\in\NB} \xi_{kj} \, x_{j} \cdot \V{J}_\text{r}(\phi_{kj}) \\ %\,, \qquad \text{WNL} 
% 	\V{J}_\text{e}(\mathbf{p}_0;\V{x}) & = \sum_{j\in\NT} \sum_{k\in\NR}  \xi_{kj} \, x_{j} \cdot \V{J}_\text{r}(\phi_{kj}) \,. %\,, \qquad \text{MTL} \,. 
% 	\label{eq:EFIM_common_MTL}
% \end{align}

\subsection{Power Allocation Formulation}\label{sec:sys_opt}

We now formulate the power allocation problems for \ac{WNL} and \ac{RNL}, aiming to achieve the optimal tradeoffs between localization accuracy and energy consumption. In particular, we minimize the total transmit power subject to a given localization requirement for the agents or the target, shown as follows.\footnote{{The proposed methods are applicable to many other formulations as shown in Section \ref{sec:speb_app}.}}

% allocate the total transmit power among the anchors or the transmit antennas for the best possible agent or target localization performance. One formulation is to minimize the total transmit power subject to given requirements on agent or target localization accuracy, shown as follows.

The power allocation problem for \ac{WNL} can be formulated as
\begin{align}
\Pact : \;\;	
\min_{\{\V{x}\}} \quad& \V{1}\T\,\V{x} \notag \\
	\text{s.t.} \quad
			& \mathcal{P}(\mathbf{p}_k;\V{x}) \leq \varrho_k \,,\;\; \forall\, k\in\NA
			\label{eq:P-con-ttl} \\
			& c_l(\V{x}) \leq 0\,,\qquad\;\,\; l = 1, 2, \ldots, L \notag
\end{align}
where $\varrho_k$ denotes the localization requirement for agent $k$ and $\{c_l(\cdot)\}$ denotes linear constraints on the \ac{PAV} $\V{x}$, e.g., the individual power constraints for anchors $\V{0} \preceq \V{x}\preceq \V{x}_\text{max}$.
%\footnote{These constraints can be any linear functions of $\V{x}$, {such as the individual power constraints for anchors $\V{0} \preceq \V{x}\preceq \V{x}_\text{max}$.}} 

Similarly, the power allocation problem for \ac{RNL} can be formulated as
\begin{align}
\Ppas : \;\;	
\min_{\{\V{x}\}} \quad&  \V{1}\T\,\V{x} \notag \\
	\text{s.t.} \quad
			& \mathcal{P}(\mathbf{p}_0;\V{x}) \leq \varrho
			\label{eq:P-accuray-mimo} \\
			& c_l(\V{x}) \leq 0\,,\qquad\;\;\,l = 1, 2, \ldots, L \notag
\end{align}
where $\varrho$ denotes the localization requirement for the target.

% Since the nodes in the wireless network are subject to power constraints, the objective is to allocate the available transmit power for the best possible localization performance. Thus, the problem of optimal power optimization subject to given accuracy requirement can be formulated as

\begin{remark}
Note that the \acp{EFIM} (\ref{eq:EFIM}) and (\ref{eq:EFIM_radar}), corresponding to the \acp{SPEB} $\mathcal{P}(\mathbf{p}_k;\V{x})$ and $\mathcal{P}(\mathbf{p}_0;\V{x})$, have a similar expression as a function of $\V{x}$. This leads to a similar structure between the localization requirement constraints (\ref{eq:P-con-ttl}) and (\ref{eq:P-accuray-mimo}), and hence $\Pact$ and $\Ppas$. Therefore, we can develop optimal power allocation algorithms for the two scenarios under a {unifying} framework.
% the difference between $\Pact$ and $\Ppas$ lies only in the localization requirement constraints (\ref{eq:P-con-ttl}) and (\ref{eq:P-accuray-mimo}), but the corresponding \acp{EFIM} given in (\ref{eq:EFIM}) and (\ref{eq:EFIM_radar}) have the identical structure. 

% the power allocation formulations for \ac{WNL} and \ac{RNL} possess the same structure except for the difference between constraints (\ref{eq:P-con-ttl}) and (\ref{eq:P-accuray-mimo}). 

% power allocation algorithms for the two scenarios can be developed under a unifying framework.

% Such common forms of the \acp{EFIM} permits unifying development of power allocation problems for \ac{WNL} and \ac{RNL}.
\end{remark}

% We next explore the properties of the SPEB and develop efficient algorithms for solving $\mathscr{P}_1$ and $\mathscr{P}_2$ in the following sections.

% \begin{proposition}[Convex \cite{}]
% The problem $\mathscr{P}$ is convex in $\{P_{kj}\}$.
% \end{proposition}

\section{SPEB Properties and SOCP Formulation}\label{sec:perfect_case}

In this section, we first explore the properties of the \ac{SPEB}, and then show that the power allocation problems can be transformed into \acp{SOCP}.

%uncover the property of the SPEB as a function of the transmit power, and then prove that . Finally, we develop a robust scheme to tackle the uncertainties in network parameters.

\subsection{SPEB Properties}\label{sec:speb_prop}

The following lemma describes the convexity property of the \ac{SPEB} given in (\ref{eq:SPEB}) as a function of the \ac{PAV} and the low rank property of the topology matrix.

% the properties of the \acp{SPEB} given in (\ref{eq:SPEB}) together with (\ref{eq:EFIM}) or (\ref{eq:EFIM_radar}) as a function of the \ac{PAV} and topology matrix.

% \begin{align*}
% 	[\,\V{1}\,\V{1}\T - \V{c}\,\V{c}\T -  \V{s}\,\V{s}\T\,](\B\phi)
% 	: = \V{1}\,\V{1}\T - \V{c}(\B\phi)\V{c}(\B\phi)\T - \V{s}(\B\phi)\V{s}(\B\phi)\T
% \end{align*}
% where $\V{c}(\B\phi) := [\, \cos\phi_{1}\;\;\cos\phi_{2}\;\;\cdots\;\;\cos\phi_{n}\,]\T$ and $\V{s}(\B\phi) := [\, \sin\phi_{1}\;\;\sin\phi_{2}\;\;\cdots\;\;\sin\phi_{n}\,]\T$.
% \begin{align*}
% 	\V{c}(\B\phi) & := \big[\, \cos(\phi_{1})\;\;\cos(\phi_{2})\;\;\cdots\;\;\cos(\phi_{n})\,\big]\T \\
% 	\V{s}(\B\phi) & := \big[\, \sin(\phi_{1})\;\;\sin(\phi_{2})\;\;\cdots\;\;\sin(\phi_{n})\,\big]\T \,.
% \end{align*}

\begin{lemma}\label{lem:speb_structure}
The \ac{SPEB} of the agent or the target is a convex function of $\PW \succeq \V{0}$. Moreover,
% \begin{itemize}
% 	\item 
	the \ac{SPEB} of agent $k$ for \ac{WNL} can be written as
	\begin{align}\label{eq:SPEB_frac}
		\mathcal{P}(\mathbf{p}_k;\V{x}) 
		%& = \frac{2 \sum_{j\in\NB} \xi_{kj}P_{kj}}{\sum_{j\in\NB,\,j'\in\NB} \xi_{kj}P_{kj} \xi_{kj'}P_{kj'} \sin^2(\phi_{kj}-\phi_{kj'})} \nonumber \\ 
		& = \frac{4\cdot\V{1}\T\,\V{R}_k\, \V{x}}{\V{x}\T \, \V{R}_k\T \,\B\Lambda_k \,\V{R}_k\, \V{x}}
	\end{align}
where the \ac{ERC} matrix $\V{R}_k = \text{diag} \{\xi_{k1},\xi_{k2},\ldots,\xi_{k\Nb}\}$ and the {topology matrix} $\B\Lambda_k$ is a symmetric matrix of $\text{rank}\{\B\Lambda_k\}\leq 3$, given by
	\begin{align}\label{eq:decomp_lambda}
		\B\Lambda_k & = \V{1}\,\V{1}\T - \V{c}(2\B\phi_k)\,\V{c}(2\B\phi_k)\T - \V{s}(2\B\phi_k)\,\V{s}(2\B\phi_k)\T
	\end{align}
	in which $\B\phi_k = [\, \phi_{k1}\;\; \phi_{k2} \;\;\cdots\;\; \phi_{k\Nb}\,]\T$;
	% \item 
	the \ac{SPEB} of the target for \ac{RNL} can be written as
	\begin{align*}%\label{eq:speb_mtl}
		\mathcal{P}(\mathbf{p}_0;\V{x}) 
		%& = \frac{2 \sum_{j\in\NB} \xi_{kj}P_{kj}}{\sum_{j\in\NB,\,j'\in\NB} \xi_{kj}P_{kj} \xi_{kj'}P_{kj'} \sin^2(\phi_{kj}-\phi_{kj'})} \nonumber \\ 
		& = \frac{4\cdot\V{1}\T\,\V{R}\,\PW}{\PW\T \,\V{R}\T \B\Lambda \,\V{R}\,\PW}
	\end{align*}
	where the \ac{ERC} matrix $\V{R} = \big[\,\V{R}_1\T~~\V{R}_2\T~~\cdots~~\V{R}_{\Nr}\T\,\big]\T$ with  $\V{R}_k=\text{diag} \{\xi_{k1},\xi_{k2},\ldots,\xi_{k\Nt}\}$ and the topology matrix $\B\Lambda$ is a symmetric matrix of $\text{rank}\{\B\Lambda\}\leq 3$, given by
	\begin{align*}%\label{eq:lambda_mimo}
		\B\Lambda = \V{1}\,\V{1}\T - \V{c}({2\B\phi})\,\V{c}({2\B\phi})\T - \V{s}({2\B\phi})\,\V{s}({2\B\phi})\T
	\end{align*}
	in which $\B\phi = \big[\,\B\phi_1\T~~\B\phi_2\T~~\cdots~~\B\phi_{\Nr}\T\,\big]\T$ with $\B\phi_k = [\, \phi_{k1}\;\;\phi_{k2}\;\;\cdots\;\;\phi_{k\Nt}\,]\T$.
% \end{itemize}
%$\V{c} = [\,\V{c}_1\T~~\V{c}_2\T~~\cdots~~\V{c}_{\Nr}\T\,]\T$ and $\V{s} = [\,\V{s}_1\T~~\V{s}_2\T~~\cdots~~\V{s}_{\Nr}\T\,]\T$ with %$\V{c}_k := \big[\, \cos(2\phi_{k1})\;\;\cos(2\phi_{k2})\;\;\cdots\;\;\cos(2\phi_{k\Nt})\,\big]\T$ and $\V{s}_k := \big[\, \sin(2\phi_{k1})\;\;\sin(2\phi_{k2})\;\;\cdots\;\;\sin(2\phi_{k\Nt})\,\big]\T$ % .
% \begin{align*}
% 	\V{c}_k & := \big[\, \cos(2\phi_{k1})\;\;\cos(2\phi_{k2})\;\;\cdots\;\;\cos(2\phi_{k\Nt})\,\big]\T \\
% 	\V{s}_k & := \big[\, \sin(2\phi_{k1})\;\;\sin(2\phi_{k2})\;\;\cdots\;\;\sin(2\phi_{k\Nt})\,\big]\T \,.
% \end{align*}
\end{lemma}

\begin{IEEEproof}
See Appendix \ref{apd:lem_speb}.
\end{IEEEproof}

\begin{remark}
% The SPEB of the target for MTL has the same properties as shown in Appendix \ref{apd:speb_prop_mrl}. 
The lemma first shows that the \ac{SPEB} is a convex function in $\V{x}$, implying that each localization requirement in (\ref{eq:P-con-ttl}) and (\ref{eq:P-accuray-mimo}) is a convex constraint on $\V{x}$. Thus, the power allocation problems $\Pact$ and $\Ppas$ are convex programs. Second, the lemma also shows the low rank (at most three) property of the topology matrix $\B{\Lambda}_k$ and $\B{\Lambda}$. 
%Note also that the \acp{SPEB} (\ref{eq:SPEB_frac}) and (\ref{eq:speb_mtl}) are written in fractional forms with the numerator linear in $\V{x}$ and denominator quadratic in $\V{x}$. 
The convexity and low rank properties can be exploited to develop efficient power allocation algorithms.

% gives a fractional form of the SPEB with the numerator linear and denominator quadratic in $\V{x}$, and it also shows the low rank of the SPEB in terms of the topology matrix $\B{\Lambda}$, i.e., rank at most three. 
\end{remark}

\subsection{Optimal Power Allocation}\label{sec:nonrobust_socp}

{We now show that the constraint (\ref{eq:P-con-ttl}) can be converted to a \ac{SOC} form using the \ac{SPEB} properties given in Lemma \ref{lem:speb_structure}. Consequently, the power allocation formulation $\Pact$ is equivalent to an \ac{SOCP}. For conciseness, we denote $\V{c}_k =\V{c}(2\B\phi_k)$ and $\V{s}_k =\V{s}(2\B\phi_k)$ in the following.}

\begin{proposition}\label{pro:SOCP_nonrobust}
The problem $\Pact$ is equivalent to the \ac{SOCP}
\begin{align}
\Pact^\text{SOCP}: \; 
\min_{\{\V{x}\}}  \quad & \V{1}\T\,\V{x}
	\notag\\
	\text{s.t.} \quad & 
				\big\|\V{A}_k\,\V{R}_k\,{\V{x}}  + \V{b}_k\big\| \leq 
				\V{1}\T \, \V{R}_k\,{\V{x}} - 2 \varrho_k^{-1}, \notag\\
				& \hspace{40mm} \forall\, k\in\NA \label{eq:P-SODP-con} \\
				& c_l(\V{x}) \leq 0\,,\qquad\;\;\,l = 1, 2, \ldots, L \notag
\end{align}
where $\V{A}_k = [\,\V{c}_k ~~ \V{s}_k ~~\V{0}\,]\T$ and $\V{b}_k = [\,0~~ 0~~ 2 \varrho_k^{-1}\,]\T$.
\end{proposition}

\begin{IEEEproof}
Let $\V{y}=\V{R}_k\,{\V{x}}$. Using (\ref{eq:SPEB_frac}) and (\ref{eq:decomp_lambda}) in Lemma \ref{lem:speb_structure}, we can rewrite $\mathcal{P}(\mathbf{p}_k;\V{x}) \leq \varrho_k$ as
	% \begin{align*}
	% 	\frac{2\cdot\V{1}\T\, \widetilde{\V{P}}_k}{\widetilde{\V{P}}_k\T ~ \B\Lambda_k ~ \widetilde{\V{P}}_k} \leq \varrho_k
	% \end{align*}
	% which is equivalent, by (\ref{eq:decomp_lambda}), to
\begin{align*}
	{4}\varrho_k^{-1}\cdot\V{1}\T\,\V{y} \leq { (\V{1}\T\,\V{y})^2 - (\V{c}_k\T\,\V{y})^2 - (\V{s}_k\T\,\V{y})^2} \,.
\end{align*}
By completing the square, we have
\begin{align*}
	%& 
	(\V{c}_k\T \,\V{y})^2 + (\V{s}_k\T\, \V{y})^2 +  4 \varrho_k^{-2} %\\
	% & \qquad \qquad 
	\leq  (\V{1}\T\,\V{y})^2 - 4 \varrho_k^{-1}\cdot\V{1}\T\, \V{y} + 4 \varrho_k^{-2}
\end{align*}
which is equivalent to (\ref{eq:P-SODP-con}) since $\V{1}\T \, \V{y} - 2 \varrho_k^{-1} \geq 0$.
%Therefore, we have shown that $\mathscr{P}$ can be converted to $\mathscr{P}^\text{SOCP}$, which is a SOCP.
\end{IEEEproof}

\begin{remark}
This \ac{SOCP} formulation is more favorable than the \ac{SDP} formulation proposed in \cite{LiSheZhaWin:J13}, since \ac{SOCP} is a subclass of \ac{SDP} and has more efficient solvers than \ac{SDP} \cite{BerNedOzd:B03}. Moreover, as later shown in Section \ref{sec:robust_case}, the \ac{SOC} form of the localization requirement given in (\ref{eq:P-SODP-con}) enables better relaxation than the \ac{SDP} formulation. 
% allows to cope with angular uncertainty than the \ac{SDP} formulation.
% better relaxation methods  for the robust case can be obtained by the formulation using \ac{SOCP} than \ac{SDP} for the robust case.
% SOCP formulation than by the SDP formulation.
% we will show in Section \ref{sec:robust_case} that 
% better relaxations than that in \cite{LiSheZhaWin:C12} for the robust case can be obtained by SOCP formulation, which exploits the low rank property of the topology matrix.
\end{remark}

Note that a similar \ac{SOCP} formulation for \ac{RNL} can be obtained since the problem $\Ppas$ has a similar structure as $\Pact$. We omit the details for brevity.

% Without loss of generality, we mainly focus on the case of WNL unless different treatment for MTL is needed. 

\subsection{Discussion}

The constraints (\ref{eq:P-SODP-con}) in $\Pact^\text{SOCP}$ are determined by the network parameters, including the inter-node angles, distances, and \acp{RC}.
% Solving allocation problem $\mathscr{P}_1^\text{SOCP}$ requires the knowledge of the network parameters, including the inter-node angles, distances, and RCs. 
However, perfect knowledge of these parameters is usually not available; especially, the angles and distances depend on the agents' positions, which are to be determined. One approach is to use estimated values of the parameters in the power allocation algorithms.\footnote{The \ac{RC} estimates can be obtained from channel estimation subsystems, and the angle and distance estimates can be obtained from agents' prior position knowledge. The prior position knowledge is available, for example, in applications such as navigation and high-accuracy localization.} Since these estimated values are subject to uncertainty, directly using them in the algorithms often fails to yield reliable or even feasible solutions. Hence, we will next develop robust methods to cope with the parameter uncertainty.

% it will be apparent in Section \ref{sec:simulation} that

\section{Robust Power Allocation Algorithms}\label{sec:robust_case}

In this section, we first introduce the uncertainty models for network parameters and formulate robust power allocation problems. We then develop an asymptotically optimal algorithm with a proven convergence rate and efficient near-optimal algorithms using relaxation methods.
 
% the optimal solution, as well as their convergence rate to the optimal solution.

\subsection{Uncertainty Models}

Based on the robust optimization framework \cite{BerBroCar:11, GhaOusLeb:98, BenNem:98}, we consider set-based uncertainty models for the network parameters in \ac{WNL} and \ac{RNL}. 

\subsubsection*{Wireless Network Localization}

Consider the unknown position of agent $k$ in an area $\mathcal{A}_k$, and the goal of robust power allocation is to guarantee the localization requirement for agent $k$ at all positions in such an area. 

\begin{figure}[t]
	\vspace{2mm}
	\centering
	
	% \psfrag{Ak}[c][][0.8]{\hspace{1mm} $\mathcal{A}_k$}
	% \psfrag{Aki-1}[c][][0.8]{\hspace{7mm} $\mathcal{A}_k^{(i-1)}$}
	% \psfrag{Aki+1}[c][][0.8]{\hspace{2mm} $\mathcal{A}_k^{(i+1)}$}
	% \psfrag{Aki}[c][][0.8]{\hspace{-2mm}$\mathcal{A}_k^{(i)}$}
	% \psfrag{R}[c][][0.8]{$\Delta$}
	% \psfrag{Phi}[c][][0.8]{$2\tilde{\phi}_{kj}^{(i)}$}
	% \psfrag{p}[c][][0.8]{\hspace{-16mm} $\hat{\V{p}}_{k}^{(i)}$}
	% \psfrag{pk}[c][][0.8]{\hspace{1mm}${\V{p}}_{k}$}
	
	\includegraphics[width=0.75\columnwidth,draft=false]{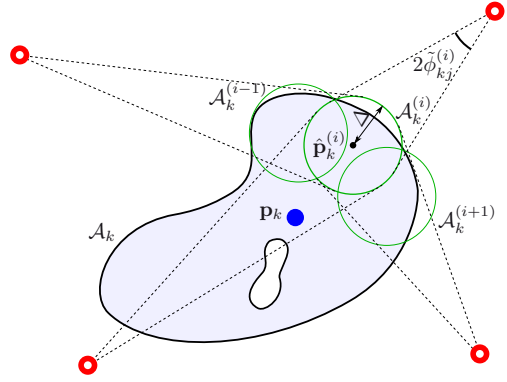}

	\caption{{Illustration of the uncertainty model for \ac{WNL}: $\big\{\mathcal{A}_k^{(i)}\big\}_{i\in\mathcal{I}_k}$ is a finite cover of the uncertainty region $\mathcal{A}_k$ for agent $k$, and each $\mathcal{A}_k^{(i)}$ is a circle with radius $\Delta$ centered at $\hat{\V{p}}_k^{(i)}$.}
	}\label{fig:robust_fig}
\end{figure}

Let $\big\{\mathcal{A}_k^{(i)}\big\}_{i\in\mathcal{I}_k}$ be a finite cover of $\mathcal{A}_k$, i.e., $\mathcal{A}_k \subseteq \cup_{i\in\mathcal{I}_k} \mathcal{A}_k^{(i)}  $, where $\mathcal{A}_k^{(i)}$ is a circle with center $\hat{\V{p}}^{(i)}_k$ and radius $\Delta$, and $\mathcal{I}_k$ is the index set of the circles (see Fig.~\ref{fig:robust_fig}). Then, for any agent's position $\V{p}_k \in \mathcal{A}_k^{(i)}$, the actual network parameters can be represented in the linear sets
\begin{align}\label{eq:uncert_phi}
	\phi_{kj} & \in \Big[\,\hat{\phi}_{kj}^{(i)} - \tilde{\phi}^{(i)}_{kj} \,,\, \hat{\phi}_{kj}^{(i)} + \tilde{\phi}^{(i)}_{kj}\,\Big] =: \Ang_{kj}^{(i)} \\
	 {d}_{kj} & \in \Big[\, \hat{d}_{kj}^{\,(i)} - \Delta \,,\, \hat{d}_{kj}^{\,(i)} + \Delta \,\Big] \quad \text{and}\quad
	{\zeta}_{kj} \in \Big[ \,\underline{\zeta}_{kj}^{(i)} \,,\, \overline{\zeta}_{kj}^{(i)}  \,\Big]  %\big[ \underline{\zeta}_{kj},\, \overline{\zeta}_{kj} \big] =:  
	\notag
\end{align}
where $\hat{\phi}_{kj}^{(i)}$ and $\hat{d}_{kj}^{\,(i)}$ are the nominal values of the topology parameters evaluated at $\hat{\V{p}}^{(i)}_k$, $\tilde{\phi}_{kj}^{(i)} = \arcsin(\Delta/\hat{d}_{kj}^{\,(i)})$ is the angular uncertainty, and the last set characterizes the uncertainty of the \ac{RC} to anchor $j$. 
According to Proposition \ref{pro:efim_wnl}, the latter two translate the uncertainty set for $\xi_{kj}$ as
\begin{align}\label{eq:uncert_xi}
	\xi_{kj} & \in  \Big[\, \underline{\xi}_{kj}^{(i)} \, ,\, \overline{\xi}_{kj}^{(i)} \, \Big] =:\Chn_{kj}^{(i)}
\end{align}
where $\underline{\xi}_{kj}^{(i)} = \underline{\zeta}_{kj}^{(i)}/(\hat{d}_{kj}^{\,(i)} + \Delta)^{2\beta}$ and $\overline{\xi}_{kj}^{(i)} = \overline{\zeta}_{kj}^{(i)} /(\hat{d}_{kj}^{\,(i)} - \Delta)^{2\beta}$.\footnote{{We assume that there is a minimum distance between anchors and agents so that the radius $\Delta < \hat{d}_{kj}$ for all $k\in\NA$ and $j\in\NB$.}} In summary, for the agent's position $\V{p}_k\in\mathcal{A}_k$, the actual network parameters lie in the set%\footnote{For \ac{WNL}, $\{\phi_{kj}, \,\xi_{kj}\}$ is a shorthand for $\{\phi_{kj}, \,\xi_{kj}:k\in\NA,\,j\in\NB\}$, and similarly for \ac{RNL}, $\{\psi_{k},\, \varphi_{j}, \,\xi_{kj}\}$ is a shorthand for $\{\psi_{k},\, \varphi_{j}, \,\xi_{kj}:k\in\NR,\,j\in\NT\}$.}
\begin{align}\label{eq:set_uncertain}
	\{\phi_{kj}, \,\xi_{kj}\}_{k\in\NA,\,j\in\NB} \in \bigcup_{i\in\mathcal{I}_k} \,\prod_{k\in\NA,\, j\in\NB} \Ang_{kj}^{(i)} \times \Chn_{kj}^{(i)} \,.
\end{align}

\subsubsection*{Radar Network Localization}

Similar to \ac{WNL}, for unknown target position $\V{p}_0 \in \mathcal{A}^{(i)}$, the actual network parameters can be represented in the linear sets
\begin{align*}
	\psi_{k} & \in \Big[\, \hat{\psi}_{k}^{(i)} - \tilde{\psi}_{k}^{(i)} \,,\, \hat{\psi}_{k}^{(i)} + \tilde{\psi}_{k}^{(i)}\, \Big] =: \Ang_{1,k}^{(i)} \\
	\varphi_{j} & \in \Big[\, \hat{\varphi}_{j}^{(i)} - \tilde{\varphi}_{j}^{(i)} \, ,\, \hat{\varphi}_{j}^{(i)} + \tilde{\varphi}_{j}^{(i)} \,\Big] =: \Ang_{2,j}^{(i)} \\ 
	 {d}_{k} & \in \Big[\, \hat{d}_{k}^{\,(i)} - \Delta \, ,\, \hat{d}_{k}^{\,(i)} + \Delta \,\Big] \quad \text{and}\quad
	{\zeta}_{kj} \in \Big[ \,\underline{\zeta}_{kj}^{(i)}  \,,\, \overline{\zeta}_{kj}^{(i)} \, \Big]  \notag 
\end{align*}
where $\hat{\psi}_k^{(i)}$, $\hat{\varphi}_k^{(i)}$, and $\hat{d}_{k}^{\,(i)}$ are the nominal values of the topology parameters evaluated at $\hat{\V{p}}_0^{(i)}$, $\tilde{\psi}_{k}^{(i)} = \arcsin(\Delta/\hat{d}_{k}^{\,(i)})$ and $\tilde{\varphi}_{j}^{(i)} = \arcsin(\Delta/\hat{d}_{j}^{\,(i)})$ are the angular uncertainty, and the last set characterizes the uncertainty of the \ac{RC} from antennas $k$ to $j$ via the target. 
According to Proposition \ref{pro:efim_rnl}, we have the uncertainty sets for $\phi_{kj}$ and $\xi_{kj}$ as 
\begin{align}\label{eq:uncert_phi_mimo}
	\phi_{kj} & \in \Big[\, \hat{\phi}_{kj}^{(i)}　
	%\frac{\hat{\phi}_{k}+\hat{\phi}_{j}}{2} 
	- \tilde{\phi}_{kj}^{(i)}　\,,\, 
	% \frac{\hat{\phi}_{k}+\hat{\phi}_{j}}{2} 
	\hat{\phi}_{kj}^{(i)}
	+ \tilde{\phi}_{kj}^{(i)}　\,\Big] =: \Ang_{kj}^{(i)} \\
	\xi_{kj} & \in \Big[ \,\underline{\xi}_{kj}^{(i)}\,,\, \overline{\xi}_{kj}^{(i)} \,\Big] =:\Chn_{kj}^{(i)}
	\label{eq:uncert_xi_mimo}
\end{align}
where $\hat{\phi}_{kj}^{(i)} = (\hat{\psi}_{k}^{(i)}+ \hat{\varphi}_{j}^{(i)})/{2}$, $\tilde{\phi}_{kj}^{(i)} = (\tilde{\psi}_{k}^{(i)}+ \tilde{\varphi}_{j}^{(i)})/{2}$, and the upper and lower bounds for $\xi_{kj}$ are given respectively by
\begin{align*}
	\underline{\xi}_{kj}^{(i)} & = \frac{4\, \underline{\zeta}_{kj}^{(i)}}{(\hat{d}_{k}^{\,(i)} + \Delta)^{2\beta} (\hat{d}_{j}^{\,(i)} + \Delta)^{2\beta}}  \bigg[\min_{\phi\,\in\mathcal{D}_{kj}^{(i)}} \cos^2\phi \; \bigg] \\
	\overline{\xi}_{kj}^{(i)}  & = \frac{4\, \overline{\zeta}_{kj}^{(i)}}{(\hat{d}_{k}^{\,(i)} - \Delta)^{2\beta} (\hat{d}_{j}^{\,(i)} - \Delta)^{2\beta}} \bigg[ \max_{\phi\,\in\mathcal{D}_{kj}^{(i)}} \cos^2\phi \; \bigg]
\end{align*}
in which $\mathcal{D}_{kj}^{(i)} : = \big[\, (\hat{\psi}_{k}^{(i)}-\hat{\varphi}_{j}^{(i)})/{2} - \tilde{\phi}_{kj}^{(i)} \,, \, (\hat{\psi}_{k}^{(i)}-\hat{\varphi}_{j}^{(i)})/{2} + \tilde{\phi}_{kj}^{(i)}\, \big]$. In summary, for the target's position $\V{p}_0\in\mathcal{A}$, the actual network parameters lie in the set
\begin{align*}
	\{\psi_{k},\, \varphi_{j}, \,\xi_{kj}\}_{k\in\NR,\,j\in\NT} \in \bigcup_{i\in\mathcal{I}} \, \prod_{k\in\NR,\,j\in\NT} \Ang_{1,k}^{(i)} \times \Ang_{2,j}^{(i)} \times \Chn_{kj}^{(i)} 
	\,.
\end{align*}

% \begin{align*}
% 	\mathcal{D}_{kj}^\phi : = \Big[ \frac{\hat{\phi}_{k}-\hat{\phi}_{j}}{2} - \varepsilon^{\phi}_{kj} , \, \frac{\hat{\phi}_{k}-\hat{\phi}_{j}}{2} + \varepsilon^{\phi}_{kj}\Big]\,.
% \end{align*}
\begin{remark}
Note that the parameter uncertainty models for \ac{WNL} and \ac{RNL} can be converted to a common form, i.e., (\ref{eq:uncert_phi}) and (\ref{eq:uncert_phi_mimo}) for $\phi_{kj}$, and  (\ref{eq:uncert_xi}) and (\ref{eq:uncert_xi_mimo}) for $\xi_{kj}$. Thus, we can develop robust formulations for the two scenarios under a unifying framework.	
\end{remark}

% \begin{remark}
% \Cb{Note that our uncertainty models are not restricted to small uncertainty conditions adopted in conventional robust optimization, and they can account for parameter uncertainty over large ranges, e.g., the uncertainty of the agent's position can be a large area.}
% \end{remark}

% can focus on the discussion of WNL based on the above estimation error model, whereas the results and formulations can be directly applied to MTL. 

\subsection{Robust Formulation}\label{sec:robust_case_form}

We now propose the robust counterparts of $\Pact$ and $\Ppas$ that guarantee the localization requirement in the presence of parameter uncertainty. The worst-case \ac{SPEB} for \ac{WNL} due to parameter uncertainty (\ref{eq:set_uncertain}) is\footnote{Note that $\mathcal{P}(\mathbf{p}_k;\V{x})$ also depends on $\phi_{kj}$ and $\xi_{kj}$ although we omit them for notational convenience.}
{\begin{align*}
	\mathcal{P}_\text{R}(\mathbf{p}_k;\V{x}) 
	% & := \! \max_{\{(\phi_{kj},\xi_{kj})\in\cup_{i\in\mathcal{I}_k} \Ang_{kj}^{(i)} \times \Chn_{kj}^{(i)} \}} \! \mathcal{P}(\mathbf{p}_k;\V{x}) \\
	& := \max_{i\in\mathcal{I}_k} \, \mathcal{P}^{(i)}_\text{R}(\V{p}_k;\V{x}) 
\end{align*}
where $\mathcal{P}^{(i)}_\text{R}(\V{p}_k;\V{x})$ is the worst-case \ac{SPEB} in $\mathcal{A}_k^{(i)}$, given by
\begin{align*}
	\mathcal{P}^{(i)}_\text{R}(\V{p}_k;\V{x})  : = \!\max_{\{(\phi_{kj},\,\xi_{kj})\in \Ang_{kj}^{(i)} \times \Chn_{kj}^{(i)}\}} \mathcal{P}(\V{p}_k;\V{x}) \,.
\end{align*}
Hence, to guarantee the localization performance in the worst case, we introduce a new constraint $\mathcal{P}_\text{R}(\mathbf{p}_k;\V{x}) \le \varrho_k$ and formulate the robust power allocation problem for \ac{WNL} as}
\begin{align}
\PactR: \;\;	
\min_{\{\V{x}\}} \quad& \V{1}\T \,\V{x} \notag\\
	\text{s.t.} \quad
			& \mathcal{P}^{(i)}_\text{R}(\mathbf{p}_k;\V{x}) \leq \varrho_k \,,\;\;   \forall\, k\in\NA {\,,\, i\in\mathcal{I}_k}	\label{eq:robust_con} \\
			& c_l(\V{x}) \leq 0\,,\qquad\quad\;\;\, l = 1, 2, \ldots, L \notag
\end{align}
{where the localization requirement (\ref{eq:robust_con}) is equivalent to $\mathcal{P}_\text{R}(\mathbf{p}_k;\V{x}) \le \varrho_k$ for $k\in\NA$.

Similarly, we can obtain the worst-case \ac{SPEB} for \ac{RNL} as
\begin{align*}
	\mathcal{P}_\text{R}(\mathbf{p}_0;\V{x}) := \max_{i\in\mathcal{I}_k} \, \mathcal{P}^{(i)}_\text{R}(\V{p}_0;\V{x}) 
\end{align*}
where
\begin{align*}
	\mathcal{P}^{(i)}_\text{R}(\V{p}_0;\V{x})  : = \!\max_{\{	(\psi_{k},\, \varphi_{j}, \,\xi_{kj}) \in \Ang_{1,k}^{(i)} \times \Ang_{2,j}^{(i)} \times \Chn_{kj}^{(i)} \}} \mathcal{P}(\V{p}_0;\V{x})
\end{align*}
and formulate the robust power allocation problem $\PpasR$ by introducing the constraint $\mathcal{P}_\text{R}(\mathbf{p}_0;\V{x}) \leq \varrho$.}
% \begin{align*}
% 	\mathcal{P}_\text{R}(\mathbf{p}_0;\V{x}) 
% 	& := \!\! \max_{\{\psi_{k}\in\mathcal{S}^{\psi}_{k},\,\varphi_{j}\in\mathcal{S}^{\varphi}_{j}, \,\xi_{kj}\in\mathcal{S}^\xi_{kj}\}} \!\! \mathcal{P}(\mathbf{p}_0;\V{x})
% \end{align*}
% and we can obtain $\PpasR$ by employing the constraint $\mathcal{P}_\text{R}(\mathbf{p}_0;\V{x}) \leq \varrho$. 
In the following we will focus on $\PactR$, and the analysis equally applies to $\PpasR$.

We next convert (\ref{eq:robust_con}) into an expression amenable for efficient optimization. Since the \ac{SPEB} is a monotonically decreasing function in $\xi_{kj}$, {the maximization over $\xi_{kj}\in\Chn^{(i)}_{kj}$ is achieved at $\xi_{kj} = \underline{\xi}_{kj}^{(i)}$. Thus, by Lemma \ref{lem:speb_structure}, we can obtain
\begin{align}%\label{eq:worst_speb}
	\label{eq:worst_speb_frac}
	\mathcal{P}^{(i)}_\text{R}(\mathbf{p}_k;\V{x}) 
	% & = \! \max_{\{\phi_{kj}\in\mathcal{S}^{\phi}_{kj}\}}\! \tr\Big\{\Big[\sum_{j\in\NB}  \underline{\xi}_{kj} \,x_{j}\, \V{J}_\text{r}(\phi_{kj})\Big]^{-1}\Big\} \\
	& = \! \max_{\{\phi_{kj}\in\Ang^{(i)}_{kj}\}} 
	% \\
	% 	&\hspace{-10mm} 
	\frac{4\cdot\V{1}\T\,\underline{\V{R}}_k^{(i)}\,\V{x}}{ \big(\V{1}\T\,\underline{\V{R}}_k^{(i)}\,\V{x}\big)^2 - \big\| [\,\V{c}_k ~~\V{s}_k\,]\T\,\underline{\V{R}}_k^{(i)}\,\V{x}\big\|^2}
	% & = \frac{4\cdot\V{1}\T\,\underline{\V{R}}_k\,\V{x}}{\min_{\{\phi_{kj}\in\mathcal{S}^{\phi}_{kj}\}} \V{x}\T\,\underline{\V{R}}_k\T \,\B\Lambda_k\, \underline{\V{R}}_k\,\V{x}}
\end{align}
where $\underline{\V{R}}_k^{(i)} = \text{diag} \big\{\underline{\xi}_{k1}^{(i)}, \underline{\xi}_{k2}^{(i)},\ldots,\underline{\xi}_{k\Nb}^{(i)}\big\}$.} Unfortunately, the remaining maximization over $\{\phi_{kj}\}$ does not permit an explicit expression due to the intricate function. %on $\{\phi_{kj}\}$.%\footnote{A brute-force way to approximate the maximization in (\ref{eq:worst_speb_frac}) is to use samples of \Cb{$\mathcal{P}^{(i)}_\text{R}(\mathbf{p}_k;\V{x})$} at different \Cb{$\phi_{kj}\in\Ang^{(i)}_{kj}$}. In view of (\ref{eq:robust_con}), increasing the approximation accuracy requires an exponential growth in the number of SOC constraints. Nevertheless, this method cannot guarantee the localization requirement (\ref{eq:robust_con}).}

% [to evaluate the function at different samples of ... ]

% $\mathcal{P}_\text{R}(\mathbf{p}_k;\V{x})$ can be sampling $\phi_{kj}$ in $\mathcal{S}_{kj}^\phi$. However, this will not only incur an exponential number of SOC constraints with the increase of approximation accuracy, but also fail to guarantee the requirement (\ref{eq:robust_con}).

% To approximate $\mathcal{P}_\text{R}(\mathbf{p}_k;\V{x})$, brute-force methods such as sampling each $\phi_{kj}$ in $\mathcal{S}_{kj}^\phi$ will generate an exponential number of SOC constraints to improve the approximation accuracy, which is computationally intractable. 

To address the angular uncertainty, we propose sequential lower and upper bounds for $\mathcal{P}^{(i)}_\text{R}(\mathbf{p}_k;\V{x})$, both of which lead to efficient optimization programs. We denote $\mathcal{M}=\{0,1,\ldots,M-1\}$ where $M \in \mathbb{Z}_+$ and 
\begin{align*}
	B(\V{x}) := \frac{1}{4}\, \mathcal{P}^{(i)}_\text{R}(\V{p}_k;\V{x}) \cdot \V{1}\T\,\underline{\V{R}}_k^{(i)}\,\V{x} \,.
\end{align*}

% \begin{lemma}[Finite Projection Bound]\label{lem:bounds_norm}
% For any $\V{y}\succeq\V{0}$,
% \begin{align}\label{eq:bounds_lemma_ineq}
% 	0 \leq \max_{m\in\mathcal{M}}\, \big\{\V{h}_{k,m}\T\,\V{y}\big\} & 
% 	\leq \max_{\{\phi_{kj}\in\mathcal{S}^{\phi}_{kj}\}} \Big\| \sum_{j\in\NB} y_j\, \V{u}(2\phi_{kj}) \Big\| \notag\\
% 	& \qquad \quad 
% 	\leq \max_{m\in\mathcal{M}}\, \big\{\V{g}_{k,m}\T\,\V{y} \big\} \,. 
% \end{align}
% % where $\V{h}_{k,m}, \, \V{g}_{k,m} \in \mathbb{R}^{\Nb}$ with the $j$th elements given by
% % \begin{align}\label{eq:bounds_hkm}
% % 	[\,\V{h}_{k,m}\,]_j & := \max_{|\epsilon|\leq 2\varepsilon_{kj}^\phi} \,\cos(2\hat{\phi}_{kj} - \vartheta_m + \epsilon ) \\
% % 	[\,\V{g}_{k,m}\,]_j & := \max_{|\epsilon|\leq 2\varepsilon_{kj}^\phi+\pi/M} \,\cos(2\hat{\phi}_{kj} - \vartheta_m + \epsilon ) \,.
% % 	\label{eq:bounds_gkm}
% % \end{align}
% \end{lemma}
% 
% \begin{IEEEproof}
% See Appendix \ref{apd:lem_bounds_norm}.
% \end{IEEEproof}

\begin{proposition}\label{pro:upper_lower_bound}
{For any given \ac{PAV} $\V{x}$ such that $\mathcal{P}^{(i)}_\text{R}(\mathbf{p}_k;\V{x}) < \infty$, if $M\geq {\pi}\sqrt{B(\V{x})}$, then $\mathcal{P}^{(i)}_\text{R}(\mathbf{p}_k;\V{x})$ is bounded below and above, respectively, by % $\underline{\mathcal{P}_\text{R}^M}(\mathbf{p}_k;\V{x}) \leq \mathcal{P}_\text{R}(\mathbf{p}_k;\V{x})
% \leq\overline{\mathcal{P}_\text{R}^M}(\mathbf{p}_k;\V{x})$,
% where
\begin{align}
	\underline{\mathcal{P}}_M^{(i)}(\mathbf{p}_k;\V{x})
	& = \max_{m\in\mathcal{M}} 
	\frac{4\cdot\V{1}\T\,\underline{\V{R}}_k^{(i)}\,\V{x}}{(\V{1}\T\,\underline{\V{R}}_k^{(i)}\,\V{x})^2 - \big(\V{h}_{k,m}^{(i)\,\text{T}}\,\underline{\V{R}}_k^{(i)}\,\V{x}\big)^2}
	\label{eq:lower_bound}\\
	\label{eq:upper_bound}
		\overline{\mathcal{P}}_M^{(i)}(\mathbf{p}_k;\V{x})
		& = \max_{m\in\mathcal{M}} \frac{4\cdot\V{1}\T\,\underline{\V{R}}_k^{(i)}\,\V{x}}{(\V{1}\T\,\underline{\V{R}}_k^{(i)}\,\V{x})^2 - \big(\V{g}_{k,m}^{(i)\,\text{T}} \,\underline{\V{R}}_k^{(i)} \,\V{x}\big)^2}
\end{align}
where $ \V{h}_{k,m}^{(i)},\, \V{g}_{k,m}^{(i)} \in \mathbb{R}^{\Nb}$ with the $j$th elements given by % $[\,\V{g}_{k,m}\,]_j := \max_{|\epsilon|\leq 2\varepsilon_{kj}^\phi+\pi/M} \,\cos(2\hat{\phi}_{kj} - \vartheta_m + \epsilon )$ and $[\,\V{h}_{k,m}\,]_j := \max_{|\epsilon|\leq 2\varepsilon_{kj}^\phi} \,\cos(2\hat{\phi}_{kj} - \vartheta_m + \epsilon )$ 
\begin{align*}%\label{eq:bounds_hkm}
	[\,\V{h}_{k,m}^{(i)}\,]_j & = \max_{|\epsilon|\leq 2\tilde{\phi}_{kj}^{(i)}} \,\cos(2\hat{\phi}_{kj}^{(i)} - \vartheta_m + \epsilon )\\
	[\,\V{g}_{k,m}^{(i)}\,]_j & = \frac{1}{\cos({\pi}/{M})}\cdot [\,\V{h}_{k,m}^{(i)} \,]_j
	% \label{eq:bounds_gkm}
\end{align*}
in which $\vartheta_m = (2m+1)\cdot\pi/M$ for $m\in\mathcal{M}$.}
\end{proposition}

\begin{IEEEproof}
See Appendix \ref{apd:lem_bounds_norm}.
\end{IEEEproof}
% $\V{h}_{k,m}$ and $\V{g}_{k,m}$ are given by (\ref{eq:bounds_hkm}) and (\ref{eq:bounds_gkm}), respectively.

% \begin{remark}
The proposed expressions parametrized by $M$ constitute a sequence of lower and upper bounds for the worst-case \ac{SPEB}. We can substitute the worst-case \ac{SPEB} in the localization requirement (\ref{eq:robust_con}) by the lower and upper bounds, leading to the robust relaxation problems $\PactRl$ and $\PactRu$, respectively.
% \end{remark}

\subsection{Asymptotically Optimal Algorithm}

We first show that both the robust relaxation problems $\PactRl$ and $\PactRu$ can be transformed into \acp{SOCP}, and then derive the convergence rates of their solutions to that of the original problem $\PactR$. 

% We next substitute the worst-case \ac{SPEB} in the localization requirement (\ref{eq:robust_con}) by the lower and upper bounds, leading to robust relaxation formulations $\PactRu$ and $\PactRl$, respectively. 

% The following proposition shows that both robust relaxation formulations can be transformed into \acp{SOCP}.

\begin{proposition}
The problem $\PactRu$ is equivalent to the \ac{SOCP}
\begin{align}
\PactRu: \;\; 
\min_{\{\V{x}\}}  \quad & \V{1}\T\,\V{x} \notag\\
	\text{s.t.} \quad & 
				{\big\|\V{A}_{k,m}^{(i)}\,\underline{\V{R}}_k^{(i)} \,{\V{x}}  + \V{b}_k\big\| \leq 
				\V{1}\T \, \underline{\V{R}}_k^{(i)}\,{\V{x}} - 2 \varrho_k^{-1},} \, \notag \\
				& \hspace{25.5mm} \forall\,m\in\mathcal{M}\,,\, k\in\NA\,,\, {i\in\mathcal{I}_k} \notag \\
				& c_l(\V{x}) \leq 0\,,\qquad\;\;\,l = 1, 2, \ldots, L \notag
\end{align}
where $\V{A}_{k,m}^{(i)} = \big[\,\V{g}_{k,m}^{(i)} ~~\V{0}\,\big]\T$ and $\V{b}_k = [\,0~~ 2\varrho_k^{-1}\,]\T$. Similarly, the problem $\PactRl$ is also equivalent to an \ac{SOCP} by letting $\V{A}_{k,m}^{(i)}=\big[\,\V{h}_{k,m}^{(i)} ~~\V{0}\,\big]\T$ in the above constraints.
\end{proposition}

\begin{IEEEproof}
For the upper bound case, we use the relaxed localization requirement $\overline{\mathcal{P}}^{(i)}_M(\mathbf{p}_k;\V{x}) \leq \varrho_k$, which can be converted to the $M$ SOC forms 
\begin{align*}
	{\big\|\V{A}_{k,m}^{(i)} \, \underline{\V{R}}_k^{(i)} \,{\V{x}}  + \V{b}_k\big\| \leq \V{1}\T \, \underline{\V{R}}_k^{(i)} \,{\V{x}} - 2 \varrho_k^{-1},\quad \forall\,m\in\mathcal{M}}
\end{align*}
by using (\ref{eq:upper_bound}). The case for $\PactRl$ can be shown similarly.
\end{IEEEproof}

\begin{remark}
Although both relaxation formulations can be solved by \acp{SOCP}, $\PactRu$ is more desirable for implementation since it guarantees the localization requirement. 
\end{remark}

% prove that the gaps between the upper and lower bounds converge to zero at the rate of $1/M$ as $M\to\infty$, and hence the optimal solution for the original problem ${\mathscr{P}}^M_{1\text{-R}}$ can be asymptotically achieved by $\overline{\mathscr{P}}^M_{1\text{-R}}$.

The next proposition proves that the gap between the lower and upper bounds for the worst-case \ac{SPEB}, i.e., (\ref{eq:lower_bound}) and (\ref{eq:upper_bound}), converges to zero as $M\to\infty$. 

% Then, we can show that the solutions of $\PactRl$ and $\PactRu$ converge asymptotically to that of the original problem $\PactR$. 
{
\begin{proposition}\label{pro:gap_upper_lower_bounds}
For any given \ac{PAV} $\V{x}\succeq \V{0}$ such that $\mathcal{P}^{(i)}_\text{R}(\mathbf{p}_k;\V{x}) < \infty$, if $M\geq \pi\sqrt{B(\V{x})}$, then 
% \begin{align*}
% 	\V{g}_k^M := \arg\max_{\{\V{g}_k:\,\vartheta\in[\,0,\,2\pi)\}} \V{g}_k\T \,\underline{\V{R}}_k\, \V{x}
% \end{align*}
% where $[\,\V{g}_k\,]_j = \max_{|\epsilon|\leq 2\varepsilon_{kj}^\phi+\pi/M} \,\cos(2\hat{\phi}_{kj} - \vartheta + \epsilon )$. 
$\overline{\mathcal{P}}^{(i)}_M(\mathbf{p}_k;\V{x}) \leq (1 + C_{k,M}^{(i)}) \, \underline{\mathcal{P}}^{(i)}_M(\mathbf{p}_k;\V{x})$, where 
\begin{align*}
	C_{k,M}^{(i)} = \frac{\sin^2({\pi}/{M}) \,(B(\V{x})-1)}{1 -  \sin^2({\pi}/{M})\, B(\V{x})}
	\,.
\end{align*}
Moreover, $C_{k,M}^{(i)}$ is monotonically decreasing with $M$ and
\begin{align}\label{eq:C_kmi}
	\lim_{M\to\infty} \frac{C_{k,M}^{(i)}}{M^{-2}} 
	= {\pi^2} (B(\V{x}) - 1) \,.
\end{align}
\end{proposition}

\begin{IEEEproof}
See Appendix \ref{apd:pro_gap_bounds}.
\end{IEEEproof}}

\begin{remark}
{This proposition implies that the gap between the lower and upper bounds goes to zero at the rate of $O(M^{-2})$.} Using this result, we can show that both the solutions of $\PactRu$ and $\PactRl$ converge to that of the original problem at the rate of $O(M^{-2})$ as follows.
\end{remark}

\begin{proposition}\label{pro:power_bounds}
Let $\V{x}^*$, $\overline{\V{x}}^M$, and $\underline{\V{x}}^M$ be the optimal solutions of $\PactR$, $\PactRu$, and $\PactRl$, respectively. Then,{
\begin{align*}
	0 & \leq {\V{1}\T \overline{\V{x}}^M - \V{1}\T \V{x}^*} \leq C_M \cdot \V{1}\T \V{x}^* \\
	0 & \leq  \V{1}\T \V{x}^* - \V{1}\T \underline{\V{x}}^M \leq C_M \cdot \V{1}\T \V{x}^*
\end{align*}
where $C_M = \max_{k\in\NA} \max_{i\in\mathcal{I}_k} C_{k,M}^{(i)}$ converges to zero at the rate of $O(M^{-2})$.}
\end{proposition}

\begin{IEEEproof}
See Appendix \ref{apd:pro_power_bounds}.
\end{IEEEproof}

\begin{remark}
Since the solutions of the proposed relaxation problems converge to that of the original problem {at the rate of $O(M^{-2})$}, {the optimal solution of $\PactR$ can be approximated by that of $\PactRu$ with a small value of $M$. For example, our simulation results show that its performance loss is less than 2\% when $M \geq 16$.} On the other hand, note that the number of SOC constraints of the relaxation problems increases linearly with $M$, resulting in the increase in computational complexity at ${O}(M^{{3}/{2}})$ \cite{LobVanBoyLeb:98}. This gives an important guideline on the performance versus complexity tradeoff for the robust power allocation algorithms in practice. 
\end{remark}

While the proposed \ac{SOCP}-based algorithms are asymptotically optimal, we next develop efficient near-optimal algorithms, which involve only a few \ac{SOC} constraints, for power allocation in dynamic networks with limited computational capability. 

% \section{Efficient Algorithms for Robust Formulation} \label{sec:efficient_algorithm}

\subsection{Efficient Algorithms}

{We next propose a relaxation method to address the angular uncertainty involved in the worst-case \ac{SPEB}, leading to efficient \ac{SOCP}-based algorithms. For notational convenience, we omit the superscript $(i)$ in this section.}

% In this section, we develop alternative relaxation methods for the proposed robust formulations, and show that these relaxation methods lead to efficient \acp{SOCP}.

% We now propose a relaxation method to address the angular uncertainty involved in the worst-case \ac{SPEB}. 
Since only the denominator in (\ref{eq:worst_speb_frac}) is a function of $\phi_{kj}$, we derive an upper bound for $\mathcal{P}_\text{R}(\mathbf{p}_k;\V{x})$ by finding a lower bound for the denominator.
% an upper bound for the quadratic form $\V{x}\T \,\underline{\V{R}}_k\T \,\big[ \V{c}(2{\B\phi}_k)\,\V{c}(2{\B\phi}_k)\T +\V{s}(2{\B\phi}_k)\,\V{s}(2{\B\phi}_k)\T\big]\, \underline{\V{R}}_k\,\V{x}$ over $\{\phi_{kj}\in\mathcal{S}^\phi_{kj}\}$. 
{Denote
%the maximum angular uncertainty $\tilde{\phi}_{k}^{(i)} := \max_{j\in\NB} \tilde{\phi}_{kj}^{(i)}$; 
the vectors $\hat{\V{c}}_k = \V{c}(2\hat{\B\phi}_k)$ and $\hat{\V{s}}_k = \V{s}(2\hat{\B\phi}_k)$, where $\hat{\B\phi}_k= [\,\hat{\phi}_{k1}~~\hat{\phi}_{k2}~~\cdots~~\hat{\phi}_{k\Nb}\,]\T$; and $\tilde{\V{s}}_k,\, \tilde{\V{c}}_k \in\mathbb{R}^{\Nb}$ with $j$th elements given, respectively, by 
\begin{align*}
	[\,\tilde{\V{s}}_{k}\,]_j & =\max_{|\epsilon|\leq \tilde{\phi}_{kj}} \big|2\sin(2\hat\phi_{kj}+\epsilon ) \sin\epsilon \,\big|\\ 		
	[\,\tilde{\V{c}}_{k}\,]_j & =\max_{|\epsilon|\leq \tilde{\phi}_{kj}} \big|2\cos(2\hat\phi_{kj}+\epsilon) \sin\epsilon \, \big| \,.
\end{align*}

\begin{proposition}\label{lem:socp_relax_3}
Let 
\begin{align}\label{eq:socp_relax_3}
	\mathcal{P}_\text{U}(\mathbf{p}_k;\V{x}) = \!\max_{e_1,\, e_2=\pm 1}\! \frac{4\cdot\V{1}\T\,\underline{\V{R}}_k\,\V{x}}{\big(\V{1}\T\,\underline{\V{R}}_k\,\V{x}\big)^2 - \big\| \hat{\V{A}}_k^{(e_1,\, e_2)} \,\underline{\V{R}}_k\,\V{x}\big\|^2} 
\end{align}
where $\hat{\V{A}}_k^{(e_1,\, e_2)} = [\,(\hat{\V{c}}_k+e_1\tilde{\V{s}}_k)~~(\hat{\V{s}}_k+e_2\tilde{\V{c}}_k) ~~\V{0}\,]\T$.
% $$\underline{\B\Lambda}_k = \V{1}\,\V{1}\T - (\hat{\V{c}}_k^{(i)}+e_1\tilde{\V{s}}_k^{(i)}) \,(\hat{\V{c}}_k^{(i)}+e_1\tilde{\V{s}}_k^{(i)})\T - (\hat{\V{s}}_k^{(i)}+e_2\tilde{\V{c}}_k^{(i)}) \,(\hat{\V{s}}_k^{(i)}+e_2\tilde{\V{c}}_k^{(i)})\T\,.$$
Then $\mathcal{P}_\text{R}(\mathbf{p}_k;\V{x}) \leq \mathcal{P}_\text{U}(\mathbf{p}_k;\V{x})$, provided that $\mathcal{P}_\text{U}(\mathbf{p}_k;\V{x})>0$.
\end{proposition}

\begin{IEEEproof}
See Appendix \ref{apd:socp_relax_3}.
\end{IEEEproof}
}
% \begin{remark}
Since $\mathcal{P}_\text{U}(\mathbf{p}_k;\V{x})$ is an upper bound for the worst-case \ac{SPEB}, we can relax the constraint (\ref{eq:robust_con}) in $\PactR$ by
\begin{align*}%\label{eq:robust_relax}
	0 < \mathcal{P}_\text{U}(\mathbf{p}_k;\V{x}) \leq \varrho_k
\end{align*}
which can be converted to the set of four SOC constraints
\begin{align}\label{eq:con_robust_3}
	\big\|\hat{\V{A}}_k^{(e_1,\, e_2)}\,\underline{\V{R}}_k\, {\V{x}} + {\V{b}}_k \big\| \leq  \V{1}\T \,\underline{\V{R}}_k\,{\V{x}} - 2   \varrho_k^{-1} , \;\; e_1, e_2=\pm 1
\end{align}
where 
%$\hat{\V{A}}_k^{(e_1,\, e_2)} = \big[\,(\hat{\V{c}}_k+e_1\tilde{\V{s}}_k)~~  (\hat{\V{s}}_k+e_2\tilde{\V{c}}_k) ~~\V{0}\,\big]\T$, % $\hat{\V{A}}_k^{(2)} := \big[\,\underline{\V{R}}_k\T\,(\hat{\V{c}}_k-\tilde{\V{s}}_k)~~\underline{\V{R}}_k\T\, (\hat{\V{s}}_k+\tilde{\V{c}}_k) ~~\V{0}\,\big]\T$, $\hat{\V{A}}_k^{(3)} := \big[\,\underline{\V{R}}_k\T\,(\hat{\V{c}}_k+\tilde{\V{s}}_k)~~\underline{\V{R}}_k\T\, (\hat{\V{s}}_k-\tilde{\V{c}}_k) ~~\V{0}\,\big]\T$, $\hat{\V{A}}_k^{(4)} := \big[\,\underline{\V{R}}_k\T\,(\hat{\V{c}}_k-\tilde{\V{s}}_k)~~\underline{\V{R}}_k\T\, (\hat{\V{s}}_k-\tilde{\V{c}}_k) ~~\V{0}\,\big]\T$, 
%and 
${\V{b}}_k = [\,0 ~~ 0 ~~2 \varrho_k^{-1}\,]\T$. 
Hence, by replacing each constraint (\ref{eq:robust_con}) in $\PactR$ with the four constraints (\ref{eq:con_robust_3}), we obtain an efficient \ac{SOCP} $\PactR^\text{SOCP}$ as a relaxation for the robust power allocation problem.
% \end{remark}

\begin{remark}
Comparing (\ref{eq:con_robust_3}) of $\PactR^\text{SOCP}$ with (\ref{eq:P-SODP-con}) of $\Pact^\text{SOCP}$, one can observe that the proposed robust relaxation retains the \ac{SOC} form as its nonrobust counterpart. Furthermore, when the parameter uncertainty vanishes, $\tilde{\V{s}}_k,\,\tilde{\V{c}}_k \to \V{0}$ and thus $\PactR^\text{SOCP}$ reduces to $\Pact^\text{SOCP}$ as (\ref{eq:con_robust_3}) reduces to (\ref{eq:P-SODP-con}).
\end{remark}

{
Similar to the \ac{WNL} case, we can derive an upper bound for the worst-case \ac{SPEB} in the form of (\ref{eq:socp_relax_3}) and formulate a corresponding robust relaxation problem for the \ac{RNL} case. Specifically for \ac{RNL}, since the $\Nr\Nt$ angles $\{\phi_{kj}\}_{k\in\NR,\,j\in\NT}$ in (\ref{eq:EFIM_radar}) are generated only by the $\Nr+\Nt$ angles $\{\psi_k,\,\varphi_j\}_{k\in\NR,\,j\in\NT}$, we can obtain a tighter bound by addressing the angular uncertainty in the transmit and receive antennas separately. In other words, we start from the uncertainty set $(\psi_{k},\, \varphi_{j}) \in \Ang_{1,k} \times \Ang_{2,j}$ instead of $\phi_{kj} \in \Ang_{kj}$.
}

Denote the matrix $\underline{\V{R}}_\Sigma =　\sum_{k=1}^{\Nr} \underline{\V{R}}_k$, where $\underline{\V{R}}_k=\text{diag} \big\{\underline{\xi}_{k1},\underline{\xi}_{k2},\ldots,\underline{\xi}_{k\Nt}\big\}$;  the vector $\hat{\B\varphi}= [\,\hat\varphi_1~~\hat\varphi_2~~\cdots~~\hat\varphi_\Nt\,]\T$;
% and $\hat{\B\phi} = [\,\hat{\B\phi}_1\T~~\hat{\B\phi}_2\T~~\cdots~~\hat{\B\phi}_{\Nr}\T\,]\T$ with $\hat{\B\phi}_k = [\, \hat{\phi}_{k1}\;\;\hat{\phi}_{k2}\;\;\cdots\;\;\hat{\phi}_{k\Nt}\,]\T$; 
the vectors 
\begin{align*}
	\hat{\V{c}} & = \underline{\V{R}}_\Sigma^{-1} \cdot \sum_{k=1}^{\Nr} \underline{\V{R}}_k\T \,\V{c}(\hat{\B\varphi}+\hat{\psi}_k\V{1}) \\
	\hat{\V{s}} & =\underline{\V{R}}_\Sigma^{-1} \cdot \sum_{k=1}^{\Nr} \underline{\V{R}}_k\T \,\V{s}(\hat{\B\varphi}+\hat{\psi}_k\V{1}) 
	% \tilde{\V{s}} & = \underline{\V{R}}_\Sigma^{-1} \cdot \sum_{k=1}^{\Nr} \underline{\V{R}}_k\T \, \tilde{\V{s}}_k\\
	% \tilde{\V{c}} & = \underline{\V{R}}_\Sigma^{-1} \cdot \sum_{k=1}^{\Nr} \underline{\V{R}}_k\T \, \tilde{\V{c}}_k
\end{align*}
% $\hat{\V{c}}=\underline{\V{R}}_\Sigma^{-1} \cdot \sum_{k=1}^{\Nr} \underline{\V{R}}_k\T \,\V{c}(\hat{\B\varphi}+\hat{\psi}_k\V{1})$, $\hat{\V{s}}=\underline{\V{R}}_\Sigma^{-1} \cdot \sum_{k=1}^{\Nr} \underline{\V{R}}_k\T \,\V{s}(\hat{\B\varphi}+\hat{\psi}_k\V{1})$,
$\tilde{\V{s}} = \underline{\V{R}}_\Sigma^{-1} \cdot \sum_{k=1}^{\Nr} \underline{\V{R}}_k\T \, \tilde{\V{s}}_k$, and $\tilde{\V{c}} = \underline{\V{R}}_\Sigma^{-1} \cdot \sum_{k=1}^{\Nr} \underline{\V{R}}_k\T \, \tilde{\V{c}}_k$, 
where $\tilde{\V{s}}_k,\,\tilde{\V{c}}_k \in\mathbb{R}^{\Nt}$ with $j$th elements given, respectively, {by
\begin{align*}
	[\,\tilde{\V{s}}_{k}\,]_j & =\max_{|\epsilon|\leq (\tilde{\psi}_{k}+ \tilde{\varphi}_{j})/2} \big|2\sin(\hat\psi_{k}+\hat\varphi_{j}+\epsilon) \sin\epsilon\, \big| \\
	[\,\tilde{\V{c}}_{k}\,]_j & =\max_{|\epsilon|\leq (\tilde{\psi}_{k}+ \tilde{\varphi}_{j})/2} \big|2\cos(\hat\psi_{k}+\hat\varphi_{j}+\epsilon) \sin\epsilon \,\big|\,.
\end{align*}

\begin{proposition}
Let
\begin{align*}
	\mathcal{P}_\text{U}(\mathbf{p}_0;\V{x}) = \!\max_{e_1,\, e_2=\pm 1}\! \frac{4\cdot\V{1}\T\,\underline{\V{R}}_\Sigma\,\V{x}}{(\V{1}\T\,\underline{\V{R}}_\Sigma\,\V{x})^2 - \big\| \hat{\V{A}}^{(e_1,\, e_2)} \,\underline{\V{R}}_\Sigma\,\V{x}\big\|^2} 
\end{align*}
% where $$\underline{\B\Lambda}= \V{1}\,\V{1}\T  - (\hat{\V{c}}+e_1\tilde{\V{s}}) \,(\hat{\V{c}}+e_1\tilde{\V{s}})\T - (\hat{\V{s}}+e_2\tilde{\V{c}}) \,(\hat{\V{s}}+e_2\tilde{\V{c}})\T \,.$$  Then $\mathcal{P}_\text{R}(\mathbf{p}_0;\V{x}) \leq \mathcal{P}^\text{U}_\text{R}(\mathbf{p}_0;\V{x})$, provided that $\mathcal{P}^\text{U}_\text{R}(\mathbf{p}_0;\V{x})>0$.
where $\hat{\V{A}}^{(e_1,\, e_2)} = [\,(\hat{\V{c}}+e_1\tilde{\V{s}})~~(\hat{\V{s}}+e_2\tilde{\V{c}}) ~~\V{0}\,]\T$.
% $$\underline{\B\Lambda}_k = \V{1}\,\V{1}\T - (\hat{\V{c}}_k^{(i)}+e_1\tilde{\V{s}}_k^{(i)}) \,(\hat{\V{c}}_k^{(i)}+e_1\tilde{\V{s}}_k^{(i)})\T - (\hat{\V{s}}_k^{(i)}+e_2\tilde{\V{c}}_k^{(i)}) \,(\hat{\V{s}}_k^{(i)}+e_2\tilde{\V{c}}_k^{(i)})\T\,.$$
Then $\mathcal{P}_\text{R}(\mathbf{p}_0;\V{x}) \leq \mathcal{P}_\text{U}(\mathbf{p}_0;\V{x})$, provided that $\mathcal{P}_\text{U}(\mathbf{p}_0;\V{x})>0$.
\end{proposition}
}
\begin{IEEEproof}
The proof follows a similar approach of Proposition \ref{lem:socp_relax_3}.
\end{IEEEproof}

\begin{remark}
{The upper bound $\mathcal{P}_\text{U}(\mathbf{p}_0;\V{x})$ for the worst-case \ac{SPEB} can be used as a relaxation for the robust power allocation problem in \ac{RNL}, leading to an efficient \ac{SOCP} $\PpasR^\text{SOCP}$.} Such a relaxation not only retains the SOC form but also naturally reduces to its nonrobust counterpart when the parameter uncertainty vanishes.
\end{remark}

\section{Discussions}\label{sec:discussion}

In this section, we provide discussions on several related issues, including 1) prior knowledge of the network parameters, 2) broader applications of the \ac{SPEB} properties, and 3) the achievability of the \ac{SPEB}.

\subsection{Prior Knowledge}\label{sec:discussion_prior}

% We have thus far investigated the case without prior knowledge of the network parameters, such as the agents' positions and \acp{RC}. 

Since the prior knowledge of the network parameters, if available, can be exploited to improve the localization accuracy,\footnote{For example, prior position knowledge can be incorporated for localization in tracking and navigation applications.} we next investigate the power allocation problem for \ac{WNL} with prior knowledge of the network parameters. The discussion is also applicable to \ac{RNL}. 

The \ac{EFIM} for the case with prior knowledge is a $2\times 2$ matrix given by \cite{SheWymWin:J10}
\begin{align}\label{eq:EFIM_prior}
	\V{J}_\text{e}(\mathbf{p}_k;\V{x}) = \V{J}_0(\V{p}_k)+\sum_{j\in\NB} x_{j} \cdot \V{J}_{kj}
	% \widetilde{\xi}_{kj} \cdot \V{J}_\text{r}(\widetilde{\phi}_{kj})
\end{align}
where $\V{J}_0(\V{p}_k)\in\mathbb{S}_+^2$ is the \ac{FIM} for the prior position knowledge of agent $k$ and $\V{J}_{kj} \in\mathbb{S}_+^2$ is given by
\begin{align}\label{eq:EFIM_avg}
	\V{J}_{kj} = \mathbb{E}_{\V{r},\B\theta} \left\{
	{\xi}_{kj} \, \V{J}_\text{r}({\phi}_{kj})\right\}
\end{align}
in which the expectation is taken with respect to the agent $k$'s prior position knowledge, the prior channel knowledge between agent $k$ and anchor $j$, and the observation noise.

% $\widetilde{\xi}_{kj} = {\widetilde{\zeta}_{kj}}/{\widetilde{d}_{kj}^{\;2\beta}}$ with $\widetilde{\zeta}_{kj}$ denoting the RC that includes the prior channel knowledge, and $\widetilde{d}_{kj}$ and $\widetilde{\phi}_{kj}$ is the distance and angle between agent $k$ and anchor $j$ evaluated at the $\widetilde{\V{p}}_k=\mathbb{E}\{\V{p}_k\}$, respectively.

% a $2\times 2$ matrice

% Under the assumption that the agent's prior position distribution is concentrated in a small area relative to the anchors (e.g., far-field like scenario), the approximate EFIM still retains the canonical form as a weighted sum of rank-one $2\times 2$ matrices, given by 

% \begin{proposition}
% When prior knowledge are available, under the ``far-field'' condition, the EFIM for the position of agent $k$ can be approximated as	
% \begin{equation}\label{eq:EFIM_prior}
% 	\V{J}_\text{e}(\mathbf{p}_k;\V{x}) = \V{J}_\text{p}(\V{p}_k)+\sum_{j\in\NB} \widetilde{\xi}_{kj}  \, x_{j} \cdot \V{J}_\text{r}(\widetilde{\phi}_{kj})
% \end{equation}
% where $\V{J}_\text{p}(\V{p}_k)\in\mathbb{S}_+^2$ is the FIM for the prior position knowledge of agent $k$, $\widetilde{\xi}_{kj}:= {\widetilde{\zeta}_{kj}}/{\widetilde{d}_{kj}^{\;2\beta}}$ with $\widetilde{\zeta}_{kj}$ denoting the RC that includes the prior channel knowledge, and $\widetilde{d}_{kj}$ and $\widetilde{\phi}_{kj}$ is the distance and angle between agent $k$ and anchor $j$ evaluated at the $\widetilde{\V{p}}_k=\mathbb{E}\{\V{p}_k\}$, respectively.
% \end{proposition}
% 
% \begin{IEEEproof}
% The derivation is given in \cite{SheWin:J10a}.
% \end{IEEEproof}

The \ac{SPEB} $\mathcal{P}(\V{p}_k;\V{x})$ for the case with prior knowledge can be obtained from the \ac{EFIM} (\ref{eq:EFIM_prior}). By employing such \ac{SPEB}, we can formulate the power allocation problem and its robust counterpart, denoted by $\wPact$ and $\wPactR$, in the analogous way as $\Pact$ and $\PactR$, respectively. %The following proposition will show that $\widetilde{\mathscr{P}}_1$ and $\widetilde{\mathscr{P}}_{1\text{-R}}$ can also be formulated as SOCPs.

\begin{proposition}\label{pro:prior_form}
The problems $\wPact$ and  $\wPactR$ can be transformed into \acp{SOCP}. 
\end{proposition}

\begin{IEEEproof}
See Appendix \ref{apd:prior_form}.
\end{IEEEproof}

\begin{remark}
The proposition shows that the power allocation problems $\wPact$ and $\wPactR$ for the case with prior knowledge can also be transformed into \acp{SOCP}. Moreover, these problems reduce to $\Pact$ and $\PactR$ when the prior knowledge vanishes since $\V{J}_0(\V{p}_k)= \V{0}$ in (\ref{eq:EFIM_prior}) and the expectation in (\ref{eq:EFIM_avg}) is only with respect to $\V{r}$.

Since the prior position knowledge $\V{J}_0(\V{p}_k)\succeq 0$ provides additional information to the \ac{EFIM} compared to the case without such knowledge, less transmit power is required to achieve the same localization requirement. In particular, if $\tr\{\V{J}_0^{-1}(\V{p}_k)\} \leq \varrho_k$ for all $k\in\NA$, then all the agents have met their localization requirement and the anchors do not need to transmit ranging signals until $\tr\{\V{J}_0^{-1}(\V{p}_k)\} > \varrho_k$ for some $k$ (e.g., due to the agent movement). 
%Moreover, the anchors can schedule their transmission for localization in a time series by incorporating the mobility knowledge of the agents.
\end{remark}

% [for the robust case/probability outage?]

% [mobility knowledge?]

% \subsection{Maximum Directional SPEB}
% 
% [Yuan: can also be cast into SOCPs.]

\subsection{Applications of SPEB Properties}\label{sec:speb_app}

Lemma \ref{lem:speb_structure} shows two important properties of the \ac{SPEB}, based on which we transformed the power allocation formulations {$\Pact$ and $\Ppas$ into \acp{SOCP} in Section \ref{sec:nonrobust_socp}. Such properties also permit efficient algorithms for other power allocation problems in network localization as discussed in the following.}

{First, the methods developed in this paper are applicable to other formulations of the power allocation problems. For instance, minimizing the maximum localization error of the agents for a given power constraint can be formulated as
\begin{align}
% \Pact : \;\;	
\max_{\{\V{x},\,\rho\}} \quad& \rho \notag \\
	\text{s.t.} \quad
	& \V{1}\T\,\V{x} \leq P_{\text{tot}} \notag \\
			& \mathcal{P}(\mathbf{p}_k;\V{x}) \leq \rho^{-1}, \quad \forall\, k\in\NA
			\notag \\
			& c_l(\V{x}) \leq 0\,,\qquad\quad\;\;l = 1, 2, \ldots, L \notag
\end{align}
where $P_{\text{tot}}$ denotes the total power constraint. Following the derivation in Proposition \ref{pro:SOCP_nonrobust}, one can see that $\mathcal{P}(\mathbf{p}_k;\V{x}) \leq \rho^{-1}$ can be converted to an \ac{SOC} form in $\V{x}$ and $\rho$, and thus the above problem is equivalent to an \ac{SOCP}. Moreover, a similar \ac{SOCP} formulation can be obtained if the objective is to minimize the total localization errors of the agents for a given power constraint.} 
%Moreover, another metric called the \ac{mDPEB} \cite{LiSheZhaWin:C12} possesses similar properties as SPEB, and hence one can transform power allocation problems involving \ac{mDPEB} into \acp{SOCP}.} 
%For example, one formulation can be minimizing the mDPEB of the agents under a total power constraint.

\begin{figure*}[t]
	\centering
	% \hspace{-5mm}
	\subfigure[]{
	%   \psfrag{AAAAAA1}[l][][0.6]{\hspace{-6mm}SOCP}
	% 	\psfrag{AAAAAA2}[l][][0.6]{\hspace{-6mm}SDP}
	% 	\psfrag{AAAAAA3}[l][][0.6]{\hspace{-6mm}Uniform}
	% 	\psfrag{t1}[l][][0.6]{\hspace{-0.6cm}$\Na=1,2,4,8$}
	% 
	% 	\psfrag{xlabel}[c][][0.8]{Number of anchors}	
	% 	\psfrag{ylabel}[c][][0.8]{Total transmit power}
	\includegraphics[width=\matfigscale\columnwidth,draft=false]{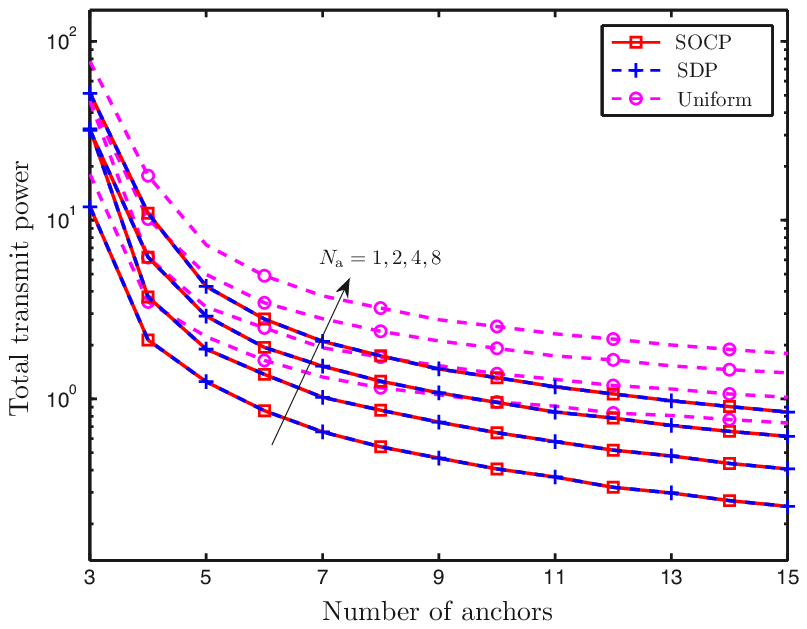}
	\label{fig:CMPSPEB}
	}
	\hspace{-3mm}
	\subfigure[]
	{
	%   \psfrag{AAAAAA1}[l][][0.6]{\hspace{-6mm}SOCP}
	% 	\psfrag{AAAAAA2}[l][][0.6]{\hspace{-6mm}SDP}
	% 	\psfrag{AAAAAA3}[l][][0.6]{\hspace{-6mm}Uniform}
	% 	\psfrag{Nb}[l][][0.6]{\hspace{-0.6cm}$\Nb=4,6,9,12$}
	% 	
	% 	\psfrag{xlabel}[c][][0.8]{Number of agents}	
	% 	\psfrag{ylabel}[c][][0.8]{Total transmit power}
	\includegraphics[width=\matfigscale\columnwidth,draft=false]{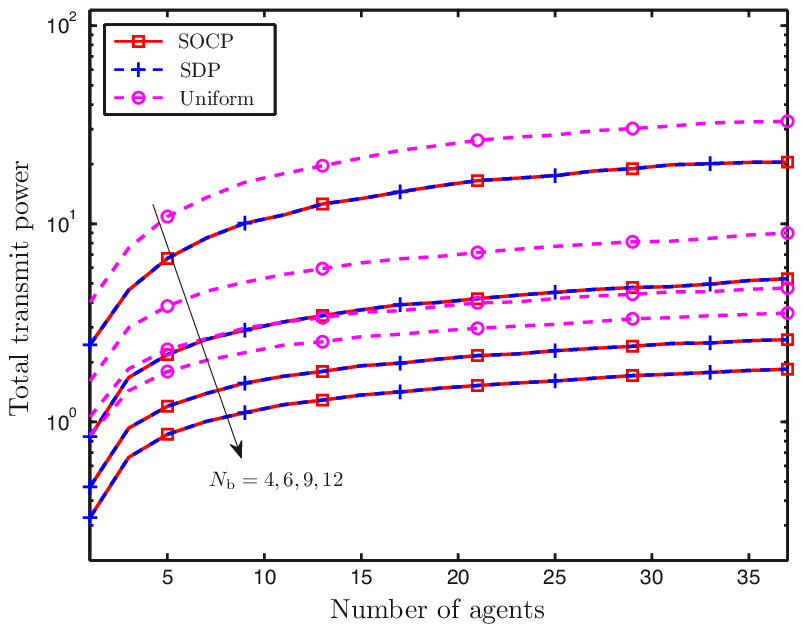}
	\label{fig:CMPSPEB_Agt}
	}

	\caption{Total transmit power as a function of the number of anchors (a) and agents (b): (a) networks with 1, 2, 4, and 8 agents; (b) networks with 4, 6, 9, and 12 anchors. %Both 1-agent and 2-agent networks are considered. 
	\label{fig:wnl_perfect}
	}
\end{figure*}

% consider the power optimization problem for the single-agent case in WNL
% \begin{align*}
% \min_{\{\V{x}\succeq \V{0}\}} \quad& \V{1}\T \,\V{x}
% 	\notag\\
% 	\text{s.t.} \quad
% 			& \mathcal{P}(\V{p}_1;\V{x}) \leq \varrho_1
% \end{align*}
% as well as its robust counterparts. 

Second, we can show by the \ac{SPEB} properties that the optimal localization performance can be achieved by activating only three anchors for the single-agent case with no individual power constraints \cite{DaiSheWin:C12}. The same claim also applies to \ac{RNL}, i.e., only three transmit antennas need to be activated for optimal target localization. This finding for the single-agent case provides important insights into the power allocation problem for network localization: only a few anchors or transmit antennas need to be activated for the optimal localization performance. % This can simplify the practical algorithm design and facilitate theoretical analysis.% e.g., we can obtain a tighter bound by setting $\Nb=3$ in Lemma \ref{lem:robust_ineq}.

% In this case, we can formulate the problem as
% \begin{align*}	
% \min_{\{\V{x}\succeq \V{0}\}} \quad & \V{1}\T\, \V{x}
% 	\notag\\
% 	\text{s.t.} \quad
% 			& \mathcal{P}(\V{p}_k;\V{x}_k) \leq \varrho_k \,, \qquad \forall\,k\in\NA	\notag
% \end{align*}
% where $\V{x} = [\,\V{x}_1\T~~\V{x}_2\T~~\cdots~~\V{x}_{\Na}\T\,]\T$ with $\V{x}_k = [\,{x}_{k1}~~{x}_{k2}~~\cdots~~{x}_{k\Nb}\,]\T$, and this problem can be transformed to conic programs using the methods proposed in previous sections.

% Third, 
% \begin{align*}
% \min_{\{\V{x}\geq\V{0},\,t\}} \quad 	& t \\
% 	\text{s.t.} \quad 
% 	&  \mathcal{P}(\V{p}_k;\V{x}) \leq t \,, \qquad \forall\,k\in\NA	\notag \\
% 			& \V{1}\T\,\V{x} \leq P_{\text{tot}} \notag 
% \end{align*}
% where $P_{\text{tot}}$ is the total available power.

\subsection{Achievability of SPEB}\label{sec:dis_speb}

The \ac{SPEB} is based on the information inequality and hence characterizes the lower bound for the mean squared position errors, which is asymptotically achievable by the maximum likelihood estimators in high SNR regimes (over $10 \sim 15$\thinspace dB) \cite{Tre:68, DarConFerGioWin:J09, GodHaiBlu:10}.\footnote{Although tighter bounds, such as Ziv-Zakai bound, apply to a wider range of SNRs \cite{ChaZakZiv:75, WeiWei:85, DarConFerGioWin:J09}, the tractability of those bounds are limited.} Wireless networks and radar networks {for high-accuracy localization} need to operate in such regimes, which can be realized for example by repeated transmissions, coded sequences, or spread spectrum. Hence, the \ac{SPEB} can be used as the performance metric for the design and analysis of power allocation for a broad range of high-accuracy localization applications. {Although the performance measure \ac{SPEB} is less meaningful in low SNR regimes, the methods and results based on the \ac{SPEB} can serve as a design guideline for localization power optimization.}

% \cite{GolSch:65, SheWin:J10a}
% \cite{GolSch:65, GolWin:J98}

\section{Numerical Results}\label{sec:simulation}

In this section, we evaluate the performance of the proposed power allocation algorithms. 
%using standard optimization tool CVX \cite{cvx}. For the robust formulation, we compare the performance of the power allocation algorithms based on the proposed \ac{SOCP} relaxation and the \ac{SDP} relaxation developed in \cite{LiSheZhaWin:C12}.  %\footnote{Both the non-robust and robust SDP schemes are developed in \cite{LiSheZhaWin:JS11}.}
%\footnote{To avoid singularity, the minimum distance between nodes is set to be 0.5.} 
For WNL, we consider a 2-D network where the anchors and agents are randomly distributed in a region of size $D \times D$. Without loss of generality, the localization requirement for the agents is normalized to $\varrho_k = 1$, $\forall\,k\in\NA$. The \acp{RC} $\{\zeta_{kj}\}_{k\in\NA,\,j\in\NB}$ are modeled as independent Rayleigh \acp{RV} with mean $\mu_{\zeta}$.\footnote{{For simplicity, we illustrate the performance of power allocation algorithms using Rayleigh distributions for \acp{RC}. Similar observations can be made with other distributions for \acp{RC}.}} We compare the normalized required total transmit power $P_\text{tot} = {\mu_{\zeta}}/{D^{2\beta}} \cdot \V{1}\T\, \V{x}$ to meet the localization requirement by different algorithms. Similarly for \ac{RNL}, the antennas and the target are randomly distributed in a region of size $D \times D$. The localization requirement for the target is normalized to $\varrho = 1$. The \acp{RC} $\{\zeta_{kj}\}_{k\in\NR,\,j\in\NT}$ are modeled as independent Rayleigh \acp{RV} with mean $\mu_{\zeta}$. We also compare the normalized required total transmit power $P_\text{tot} = {\mu_{\zeta}}/{D^{4\beta}} \cdot \V{1}\T \, \V{x}$.\footnote{Note that the power loss for \ac{WNL} and \ac{RNL} is proportional to $d_{kj}^{\,2\beta}$ and $d_{k}^{\,2\beta} d_{j}^{\,2\beta}$, which scale as $D^{2\beta}$ and $D^{4\beta}$, respectively.}

\subsection{Wireless Network Localization}

%[Implementation meaning in real environments]

% We first compare the total transmit power required by SOCP, SDP, and uniform optimization as 
% \subsubsection*{Perfect parameter case}
We first compare the performance of SOCP-based, SDP-based, and uniform power allocation algorithms for \ac{WNL} with perfect network parameters. The required total transmit power as a function of the number of anchors and agents is shown in Fig.~\ref{fig:wnl_perfect}. First, for a given number of agents, the required power decreases with the number of anchors as shown in Fig.~\ref{fig:CMPSPEB} since more degrees of freedom are available for power allocation. On the other hand, for a given number of anchors, the required power increases with the number of agents as shown in Fig.~\ref{fig:CMPSPEB_Agt} since more constraints are imposed for the localization requirement of additional agents. Second, the SOCP- and SDP-based algorithms yield identical solutions as they both achieve the global optimum, significantly outperforming the uniform allocation algorithm, e.g., reducing the required power by more than $40\%$. Third, the concavity of the curves in Fig.~\ref{fig:CMPSPEB_Agt} implies that less incremental power is required for additional agents as the number of agents increases. This agrees with the intuition because due to anchor broadcasting, each new agent can utilize the transmit power intended for the existing agents and thus less additional power is needed to meet its localization requirement.

\begin{figure}[t]
	\centering
	% \hspace{-5mm}
	\subfigure[]
	{
	% \psfrag{AAAAAA1}[l][][0.6]{\hspace{-6mm}$M=4$}
	% \psfrag{AAAAAA2}[l][][0.6]{\hspace{-6.5mm}$M=8$}
	% \psfrag{AAAAAA3}[l][][0.6]{\hspace{-6.5mm}$M=16$}	
	% \psfrag{AAAAAA4}[l][][0.6]{\hspace{-6.5mm}Optimum}
	% 
	% \psfrag{t3}[l][][0.6]{\hspace{-1mm}$\Na=1$}
	% \psfrag{t2}[l][][0.6]{\hspace{-6mm}$\Na=2$}
	% 
	% \psfrag{xlabel}[c][][0.8]{NUSS}
	% \psfrag{ylabel}[c][][0.8]{Total transmit power}
	\includegraphics[width=\matfigscale\columnwidth]{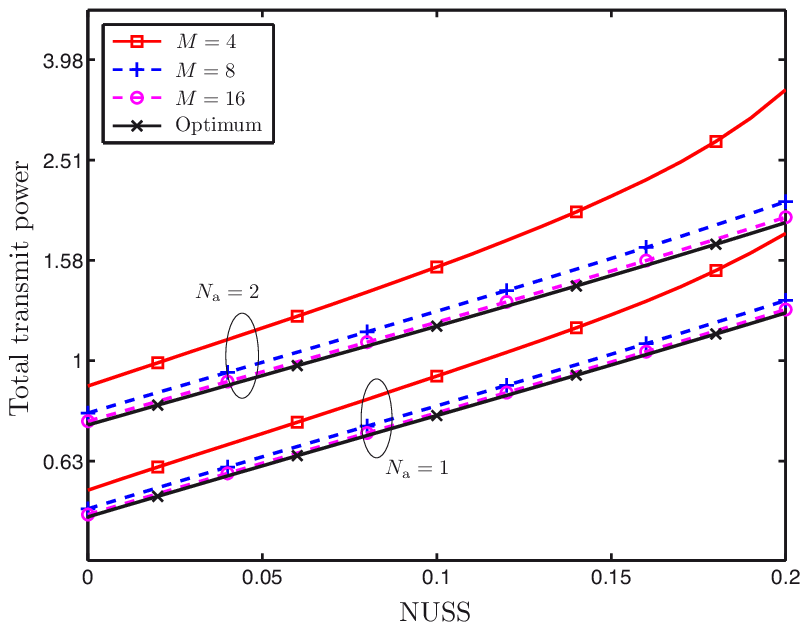}
	\label{fig:wnl_optimal_a}
	}
	% \hspace{-5mm}
	\subfigure[]{
	% \psfrag{AAAAAAAA1}[l][][0.58]{\hspace{-9mm}Upper Bound}
	% 	\psfrag{AAAAAAAA2}[l][][0.58]{\hspace{-9mm}Lower Bound}
	% 	\psfrag{AAAAAAAA3}[l][][0.58]{\hspace{-9mm}Optimum}
	% 	
	% 	\psfrag{t3}[l][][0.6]{\hspace{-1mm}$\Na=1$}
	% 	\psfrag{t2}[l][][0.6]{\hspace{-5mm}$\Na=2$}
	% 	
	% 	\psfrag{xlabel}[c][][0.8]{$M$}
	% 	\psfrag{ylabel}[c][][0.8]{Total transmit power}
	\includegraphics[width=\matfigscale\columnwidth]{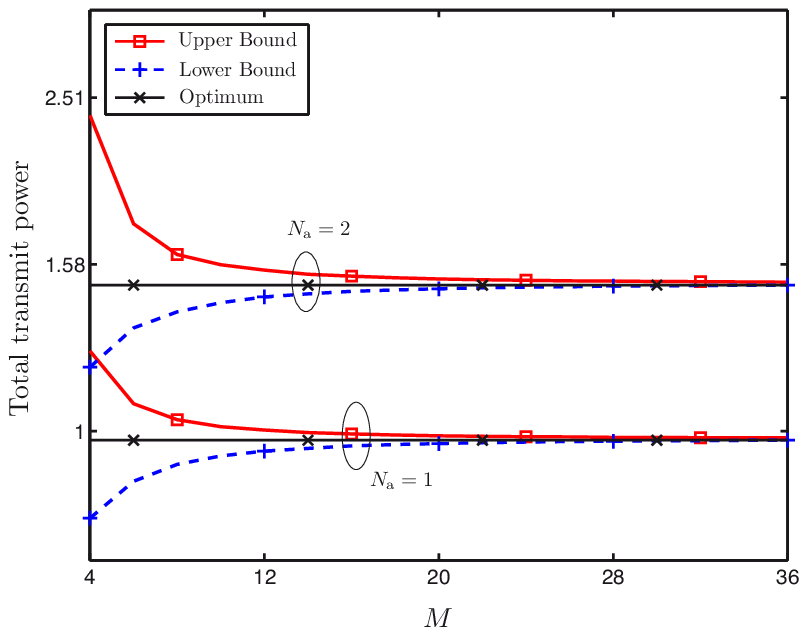}
	\label{fig:wnl_optimal_b}
	}
	
	\caption{{Total transmit power based on asymptotically optimal algorithms as a function of the NUSS (a) and parameter $M$ (b) for networks with eight anchors and one/two agents: (a) $M=$4, 8, 16; (b) NUSS$=$ 0.15.
	}}\label{fig:wnl_optimal}
\end{figure}

% \subsubsection*{Asymptotically optimal algorithms}
We next consider the case with network parameter uncertainty and compare the solutions of the asymptotically optimal algorithms to the optimal solution for a network with eight anchors and one/two agents. We denote {$\varepsilon = 2\Delta/D$ as the \emph{normalized uncertainty set size} (NUSS),} where the true position of each agent can be anywhere in the circle centered at its nominal position with radius $\Delta$. Thus, the maximum uncertainty in $d_{kj}$ is $\varepsilon D/2$ and in ${\phi}_{kj}$ is  $\arcsin(\varepsilon D/2 d_{kj})$.
% angular uncertainty is equal to $\varepsilon^{\phi}_{kj} = \varepsilon^{\phi}$
% The actual position of each agent lies in an uncertainty circle with center radius $d$
% The maximum angular uncertainty is equal to $\varepsilon^{\phi}_{kj} \equiv \varepsilon^{\phi}$, defined as the \emph{uncertainty set size} (USS), whereas the maximum distance uncertainty is equal to $\varepsilon^{d}_{kj} \equiv \varepsilon^{\phi}\cdot D/2$. 
The required total transmit power as a function of the NUSS and parameter $M$ is shown in Fig.~\ref{fig:wnl_optimal}. First, the required power increases with the NUSS as shown in Fig.~\ref{fig:wnl_optimal_a}.  This is because a larger NUSS translates to a larger range of possible network parameters and consequently a larger worst-case \ac{SPEB}, thus requiring more transmit power to guarantee the localization requirement. Second, Fig.~\ref{fig:wnl_optimal_b} depicts the convergence behaviors of $\PactRu$ and $\PactRl$ as a function of $M$ for the NUSS equal to 0.15. In particular, the solutions of both problems approach the optimal solution as $M$ increases, which agrees with Proposition \ref{pro:power_bounds}. {For example, when $\Na=1$, the gaps between the solutions of $\PactRu$ and the optimal solution are about 30\%, 5\%, and 2\% for $M=4$, $8$, and $16$, respectively.} %Note that, although $\PactRl$ does not guarantee the localization requirements of the agents, it yields a solution closer to the optimal solution than $\PactRu$.

\begin{figure}[t]
	\centering
	% \hspace{-5mm}
	\subfigure[]
	{
	% \psfrag{AAAAAA1}[l][][0.6]{\hspace{-6mm}SOCP}
	% \psfrag{AAAAAA2}[l][][0.6]{\hspace{-6mm}SDP}
	% \psfrag{AAAAAA3}[l][][0.6]{\hspace{-6mm}Optimum}
	% 
	% \psfrag{t3}[l][][0.6]{\hspace{-1mm}$\Na=1$}
	% \psfrag{t2}[l][][0.6]{\hspace{-7mm}$\Na=2$}
	% 
	% \psfrag{xlabel}[c][][0.8]{NUSS}
	% \psfrag{ylabel}[c][][0.8]{Total transmit power}
	\includegraphics[width=\matfigscale\columnwidth]{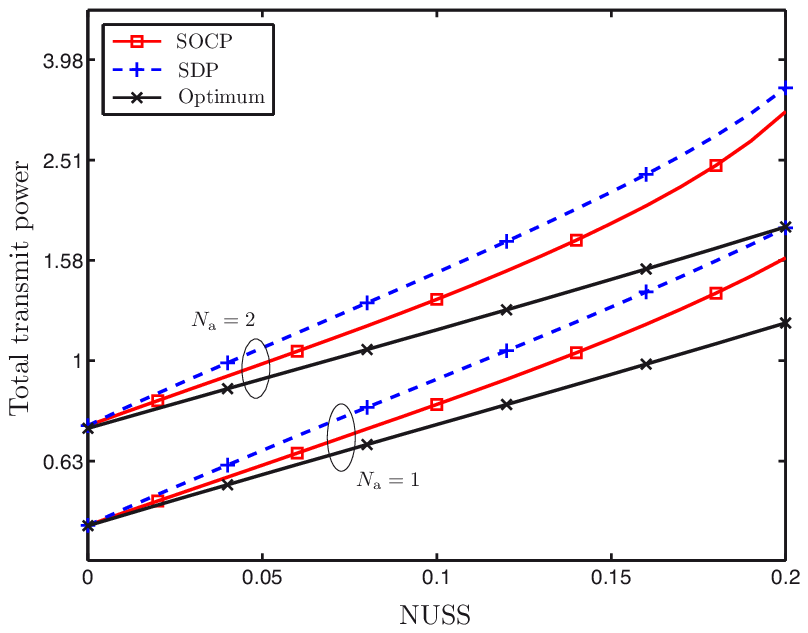}
	\label{fig:wnl_algs_a}
	}
	% \hspace{-5mm}
	\subfigure[]{
	% \psfrag{AAAAAAAA1}[l][][0.6]{\hspace{-8mm}SOCP}
	% \psfrag{AAAAAAAA2}[l][][0.6]{\hspace{-8mm}SDP}
	% \psfrag{AAAAAAAA3}[l][][0.6]{\hspace{-8mm}$M=4$}
	% \psfrag{AAAAAAAA4}[l][][0.6]{\hspace{-8mm}$M=8$}
	% \psfrag{AAAAAAAA5}[l][][0.6]{\hspace{-8mm}Non-Robust}
	% 
	% 
	% \psfrag{t3}[l][][0.7]{\hspace{0mm}$\Na=1$}
	% \psfrag{t2}[l][][0.7]{\hspace{-7mm}$\Na=2$}
	% 
	% \psfrag{xlabel}[c][][0.8]{NUSS}
	% \psfrag{ylabel}[c][][0.8]{Worst-case SPEB}
	\includegraphics[width=\matfigscale\columnwidth]{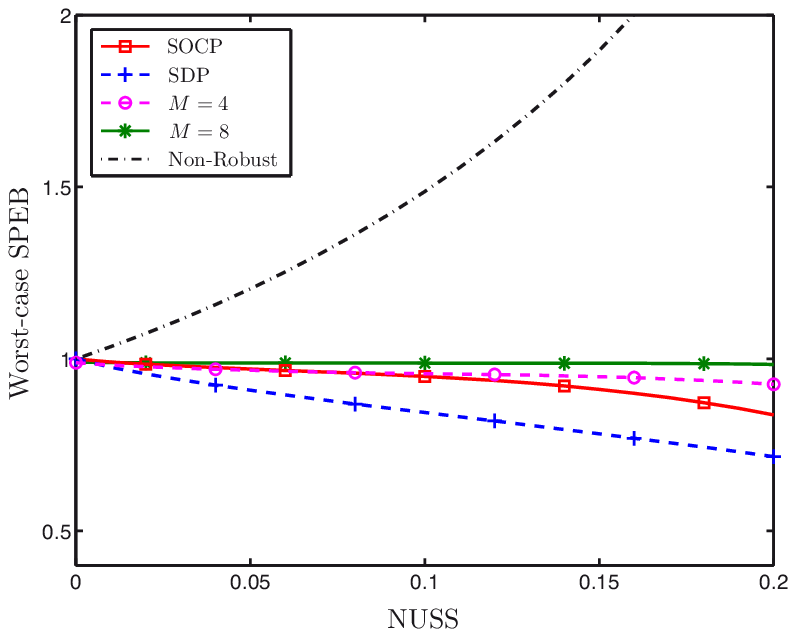}
	\label{fig:wnl_algs_b}
	}
	
	\caption{{Total transmit power (a) and worst-case SPEB (b) as a function of the NUSS: (a) networks with eight anchors and one/two agents; (b) a network with eight anchors and one agent.}}\label{fig:wnl_algs}
\end{figure}

% \subsubsection*{Efficient SOCP-based algorithms}
We then evaluate the performance of the proposed efficient \ac{SOCP}-based algorithm for a network with eight anchors and one/two agents. Figure~\ref{fig:wnl_algs} shows the required total transmit power and the worst-case \ac{SPEB} as a function of the NUSS. First, all the algorithms require more power when the NUSS increases, as shown in Fig.~\ref{fig:wnl_algs_a}. Second, the \ac{SOCP}-based algorithm outperforms the SDP-based algorithm developed in \cite{LiSheZhaWin:J13}, e.g., the gaps between their solutions and the optimal solution are 15\% and 33\%, respectively, for the NUSS equal to 0.15.\footnote{The advantage of the SOCP-based algorithm comes from the fact that it copes with the angular uncertainty altogether, while the SDP-based one copes with such uncertainty individually (cf. (\ref{eq:apd_angular_uncert}) in Appendix \ref{apd:socp_relax_3} and (7) of \cite{LiSheZhaWin:J13}).} Third, the gap between the solution of the SOCP-based algorithm and the optimal solution increases with the NUSS and vanishes when the NUSS is zero. This performance loss is expected since larger uncertainty requires more conservative relaxation. Fourth, as shown in Fig.~\ref{fig:wnl_algs_b}, the worst-case \ac{SPEB} by the nonrobust algorithm increases with the NUSS, significantly violating the localization requirement. This manifests the necessity of robust formulations to guarantee the localization requirement in the presence of parameter uncertainty.

% the robust SOCP scheme yields a close performance to the robust SDP scheme since they employed different relaxation methods; both schemes significantly outperform the non-robust one when the uncertainty size exceeds 0.1, e.g., more than $40\%$ when the uncertainty size is $0.2$. Third, the performance of the non-robust schemes degrades much faster than that of the uniform scheme with the uncertainty size. This can be explained as follows: although not efficient, the uniform scheme is robust in the sense that it is not affected by the parameter uncertainties, while the non-robust schemes that optimize the performance for given parameters is likely to perform poorly in the worst case when the uncertainty size is large.

\subsection{Radar Network Localization}

\begin{figure}[t]
	\centering
	% \hspace{-5mm}
	% \subfigure[]{
	% \psfrag{AAAAAA1}[l][][0.6]{\hspace{-6mm}SOCP}
	% \psfrag{AAAAAA2}[l][][0.6]{\hspace{-6mm}SDP}
	% \psfrag{AAAAAA3}[l][][0.6]{\hspace{-6mm}Uniform}
	% 	
	% \psfrag{t1}[l][][0.6]{\hspace{-12mm} $\Nr = 1$}
	% \psfrag{t2}[l][][0.6]{\hspace{-9mm} $\Nr = 2$}
	% \psfrag{t3}[l][][0.6]{\hspace{-5mm} $\Nr = 4$}
	% \psfrag{t4}[l][][0.6]{\hspace{-10mm} $\Nr = 8$}
	% 
	% \psfrag{xlabel}[c][][0.8]{Number of transmit antennas}	
	% \psfrag{ylabel}[c][][0.8]{Total transmit power}
	% \includegraphics[width=0.95\columnwidth,draft=false]{Figures/mimo_tx_nb.eps}
	% \label{fig:mimo_tx_nb}
	% }
	% \hspace{-5mm}
	% \subfigure[]
	% {
	% \psfrag{AAAAAA1}[l][][0.6]{\hspace{-6mm}SOCP}
	% \psfrag{AAAAAA2}[l][][0.6]{\hspace{-6mm}SDP}
	% \psfrag{AAAAAA3}[l][][0.6]{\hspace{-6mm}Uniform}
	% 
	% \psfrag{t1}[l][][0.6]{\hspace{-10mm} $\Nt = 1,2,4,8$}
	% 
	% \psfrag{xlabel}[c][][0.8]{Number of receive antennas}	
	% \psfrag{ylabel}[c][][0.8]{Total transmit power}
	\includegraphics[width=\matfigscale\columnwidth,draft=false]{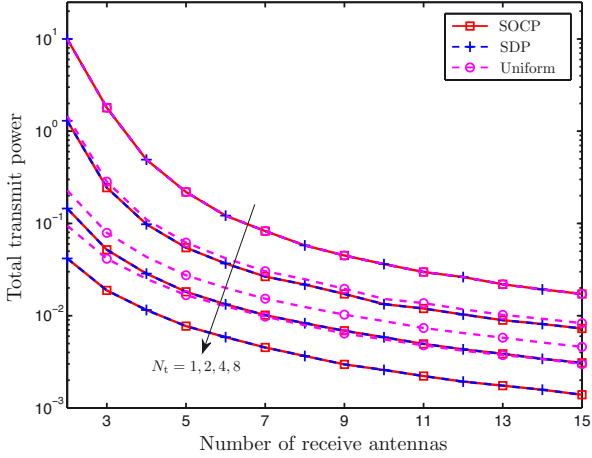}
	% \label{fig:mimo_rx_nb}
	% }

	\caption{Total transmit power as a function of the number of transmit antennas: networks with 1, 2, 4, and 8 receive antennas.
	% ; (b) networks with 1, 2, 4, and 8 transmit antennas. %Both 1-agent and 2-agent networks are considered.
	\label{fig:mtl_perfect}
	}
\end{figure}

% \subsubsection*{Perfect parameter case}
We next compare the performance of the SOCP-based, SDP-based, and uniform power allocation algorithms for \ac{RNL} with perfect network parameters. The required total transmit power as a function of the number of transmit and receive antennas is shown in Fig.~\ref{fig:mtl_perfect}. First, for a given number of receive antennas, the required power decreases with the number of transmit antennas since more degrees of freedom are available for power allocation. On the other hand, for a given number of transmit antennas, the required power decreases with the number of receive antennas since more independent copies of signals are received. Second, the SOCP- and SDP-based algorithms yield identical solutions, significantly outperforming the uniform allocation when there are more than one transmit antenna, e.g., reducing the required power by 30\% when there are four transmit antennas. Third, the performance improvement of the SOCP-based algorithm over the uniform allocation increases with the number of transmit antennas. In particular, there is no improvement for the case with one transmit antenna (i.e., the three curves overlap), while the SOCP-based algorithm reduces over 70\% of the power when there are eight transmit antennas. Fourth, for a given number of transmit antennas, the required power reduction by the SOCP-based algorithm does not depend on the number of receive antennas, e.g., the reduction is 15\%, 35\%, and 50\% for 2, 4, and 8 transmit antennas, respectively, regardless of the number of receive antennas. This implies that different numbers of receive antennas provide the same gain from independent signals for optimal and uniform power allocation.

% only increase the receive signal power and diversity, but not provide the additional advantage for power allocation as the number of the transmit antennas is fixed.

\begin{figure}[t]
	\centering
	% \hspace{-5mm}
	% \subfigure[]
	% {
	% \psfrag{AAAAAA1}[l][][0.6]{\hspace{-6mm}SOCP}
	% \psfrag{AAAAAA2}[l][][0.6]{\hspace{-6mm}SDP}
	% \psfrag{AAAAAA3}[l][][0.6]{\hspace{-6mm}Optimum}
	% \psfrag{t1}[l][][0.6]{\hspace{-6mm}$6\times 2$}
	% \psfrag{t2}[l][][0.6]{\hspace{-7mm}$6\times 6$}
	% \psfrag{t3}[l][][0.6]{\hspace{-2mm}$6\times 10$}
	% 
	% \psfrag{xlabel}[c][][0.8]{NUSS}
	% \psfrag{ylabel}[c][][0.8]{Total transmit power}
	\includegraphics[width=\matfigscale\columnwidth]{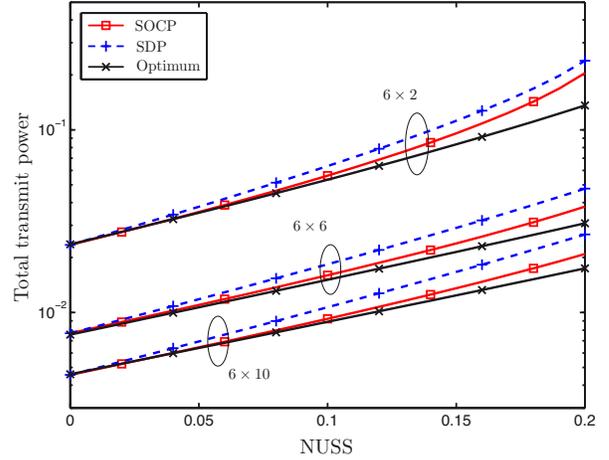}
	% \label{fig:mtl_algs_a}
	% }
	% % \hspace{-5mm}
	% \subfigure[]{
	% \psfrag{AAAAAAAA1}[l][][0.6]{\hspace{-8mm}SOCP}
	% \psfrag{AAAAAAAA2}[l][][0.6]{\hspace{-8mm}SDP}
	% \psfrag{AAAAAAAA3}[l][][0.6]{\hspace{-8mm}$M=4$}
	% \psfrag{AAAAAAAA4}[l][][0.6]{\hspace{-8mm}$M=8$}
	% \psfrag{AAAAAAAA5}[l][][0.6]{\hspace{-8mm}Non-Robust}
	% 
	% \psfrag{xlabel}[c][][0.8]{NUSS}
	% \psfrag{ylabel}[c][][0.8]{Worst-case SPEB}
	% \includegraphics[width=0.95\columnwidth]{Figures/speb_mtl_2}
	% \label{fig:mtl_algs_b}
	% }
	
	\caption{{Total transmit power as a function of the NUSS: networks with $6\times 2$, $6\times 6$, and $6\times 10$ transmit-receive antennas.
	% ; (b) a network with $6\times 2$ transmit-receive antennas.
	}}\label{fig:mtl_algs}
\end{figure}

% \subsubsection*{Efficient SOCP-based algorithm}
We finally compare the solution of the proposed efficient SOCP-based algorithm to the optimal solution for a network with $6\times2$, $6\times6$, and $6\times10$ transmit-receive antenna pairs. Figure~\ref{fig:mtl_algs} shows the required total transmit power as a function of the NUSS, from which we can make similar observations as those for the \ac{WNL} case. First, all the algorithms require more power when the NUSS increases, as shown in Fig.~\ref{fig:mtl_algs}. Second, the SOCP-based algorithm outperforms the SDP-based algorithm, e.g., the gaps between their solutions and the optimal solution are 12\% and 35\%, respectively, for the NUSS equal to 0.15. Third, the gap between the solution of the SOCP-based algorithm to the optimal solution increases with the NUSS and vanishes when the NUSS is zero. %Fourth, Fig.~\ref{fig:mtl_algs_b} shows that the worst-case \ac{SPEB} by the non-robust algorithm increases with the NUSS, significantly violating the localization requirement.

% e.g., over 2 times of the \ac{LAR} when the NUSS is larger than 0.18, implying the necessity of robust methods to guarantee the performance in the presence of parameter uncertainty.

% \subsection{Computation Complexity}
% 
% Finally, in Table \ref{table:CpTime}, we compare the computation time, using the CVX solver, of the SOCP and SDP schemes for different numbers of anchors and agents. First, the computation time for both schemes in the case considered in the simulation is insensitive to the number of anchors, while it increases with the number of agents due to the additional SOC constraints. %\footnote{The computation time should increase linearly with the number of agents when the functions $h_i(\cdot)$'s do not invovle joint power constraints of different agents, in which case $\mathscr{P}^\text{SOCP}$ can be decomposed into $\Na$ parallel programs.} 
% Second, the computation time of the SOCP is significantly less than that of the SDP, e.g., by about $26\%$ and $13\%$ less for the 1-agent and 2-agent cases, respectively. 
% 
% [need to update with the broadcast model]
% 
% In Table ...  [add the case of MTL (it should increase differently with Nt and Nr)]

\section{Conclusion}

In this paper, we established a unifying optimization framework for power allocation in both active and passive localization networks. We first determined two functional properties, i.e., convexity and low rank, of the \ac{SPEB}. Based on these properties, we showed that the power allocation problems can be transformed into SOCPs, which are amenable for efficient optimization. Moreover, we proposed a robust formulation to tackle the uncertainty in network parameters, and then developed both asymptotically optimal and efficient near-optimal algorithms. These algorithms 
% We further employed relaxation methods to develop efficient near-optimal algorithms, which 
retain the SOCP form and naturally reduce to their nonrobust counterparts when the uncertainty vanishes. Our simulation results showed that the proposed power allocation algorithms significantly outperform the uniform allocation algorithm. The results also manifested the necessity of the robust formulation to guarantee the localization requirement in the presence of parameter uncertainty. 
%as the worst-case \ac{SPEB} by non-robust algorithms significantly violates the \ac{LAR}. 
The performance comparison of the asymptotically optimal and efficient near-optimal algorithms provides important insights into robust algorithm design under the performance versus complexity tradeoffs.

% and quantified the performance gaps of efficient algorithms to the optimal solution. 

% the significant performance improvement provided by the proposed power allocation schemes over the uniform allocation, 

% The functional properties of the SPEB reveal the essence of the localization problems, enabling the design and analysis of efficient power allocation schemes.

\appendices

\section{Proof of Lemma \ref{lem:speb_structure}}\label{apd:lem_speb}

\begin{IEEEproof}
Note that $\mathcal{P}(\V{p_k};\V{x}) = \tr\{\V{J}_\text{e}^{-1}(\V{p}_k;\V{x})\}$ is a non-increasing convex function of $\V{J}_\text{e}(\V{p}_k;\V{x}) \in \mathbb{S}_{+}^n$ \cite{BoyVan:B04} and $\V{J}_\text{e}(\V{p}_k;\V{x})$ is a linear function of $\V{x} \succeq \V{0}$. By the convexity property of the composition functions, we can conclude that the \ac{SPEB} is a convex function of $\PW \succeq \V{0}$. 

We next show that the topology matrix $\B\Lambda_k$ can be written as (\ref{eq:decomp_lambda}). Based on (\ref{eq:SPEB}) and (\ref{eq:EFIM}), we can derive the \ac{SPEB} of agent $k$ as (\ref{eq:SPEB_frac}), where the elements of $\B\Lambda_k$ can be written as %. \addtocounter{equation}{1}
\begin{align*}%\label{eq:lambda_matrix}
	[\B\Lambda_k]_{ij} 
		& = 2\sin^2(\phi_{ki}-\phi_{kj}) \\
		& \stackrel{\text{(a)}}{=} 1 - \cos(2\phi_{ki})\cos(2\phi_{kj}) - \sin(2\phi_{ki})\sin(2\phi_{kj})
\end{align*}
where (a) follows from the sum and difference formulas of the trigonometric functions.
% \begin{align*}%\label{eq:lambda_matrix}
% 	\B\Lambda_k = \Matrix{ccccc}{\sin^2(\phi_{k1}-\phi_{k1}) & \sin^2(\phi_{k1}-\phi_{k2}) & \cdots & \sin^2(\phi_{k1}-\phi_{k\Nb}) \\ 
% 	\sin^2(\phi_{k2}-\phi_{k1}) &  \sin^2(\phi_{k2}-\phi_{k2}) & \cdots & \sin^2(\phi_{k2}-\phi_{k\Nb}) \\ 
% 	\vdots  & &\ddots \\
% 	\sin^2(\phi_{k\Nb}-\phi_{k1}) &  \sin^2(\phi_{k\Nb}-\phi_{k2}) & \cdots & \sin^2(\phi_{k\Nb}-\phi_{k\Nb})}
% \end{align*}
% Note that by the sum and difference formulas of the trigonometric functions, we have
% \begin{align*}
% 	\sin^2(\phi_{ki}-\phi_{kj}) 
% 	% = \frac{1-\cos2(\alpha-\beta)}{2} 
% 	= \frac{1 - \cos(2\phi_{ki})\cos(2\phi_{kj}) - \sin(2\phi_{ki})\sin(2\phi_{kj})}{2} \,.
% \end{align*}
After some rearrangement, we can obtain the expression (\ref{eq:decomp_lambda}) for $\B\Lambda_k$. We omit the proof for the \ac{RNL} case since it can be derived in a similar way.
\end{IEEEproof}

\section{Proof of Proposition \ref{pro:upper_lower_bound}} \label{apd:lem_bounds_norm}

We first present the following lemma for the proof of Proposition \ref{pro:upper_lower_bound}.

\begin{lemma}[Finite Projection Bound]\label{lem:bounds_norm}
For any $\V{y}\succeq\V{0}_{\Nb}$, let {$S(\V{y}) := \max_{\{\phi_{kj}\in\Ang^{(i)}_{kj}\}} \big\| \sum_{j\in\NB} y_j\, \V{u}(2\phi_{kj}) \big\|$}, then
\begin{align*}%\label{eq:bounds_lemma_ineq}
	0 \leq \max_{m\in\mathcal{M}}\, \big\{\V{h}_{k,m}^{(i)\,\text{T}}\,\V{y}\big\} & 
	\leq % \max_{\{\phi_{kj}\in\mathcal{S}^{\phi}_{kj}\}} \Big\| \sum_{j\in\NB} y_j\, \V{u}(2\phi_{kj}) \Big\| \notag\\
	% 	& \qquad \quad 
	S(\V{y}) \leq \max_{m\in\mathcal{M}}\, \big\{\V{g}_{k,m}^{(i)\,\text{T}}\,\V{y} \big\} \,. 
\end{align*}
% where $\V{h}_{k,m}, \, \V{g}_{k,m} \in \mathbb{R}^{\Nb}$ with the $j$th elements given by
% \begin{align}\label{eq:bounds_hkm}
% 	[\,\V{h}_{k,m}\,]_j & := \max_{|\epsilon|\leq 2\varepsilon_{kj}^\phi} \,\cos(2\hat{\phi}_{kj} - \vartheta_m + \epsilon ) \\
% 	[\,\V{g}_{k,m}\,]_j & := \max_{|\epsilon|\leq 2\varepsilon_{kj}^\phi+\pi/M} \,\cos(2\hat{\phi}_{kj} - \vartheta_m + \epsilon ) \,.
% 	\label{eq:bounds_gkm}
% \end{align}
\end{lemma}

% \begin{IEEEproof}
% See Appendix \ref{apd:lem_bounds_norm}.
% \end{IEEEproof}

\begin{IEEEproof}
Note that for a given $\V{y}\succeq\V{0}$,
\begin{align}
	% & \hspace{-5mm} \max_{\{\phi_{kj}\in\mathcal{S}^{\phi}_{kj}\}} \Big\| \sum_{j\in\NB} y_j \, \V{u}(2\phi_{kj}) \Big\| \\
	S(\V{y}) & = \max_{\vartheta\in[\,0,\,2\pi)} \max_{\{\phi_{kj}\in\Ang^{(i)}_{kj}\}} \V{u}\T(\vartheta) \cdot\sum_{j\in\NB} y_j \, \V{u}(2\phi_{kj}) \notag \\ 
	% & = \max_{\vartheta\in[\,0,\,2\pi)} \,\sum_{j\in\NB} y_j \max_{\{\phi_{kj}\in\mathcal{S}^{\phi}_{kj}\}} \cos(2\phi_{kj}-\vartheta)\\
	& \stackrel{\text{(a)}}{=} \max_{\vartheta\in[\,0,\,2\pi)} \sum_{j\in\NB} y_j \max_{|\epsilon_{kj}|\leq 2\tilde{\phi}_{kj}^{(i)}} \cos(2\hat{\phi}_{kj}^{(i)}+\epsilon_{kj}-\vartheta)
	% \,. 
\!	\label{eq:S_dev_1}
\end{align}
where (a) follows from ${\phi}_{kj} = \hat{\phi}_{kj}^{(i)} + \epsilon_{kj}/2$ in which $|\epsilon_{kj}|\leq 2\tilde{\phi}_{kj}^{(i)}$ according to (\ref{eq:uncert_phi}). 

For the lower bound, since $\{\vartheta_m:m\in\mathcal{M} \} \subset [\,0,\,2\pi)$, the maximum over $\vartheta\in[\,0,\,2\pi)$ in (\ref{eq:S_dev_1}) can be bounded below by the maximum over $\vartheta\in\{\vartheta_m:m\in\mathcal{M} \}$, and thus
\begin{align*}
	% & \hspace{-5mm} \max_{\{\phi_{kj}\in\mathcal{S}^{\phi}_{kj}\}} \Big\|\sum_{j\in\NB} y_j \, \V{u}(2\phi_{kj}) \Big\|  \\
	S(\V{y}) &\geq \max_{m\in\mathcal{M}} \,\sum_{j\in\NB} y_j \max_{|\epsilon_{kj}|\leq 2\tilde{\phi}_{kj}^{(i)}} \cos(2\hat{\phi}_{kj}^{(i)}+\epsilon_{kj}-\vartheta_m) \\
	& 
	= \max_{m\in\mathcal{M}}\, \big\{\V{h}_{k,m}^{(i)\,\text{T}}\,\V{y}\big\}
	\,. 
\end{align*}
%where the maximum is achieved when $m=m^*$, i.e.,
% \begin{align*}
% 	m^* = \arg\max_{m\in\mathcal{M}}\, \big\{\V{h}_{k,m}^{(i)\,\text{T}}\,\V{y}\big\} \,.
% \end{align*}

{For the upper bound, let $\vartheta^*$ and $\{\epsilon_{kj}^*\}$ be the optimal angles that achieve the maximum in (\ref{eq:S_dev_1}),
and let $m^* \in \mathcal{M}$ such that $\vartheta^* \in [\,\vartheta_{m^*}-{\pi}/{M},\, \vartheta_{m^*}+{\pi}/{M})$. Then, by the definition of $S(\V{y})$, we have
\begin{align*}
	S(\V{y}) \, \V{u}(\vartheta^*) = \sum_{j\in\NB} y_j  \V{u}(2\hat{\phi}_{kj}^{(i)}+\epsilon_{kj}^*)
\end{align*}
and multiplying both sides by $\V{u}\T(\vartheta_{m^*})$ leads to
\begin{align*}
	S(\V{y}) & \cos(\vartheta^*-\vartheta_{m^*}) \\
	& = \sum_{j\in\NB} y_j \cos(2\hat{\phi}_{kj}^{(i)} +\epsilon_{kj}^* -\vartheta_{m^*}) \\
	& \stackrel{\text{(a)}}{\leq} \sum_{j\in\NB} y_j \!\max_{|\epsilon_{kj}|\leq 2\tilde{\phi}_{kj}^{(i)}} \cos(2\hat{\phi}_{kj}^{(i)}+\epsilon_{kj}-\vartheta_{m^*}) \\
	& = \V{h}_{k,m^*}^{(i)\,\text{T}}\,\V{y} 
	\leq \max_{m\in\mathcal{M}}\, \big\{\V{h}_{k,m}^{(i)\,\text{T}}\,\V{y}\big\}
\end{align*}
where (a) follows from $|\epsilon_{kj}^*| \leq  2\tilde{\phi}_{kj}^{(i)}$. On the other hand, since $|\vartheta^*-\vartheta_{m^*}| \leq \pi/M$, we have
\begin{align*}
	S(\V{y}) \, \cos(\vartheta^*-\vartheta_{m^*}) \geq S(\V{y}) \, \cos(\pi/M) 
\end{align*}
which in combination with the above leads to
\begin{align*}
	S(\V{y}) \leq \max_{m\in\mathcal{M}}\, \big\{\V{g}_{k,m}^{(i)\,\text{T}}\,\V{y}\big\} \,.
\end{align*}

Finally, note that}
\begin{align*}
	\sum_{m\in\mathcal{M}} \V{h}_{k,m}^{(i)\,\text{T}}\,\V{y}
	& = \sum_{j\in\NB} y_j \!\! \sum_{m\in\mathcal{M}} \max_{\{|\epsilon_{kj}|\leq 2\tilde{\phi}_{kj}^{(i)}\}}\! \cos(2\hat{\phi}_{kj}^{(i)}+\epsilon_{kj}-\vartheta_m) \\
	& \geq \sum_{j\in\NB} y_j \sum_{m\in\mathcal{M}}  \cos(2\hat{\phi}_{kj}^{(i)}-\vartheta_m) = 0
\end{align*}
which implies that $\max_{m\in\mathcal{M}}\, \big\{\V{h}_{k,m}^{(i)\,\text{T}}\,\V{y}\big\}\geq 0$.
\end{IEEEproof}

We next give the proof of the proposition.

\begin{IEEEproof}
Let $\V{y}=\underline{\V{R}}_k^{(i)}\,\V{x}$. {First, by the definition of $S(\V{y})$, the worst-case \ac{SPEB} in (\ref{eq:worst_speb_frac}) can be rewritten as 
\begin{align}\label{eq:speb_i_sy}
	\mathcal{P}^{(i)}_\text{R}(\mathbf{p}_k;\V{x}) = \frac{4\cdot\V{1}\T\,\V{y}}{(\V{1}\T\,\V{y})^2 - S(\V{y})^2}\,.
\end{align}
Then, by the definition of $\V{h}_{k,m}^{(i)}$ and $\V{g}_{k,m}^{(i)}$, we have that
\begin{align}\label{eq:gkm_ineq}
	% \cos\Big(\frac{\pi}{M}\Big) \cdot
	\cos({\pi}/{M}) \cdot 
	\max_{m\in\mathcal{M}} \big\{\V{g}_{k,m}^{(i)\,\text{T}}\,\V{y} \big\} 
	& = \max_{m\in\mathcal{M}}\, \big\{\V{h}_{k,m}^{(i)\,\text{T}}\,\V{y} \big\} \\ 
	& \stackrel{\text{(a)}}{\leq} S(\V{y}) %\notag \\
	%& 
	\stackrel{\text{(b)}}{=} \V{1}\T\,\V{y} \cdot \sqrt{1 - 1/B(\V{x})}  \notag
\end{align}
where (a) is due to Lemma \ref{lem:bounds_norm} and (b) follows from  (\ref{eq:speb_i_sy}) and the definition of $B(\V{x})$.

Note also that when $M \geq \pi \sqrt{B(\V{x})}$, we have
\begin{align*}
	\cos({\pi}/{M}) 
	>  \sqrt{1 - {\pi^2}/{M^2}}
	\geq \sqrt{1- 1/B(\V{x})} \,.
\end{align*}
Hence, (\ref{eq:gkm_ineq}) implies that
\begin{align*}
	\max_{m\in\mathcal{M}}\, \big\{\V{g}_{k,m}^{(i)\,\text{T}}\,\V{y} \big\} 
	& \leq \frac{1}{\cos(\pi/M)} \,\V{1}\T\,\V{y} \cdot \sqrt{1 - 1/B(\V{x})}  \notag \\
	& < \V{1}\T\,\V{y} \,. 
\end{align*} 

Therefore, by Lemma \ref{lem:bounds_norm}, the denominator of (\ref{eq:speb_i_sy}) can be bounded as
\begin{align*}
	0 < (\V{1}\T\,\V{y})^2 - \Big(\max_{m\in\mathcal{M}}\, \big\{&\V{g}_{k,m}^{(i)\,\text{T}}\,\V{y} \big\}\Big)^2 
	\leq  (\V{1}\T\,\V{y})^2 - S(\V{y})^2 \\
	& \leq (\V{1}\T\,\V{y})^2 - \Big(\max_{m\in\mathcal{M}}\, \big\{\V{h}_{k,m}^{(i)\,\text{T}}\,\V{y} \big\}\Big)^2
\end{align*}
which leads to the claim of the proposition.}
\end{IEEEproof}

\section{Proof of Proposition \ref{pro:gap_upper_lower_bounds}}
\label{apd:pro_gap_bounds}

\begin{IEEEproof}
{Let $\V{y}=\underline{\V{R}}^{(i)}_k\,\V{x}$. To prove the inequality, it is sufficient to show that $\forall\, m \in \mathcal{M}$
\begin{align*}
	\frac{4\cdot\V{1}\T\,\V{y}}{(\V{1}\T\,\V{y})^2 - \big(\V{g}_{k,m}^{(i)\,\text{T}}\,\V{y}\big)^2} 
	% &\\ & \hspace{-20mm} 
	\leq \big(1+{C_{k,M}^{(i)}}\big) \, \frac{4\cdot\V{1}\T\,\V{y}}{(\V{1}\T\,\V{y})^2 - \big(\V{h}_{k,m}^{(i)\,\text{T}}\,\V{y}\big)^2}
\end{align*}
which is equivalent to
\begin{align}\label{eq:derive_speb_bound_1}
	{C_{k,M}^{(i)}} 
	& \geq \frac{\big(\V{g}_{k,m}^{(i)\,\text{T}}\,\V{y}\big)^2 - \big(\V{h}_{k,m}^{(i)\,\text{T}}\,\V{y}\big)^2}{(\V{1}\T\,\V{y})^2 - \big(\V{g}_{k,m}^{(i)\,\text{T}}\,\V{y}\big)^2}\,.
\end{align}

Since $\V{h}_{k,m}^{(i)} = \cos(\pi/M)\,\V{g}_{k,m}^{(i)}$ and $(\V{h}_{k,m}^{(i)\,\text{T}}\,\V{y})^2 \leq S(\V{y})^2 = (\V{1}\T\,\V{y})^2 (1 - 1/B(\V{x}))$, the \ac{RHS} of (\ref{eq:derive_speb_bound_1}) can be bounded above as
\begin{align*}
	\frac{\big(\V{g}_{k,m}^{(i)\,\text{T}}\,\V{y}\big)^2 - \big(\V{h}_{k,m}^{(i)\,\text{T}}\,\V{y}\big)^2}{(\V{1}\T\,\V{y})^2 - \big(\V{g}_{k,m}^{(i)\,\text{T}}\,\V{y}\big)^2} 
	& = \frac{\sin^2({\pi}/{M}) \, \big(\V{h}_{k,m}^{(i)\,\text{T}}\,\V{y}\big)^2}{\cos^2({\pi}/{M})\, (\V{1}\T\,\V{y})^2 - \big(\V{h}_{k,m}^{(i)\,\text{T}}\,\V{y}\big)^2} \notag \\
	& \leq \frac{\sin^2({\pi}/{M}) [\,1-1/B(\V{x})\,]}{1/B(\V{x}) - \sin^2({\pi}/{M})}
\end{align*}
where the denominators are always positive when $M \geq \pi\, \sqrt{B(\V{x})}$, as proven in (\ref{eq:gkm_ineq}). This leads to the inequality (\ref{eq:derive_speb_bound_1}).

Moreover, it is straightforward to verify that ${C_{k,M}^{(i)}}$ is monotonically decreasing with $M$ to zero at the rate of $O(M^{-2})$ as shown in (\ref{eq:C_kmi}).}
\end{IEEEproof}

\section{Proof of Proposition \ref{pro:power_bounds}}
\label{apd:pro_power_bounds}

\begin{IEEEproof}
Since $\underline{\mathcal{P}}^{(i)}_M (\mathbf{p}_k;\V{x}) \leq {\mathcal{P}}^{(i)} (\mathbf{p}_k;\V{x}) \leq \overline{\mathcal{P}}^{(i)}_M (\mathbf{p}_k;\V{x})$, the feasible sets satisfy
\begin{align*}
	& \hspace{-8mm}
	\bigcap_{k\in\NA,\, i\in\mathcal{I}_k} \! \big\{\V{x}: \underline{\mathcal{P}}^{(i)}_M (\mathbf{p}_k;\V{x}) \leq \varrho_k\big\} 
	\\
	& \supseteq \bigcap_{k\in\NA,\, i\in\mathcal{I}_k}\! \big\{\V{x}:  {\mathcal{P}}^{(i)} (\mathbf{p}_k;\V{x}) \leq \varrho_k\big\}	\\
	& \supseteq \bigcap_{k\in\NA,\, i\in\mathcal{I}_k} \! \big\{\V{x}: \overline{\mathcal{P}}^{(i)}_M (\mathbf{p}_k;\V{x}) \leq \varrho_k\big\} 
\end{align*}
and consequently the optimal solutions satisfy $\V{1}\T \underline{\V{x}}^M \leq \V{1}\T \V{x}^* \leq \V{1}\T \overline{\V{x}}^M$. Hence, we have
\begin{align}\label{eq:power_bounds_1}
	0 &\leq {\V{1}\T \overline{\V{x}}^M - \V{1}\T \V{x}^*} \leq {\V{1}\T \overline{\V{x}}^M - \V{1}\T \underline{\V{x}}^M} % \\
	% 	0 &\leq  {\V{1}\T \V{x}^* - \V{1}\T \underline{\V{x}}^M} \leq {\V{1}\T \overline{\V{x}}^M - \V{1}\T \underline{\V{x}}^M} \label{eq:power_bounds_2}
	\,.
\end{align}

Note that
\begin{align*}
	\overline{\mathcal{P}}^{(i)}_M \big(\mathbf{p}_k;(1+C_{M}) \,\underline{\V{x}}^M\big) 
	& \stackrel{\text{(a)}}{=} \frac{\overline{\mathcal{P}}^{(i)}_M (\mathbf{p}_k; \,\underline{\V{x}}^M)}{1+C_{M}} \\
	& \stackrel{\text{(b)}}{\leq} \underline{\mathcal{P}}^{(i)}_M (\mathbf{p}_k;\underline{\V{x}}^M) \leq \varrho_k
\end{align*}
where (a) is due to the power scaling property of the \ac{SPEB} and (b) follows from Proposition \ref{pro:gap_upper_lower_bounds}. Thus, $(1+C_M) \,\underline{\V{x}}^M$ is in the feasible set of $\PactRu$, i.e.,
\begin{align*}
	\big(1+{C_M}\big) \,\underline{\V{x}}^M \in\! \bigcap_{k\in\NA,\,i\in\mathcal{I}_k} \! \big\{\V{x}: \overline{\mathcal{P}}^{(i)}_M(\mathbf{p}_k;\V{x}) \leq \varrho_k\big\} \,.
\end{align*}
On the other hand, since $\overline{\V{x}}^M$ is the optimal solution of $\PactRu$, we have $\V{1}\T \overline{\V{x}}^M \leq (1+ C_M) \V{1}\T \underline{\V{x}}^M$. Therefore, the \ac{RHS} of (\ref{eq:power_bounds_1}) is bounded above as
\begin{align*}
	{\V{1}\T \overline{\V{x}}^M - \V{1}\T \underline{\V{x}}^M} 
	\leq {C}_{M} \cdot {\V{1}\T \underline{\V{x}}^M} \leq {C}_{M} \cdot {\V{1}\T \V{x}^*}	\,.
\end{align*}

The case for the lower bound can be shown similarly, since ${\V{1}\T \V{x}^* - \V{1}\T \underline{\V{x}}^M} \leq {\V{1}\T \overline{\V{x}}^M - \V{1}\T \underline{\V{x}}^M}$.
\end{IEEEproof}

\section{Proof of Proposition \ref{lem:socp_relax_3}} \label{apd:socp_relax_3}

\begin{IEEEproof}
Let $\V{y}=\underline{\V{R}}_k\,\V{x}$. We next derive an upper bound for $(\V{c}_k\T\,\V{y})^2 + (\V{s}_k\T\,\V{y})^2$ over $\{\phi_{kj}\in\Ang_{kj}\}$ in (\ref{eq:worst_speb_frac}), which leads to the upper bound (\ref{eq:socp_relax_3}) for the worst-case \ac{SPEB}. 
% Note that finding a lower bound for the denominator in  can be accomplished by maximizing $(\V{c}_k\T\,\V{y})^2 + (\V{s}_k\T\,\V{y})^2$ over $\{\phi_{kj}\in\Ang_{kj}\}$.
% $\V{y}\T \,\big[\, \V{c}(2{\B\phi}_k)\,\V{c}(2{\B\phi}_k)\T +\V{s}(2{\B\phi}_k)\,\V{s}(2{\B\phi}_k)\T\big]\,\V{y}$. 
% We next derive upper bounds for $(\V{c}_k\T\,\V{y})^2$ and $(\V{s}_k\T\,\V{y})^2$ separately.

% We can express the actual angle based on (\ref{eq:uncert_phi}) as $\phi_{kj}=\hat{\phi}_{kj}+\epsilon_{kj}^\phi$ with $|\epsilon_{kj}^\phi|\leq \varepsilon_{kj}^\phi$. 
{
Since $\V{y}\succeq \V{0}$, we have
\begin{align}
	& \hspace{-1mm} \max_{\{\phi_{kj}\in\Ang_{kj}\}} \big|(\V{c}_k-\hat{\V{c}}_k)\T \V{y}\big| \notag \\
	& \stackrel{\text{(a)}}{=} \max_{\{|\epsilon_{kj}|\leq \tilde{\phi}_{kj}\}} \bigg| \sum_{j\in\NB} y_j \cdot \big(\cos(2\hat\phi_{kj}+2\epsilon_{kj})- \cos(2\hat\phi_{kj}) \big) \bigg| \notag \\
	& \stackrel{\text{(b)}}{\leq} \sum_{j\in\NB} y_j \!\max_{\{|\epsilon_{kj}|\leq \tilde{\phi}_{kj}\}} \! \big| \cos(2\hat\phi_{kj}+2\epsilon_{kj})- \cos(2\hat\phi_{kj}) \big| = \tilde{\V{s}}_k\T \,\V{y} \notag
	% \label{eq:socp3_proof}
\end{align}
where (a) follows from ${\phi}_{kj} = \hat{\phi}_{kj} + \epsilon_{kj}$ in which $|\epsilon_{kj}|\leq \tilde{\phi}_{kj}$ according to (\ref{eq:uncert_phi}) and (b) follows from the triangular inequality. Thus, applying the triangular inequality again gives
\begin{align*}
	% & \hspace{-5mm} 
	\max_{\{\phi_{kj}\in\Ang_{kj}\}} \big|\V{c}_k\T\, \V{y}\big|
	& \leq \big|\hat{\V{c}}_k\T \,\V{y}\big| +  \max_{\{\phi_{kj}\in\Ang_{kj}\}} \big|(\V{c}_k-\hat{\V{c}}_k)\T \V{y}\big| \\
	& \leq \big|\hat{\V{c}}_k\T \,\V{y}\big| + \tilde{\V{s}}\T_k\, \V{y} \\
	& \leq \max_{\{e_1=\pm 1\}} \big\{ \big|(\hat{\V{c}}_k  + e_1 \tilde{\V{s}}_k) \T \V{y}\big| \big\}
\end{align*}
Similarly, we can obtain
\begin{align*}
	\max_{\{\phi_{kj}\in\Ang_{kj}\}} \big|\V{s}_k\T \, \V{y}\big| & \leq \max_{\{e_2=\pm 1\}} \big\{ \big|(\hat{\V{s}}_k + e_2 \tilde{\V{c}}_k) \T \V{y}\big| \big\} \,.
\end{align*}

Combining the above two, we have 
\begin{align}\label{eq:apd_angular_uncert}
	\max_{\{\phi_{kj}\in\Ang_{kj}\}}& 
	\big\{ (\V{c}_k\T\,\V{y})^2 + (\V{s}_k\T\,\V{y})^2 \big\} \\
	& \hspace{-5mm}\leq \max_{\{e_{1},e_2=\pm 1\}}\!\! \Big\{ \big[\,(\hat{\V{c}}_k + e_1 \tilde{\V{s}}_k) \T \V{y}\,\big]^2 \!+  \big[\,(\hat{\V{s}}_k + e_2 \tilde{\V{c}}_k) \T \V{y}\,\big]^2 \Big\}
	\notag
\end{align}
which leads to the upper bound (\ref{eq:socp_relax_3}).}
% Note that from (\ref{eq:socp3_proof}) we can see that relaxation due to the angle error is on the order of $2\sin(\varepsilon_{k}^\phi)$.
\end{IEEEproof}

\section{Proof of Proposition \ref{pro:prior_form}} \label{apd:prior_form}

\begin{IEEEproof}
By eigenvalue decomposition, the \ac{FIM} $\V{J}_0(\V{p}_k) \in\mathbb{S}^2_+$ and the \ac{EFIM} $\V{J}_{kj}\in\mathbb{S}^2_+$ can be written, respectively, as
\begin{align*}
	\V{J}_0(\V{p}_k) & = \mu_k^{(1)} \, \V{J}_\text{r}(\vartheta_k) + \mu_k^{(2)} \, \V{J}_\text{r}(\vartheta_k+\pi/2) \\
	\V{J}_{kj} & = \xi_{kj}^{(1)} \, \V{J}_\text{r}(\widetilde{\phi}_{kj}) +\xi_{kj}^{(2)} \, \, \V{J}_\text{r}(\widetilde{\phi}_{kj}+\pi/2)
\end{align*}
where $\mu_k^{(1)},\, \mu_k^{(2)}, \, \xi_{kj}^{(1)},\, \xi_{kj}^{(2)} \geq 0$ are the eigenvalues and $\vartheta_k,\, \vartheta_k+\pi/2,\, \widetilde{\phi}_{kj},\, \widetilde{\phi}_{kj}+\pi/2$ are the angles of corresponding eigenvectors. Then, the \ac{SPEB} can be written as
\begin{align*}%\label{eq:speb_prior}
	\mathcal{P}(\mathbf{p}_k;\V{x}) 
	%& = \frac{2 \sum_{j\in\NB} \xi_{kj}P_{kj}}{\sum_{j\in\NB,\,j'\in\NB} \xi_{kj}P_{kj} \xi_{kj'}P_{kj'} \sin^2(\phi_{kj}-\phi_{kj'})} \nonumber \\ 
	& = \frac{4\cdot\V{1}\T\,\widetilde{\V{R}}_k\, \widetilde{\V{x}}}{ \big(\V{1}\T\,\widetilde{\V{R}}_k\, \widetilde{\V{x}}\big)^2 - \Big\| \big[\,\V{c}(2\widetilde{\B\phi}_k) ~~\V{s}(2\widetilde{\B\phi}_k)\,\big]\T\,\widetilde{\V{R}}_k\, \widetilde{\V{x}}\Big\|^2}
\end{align*}
where 
\begin{align*}
	\widetilde{\V{x}} &= \big[\,\V{x}\T~~~\V{x}\T~~~1~~~1\,\big]\T \\
	\widetilde{\B\phi}_k &= \Big[ \,\widetilde{\phi}_{k1} ~~~\cdots~~~\widetilde{\phi}_{k\Nb}~~~
	%\widetilde{\phi}_{k1}+\frac{\pi}{2}~~~\cdots~~~\widetilde{\phi}_{k\Nb}+\frac{\pi}{2} 
	\\
	% & \hspace{52.8mm} ~~~\vartheta_k~~~\vartheta_{k}+\frac{\pi}{2}\,\Big]\T \\
	& \qquad 	\widetilde{\phi}_{k1}+\frac{\pi}{2}~~~\cdots~~~\widetilde{\phi}_{k\Nb}+\frac{\pi}{2} ~~~\vartheta_k~~~\vartheta_{k}+\frac{\pi}{2}\,\Big]\T \\
% \end{align*}
% $\widetilde{\V{x}} = \big[\,\V{x}\T~~~\V{x}\T~~~1~~~1\,\big]\T$, $\widetilde{\B\phi}_k = \big[ \,\widetilde{\phi}_{k1} ~~~\cdots~~~\widetilde{\phi}_{k\Nb}~~~
% \widetilde{\phi}_{k1}+{\pi}/{2}~~~\cdots~~~\widetilde{\phi}_{k\Nb}+{\pi}/{2} ~~~\vartheta_k~~~\vartheta_{k}+{\pi}/{2}\,\big]\T$, and 
% \begin{align*}
	\widetilde{\V{R}}_k & = \text{diag} \Big\{{\xi}_{k1}^{(1)},\,\ldots,\,{\xi}_{k\Nb}^{(1)},\,{\xi}_{k1}^{(2)},\,\ldots,{\xi}_{k\Nb}^{(2)},\,\mu_k^{(1)},\mu_k^{(2)}\Big\} \,.
	%\\
	% \widetilde{\B\Lambda}_k &= \V{1}\,\V{1}\T- \V{c}(2\widetilde{\B\phi}_k)\, \V{c}(2\widetilde{\B\phi}_k)\T - \V{s}(2\widetilde{\B\phi}_k)\, \V{s}(2\widetilde{\B\phi}_k)\T
\end{align*}
Therefore, $\mathcal{P}(\V{p}_k;\V{x})\leq \varrho_k$ can be converted to an \ac{SOC} form after some algebra. The robust case can be proved in an analogous way.
\end{IEEEproof}

\section*{Acknowledgment}\label{Sec:Achn}

The authors gratefully acknowledge G.~J.~Foschini, L.~A.~Shepp, and Z.-Q.~Luo for their insightful discussion. The authors would also like to thank S.~Mazuelas, L.~Ruan, and T.~Wang for their careful reviewing and valuable suggestion.

\bibliographystyle{IEEEtran}
% \bibliography{IEEEabrv,StringDefinitions,BiblioCV,WGroup,temp_bib}


% \newpage

\vspace*{-0.9cm}

\begin{IEEEbiography}
[{\includegraphics[width=1in, height=1.25in, clip, keepaspectratio]{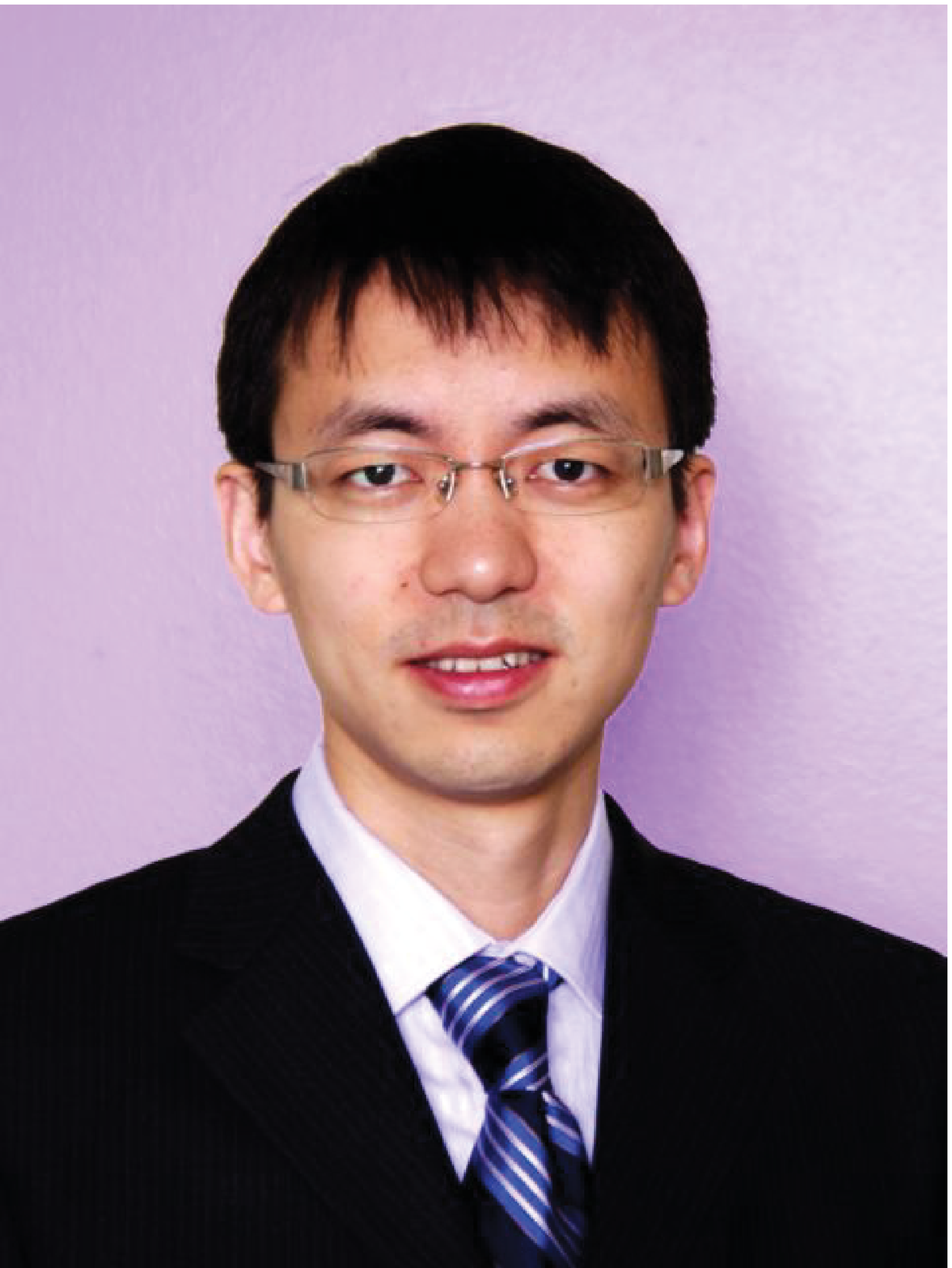}}]{Yuan Shen} (S'05) received the B.S.\ degree (\emph{with highest honor}) in electrical engineering from Tsinghua University, China, in 2005, and the S.M.\ degree in electrical engineering from %MIT
the Massachusetts Institute of Technology (MIT) 
in 2008, and is currently pursuing the Ph.D.\ degree in electrical engineering at MIT.

Since 2005, he has been a research assistant with Wireless Communications and Network Science Laboratory at MIT. 
He was with the Wireless Communications Laboratory at The Chinese University of Hong Kong in summer 2010, 
the Hewlett-Packard Labs in winter 2009,
and the Corporate R\&D of Qualcomm Inc.~in summer 2008. 
%, and the Intelligent Sensing Laboratory at Tsinghua University from 2003 to 2005. 
% His research interests include statistical inference, network science, communication theory, and information theory. 
His current research focuses on network localization and navigation, network inference techniques, wireless secrecy techniques, and localization network optimization.

Mr.~Shen served as a TPC member for the IEEE Globecom (2010--2013), ICC  (2010--2014), WCNC (2009--2014), ICUWB (2011--2013), and ICCC (2012). He was a recipient of the Marconi Society Paul Baran Young Scholar Award (2010), the MIT EECS Ernst A.~Guillemin Best S.M.~Thesis Award (2008), the Qualcomm Roberto Padovani Scholarship (2008), and the MIT Walter A.~Rosenblith Presidential Fellowship (2005). His papers received the IEEE Communications Society Fred W.~Ellersick Prize (2012) and three Best Paper Awards from the IEEE Globecom (2011), the IEEE ICUWB (2011), and the IEEE WCNC (2007).
\end{IEEEbiography}

\vspace*{-1cm}
\begin{IEEEbiography}
[{\includegraphics[width=1in, height=1.25in, clip, keepaspectratio]{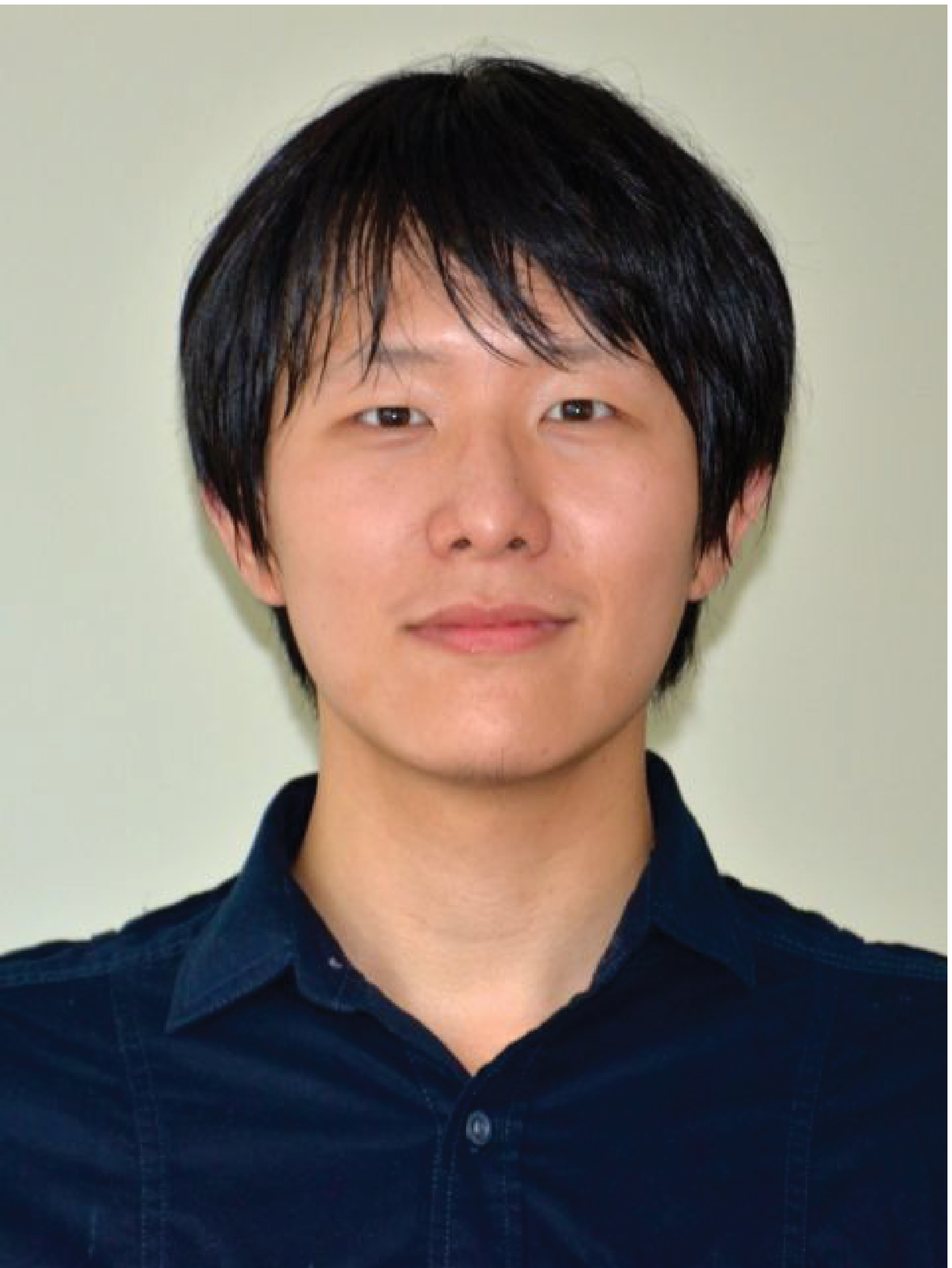}}]{Wenhan Dai} (S'12) received the B.S.\ degrees in electrical engineering and in mathematics from Tsinghua University, Beijing, China, in 2011, and is currently pursuing the master's degree in aeronautics and astronautics at the Massachusetts Institute of Technology (MIT), Cambridge, MA, USA.

Since 2011, he has been a research assistant with Wireless Communications and Network Science Laboratory, MIT. His research interests include communication theory, stochastic optimization, and their application to wireless communication and network localization. His current research focuses on resource allocation for network localization, cooperative network operation,  and ultra-wide bandwidth communications. 

He served as a reviewer for {\scshape IEEE Transactions on Wireless Communications} and {\scshape IEEE Journal on Selected Areas in Communications}. He received the academic excellence scholarships from 2008 to 2010 and the Outstanding Thesis Award in 2011 from Tsinghua University.

% received the B.S.\ degrees in electrical engineering and in mathematics from Tsinghua University, China, in 2011. Since 2011, he has been with Wireless Communications and Network Science Laboratory at MIT, where he is now a master student. His research interest include communication theory, stochastic optimization, and their application to wireless communication and network localization.
\end{IEEEbiography}

\vspace*{-1.6cm}

\begin{IEEEbiography}
	[{\includegraphics[width=1in,height=1.25in,clip,keepaspectratio]{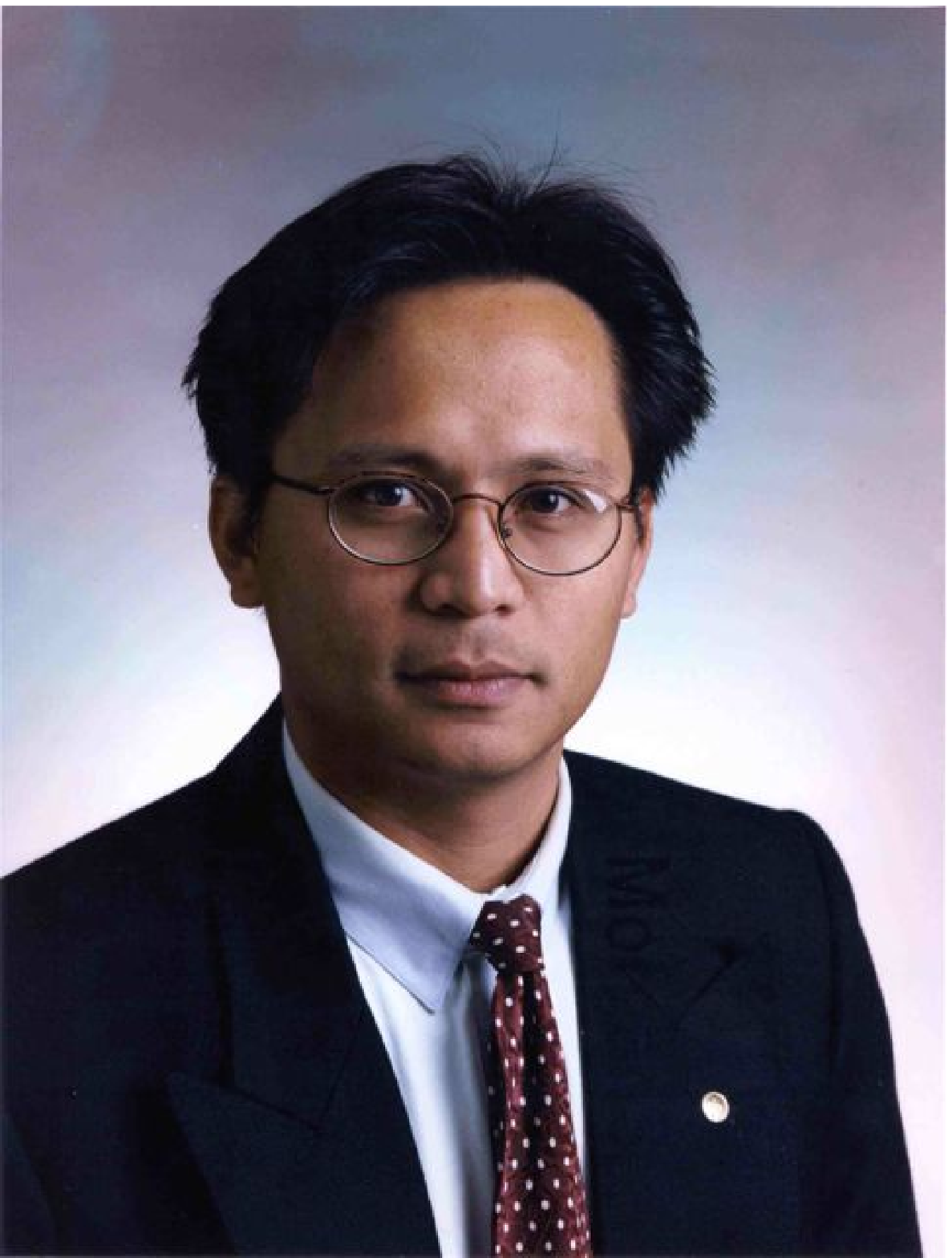}}]
	{Moe Z.~Win}
(S'85-M'87-SM'97-F'04) 
received 
	both the Ph.D.\ in 
		{Electrical Engineering}
	and the M.S.\ in 
		{Applied Mathematics}
		as a Presidential Fellow at the University of Southern California (USC) in 1998.
He received the M.S.\ in Electrical Engineering from USC in 1989 
	and the B.S.\ ({\em magna cum laude}) in Electrical Engineering from Texas A\&M University in 1987.

%%%%%%%%%%%%%%%%%%%%%%%%%%%%%%%%%%%%%%%%%%%%%%%%%%%%%%%%%%%%%%%%%%
%
%	E X P E R I E N C E
%
%%%%%%%%%%%%%%%%%%%%%%%%%%%%%%%%%%%%%%%%%%%%%%%%%%%%%%%%%%%%%%%%%%
He is a Professor at the Massachusetts Institute of Technology (MIT)
and the founding director of the Wireless Communication and Network Sciences Laboratory. 
%Dr.\ Win is a Professor at the Laboratory for Information
%\& Decision Systems (LIDS), Massachusetts Institute of Technology (MIT).
%%Dr.\ Win joined the Laboratory for Information and Decision Systems (LIDS)
%%%and the Department of Aeronautics and Astronautics,
%%at the Massachusetts Institute of Technology in 2002, where he is an Associate Professor.
Prior to joining MIT, he was with AT\&T Research Laboratories for five years 
and with the Jet Propulsion Laboratory for seven years.
%%%%%%%%%%%%%%%%%%%%%%%%%%%%%%%%%%%%%%%%%%%%%%%%%%%%%%%%%%%%%%%%%%
%
%	R E S E A R C H	I N T E R E S T S
%
%%%%%%%%%%%%%%%%%%%%%%%%%%%%%%%%%%%%%%%%%%%%%%%%%%%%%%%%%%%%%%%%%%
His research encompasses fundamental theories, algorithm design, and
experimentation for a broad range of real-world problems.
%His research encompasses developing fundamental theories, designing algorithms, and
%conducting experimentation for a broad range of real-world problems.
%His main research interests are the applications of mathematical and
%statistical theories to communication, detection, and estimation problems.
His current research topics include
%	location-aware networks, 
%	intrinsically secure wireless networks,
%	aggregate interference in heterogeneous networks,
	network localization and navigation, 
	network interference exploitation, 
	intrinsic wireless secrecy,
	adaptive diversity techniques,
		{and ultra-wide bandwidth systems.}
		
%%%%%%%%%%%%%%%%%%%%%%%%%%%%%%%%%%%%%%%%%%%%%%%%%%%%%%%%%%%%%%%%%%
%
%	M A J O R		R E C O G N I T I O N S	
%
%%%%%%%%%%%%%%%%%%%%%%%%%%%%%%%%%%%%%%%%%%%%%%%%%%%%%%%%%%%%%%%%%%
Professor Win is 
	an elected Fellow of the AAAS, the IEEE, and the IET, 
	and was an IEEE Distinguished Lecturer.
%	elected Fellow of the IEEE, cited for ``contributions to wideband wireless transmission.''
He was honored with two IEEE Technical Field Awards: 
	the IEEE Kiyo Tomiyasu Award (2011) 
		{and}
	the IEEE Eric E. Sumner Award 
	{(2006, jointly with R.\ A.\ Scholtz).}
	%%%%%%%%%%%%%%%%%%%%%%%%%%%%%%%%%%%%%%%%%%%%%%%%%%%%%%%%%%%%%%%%%%
%
%	P A P E R		A W A R D S	
%
%%%%%%%%%%%%%%%%%%%%%%%%%%%%%%%%%%%%%%%%%%%%%%%%%%%%%%%%%%%%%%%%%%
Together with students and colleagues, his papers have received numerous awards, including
	{the IEEE Communications Society's Stephen O.\ Rice Prize (2012),
			the IEEE Aerospace and Electronic Systems Society's M.\ Barry Carlton Award (2011),
			the IEEE Communications Society's Guglielmo Marconi Prize Paper Award (2008),
    			and the IEEE Antennas and Propagation Society's Sergei A.\ Schelkunoff Transactions Prize Paper Award (2003).}
%%%%%%%%%%%%%%%%%%%%%%%%%%%%%%%%%%%%%%%%%%%%%%%%%%%%%%%%%%%%%%%%%%
%
%	F E L L O W S H I P S
%
%%%%%%%%%%%%%%%%%%%%%%%%%%%%%%%%%%%%%%%%%%%%%%%%%%%%%%%%%%%%%%%%%%
%Highlights of his international research and education collaborations include
%Highlights of his worldwide educational and research collaborations are
Highlights of his international scholarly initiatives are
	the Copernicus Fellowship (2011),
	the Royal Academy of Engineering Distinguished Visiting Fellowship (2009),
%	Institute of Advanced Study Natural Sciences and Technology Fellowship (2004),
	and
	the Fulbright
		{Fellowship (2004).}
%%%%%%%%%%%%%%%%%%%%%%%%%%%%%%%%%%%%%%%%%%%%%%%%%%%%%%%%%%%%%%%%%%
%
%	O T H E R		R E C O G N I T I O N S
%
%%%%%%%%%%%%%%%%%%%%%%%%%%%%%%%%%%%%%%%%%%%%%%%%%%%%%%%%%%%%%%%%%%
Other recognitions include
	{the International Prize for Communications Cristoforo Colombo (2013),
				the {\it Laurea Honoris Causa} from the University of Ferrara (2008),	
				the Technical Recognition Award of the IEEE ComSoc Radio Communications Committee (2008),
        				and the U.S. Presidential Early Career Award for Scientists and Engineers (2004).}	

%%%%%%%%%%%%%%%%%%%%%%%%%%%%%%%%%%%%%%%%%%%%%%%%%%%%%%%%%%%%%%%%%%
%
%	E L E C T E D		P O S I T I O N S
%
%%%%%%%%%%%%%%%%%%%%%%%%%%%%%%%%%%%%%%%%%%%%%%%%%%%%%%%%%%%%%%%%%%
Dr.\ Win is an elected Member-at-Large on the IEEE Communications Society Board of Governors (2011--2013).
%Professor Win has been actively involved in various aspects of IEEE. 
%His active involvement in IEEE includes 
He was
    the Chair (2004--2006) and Secretary (2002--2004) for
        the Radio Communications Committee of the IEEE Communications Society.
%%%%%%%%%%%%%%%%%%%%%%%%%%%%%%%%%%%%%%%%%%%%%%%%%%%%%%%%%%%%%%%%%%
%
%	C O N F E R E N C E S
%
%%%%%%%%%%%%%%%%%%%%%%%%%%%%%%%%%%%%%%%%%%%%%%%%%%%%%%%%%%%%%%%%%%
%Professor Win has been actively involved in organizing and chairing
%%sessions, and has served as a member of the Technical Program Committee in
%a number of international conferences. 
{Over the last decade, he has organized and chaired numerous international conferences.} 		
%%%%%%%%%%%%%%%%%%%%%%%%%%%%%%%%%%%%%%%%%%%%%%%%%%%%%%%%%%%%%%%%%%
%
%	E D I T O R I A L S
%
%%%%%%%%%%%%%%%%%%%%%%%%%%%%%%%%%%%%%%%%%%%%%%%%%%%%%%%%%%%%%%%%%%
He is currently
	an Editor-at-Large for the 
	{\scshape IEEE Wireless Communications Letters}
and is serving on 
	the Editorial Advisory Board for the 
	{\scshape IEEE Transactions on Wireless Communications}.
\end{IEEEbiography}

\end{document}